\documentclass[a4paper,11pt]{article}
\usepackage{jcappub} 
\usepackage{lineno}

\usepackage{graphicx}

\usepackage{xcolor}
\usepackage{longtable}
\usepackage{subfigure}
\usepackage{tikz}
\usepackage{csquotes}
\usepackage{appendix}

\usepackage{tabularx}
\usepackage{amssymb}
\usepackage{orcidlink}

\usepackage{amsmath}

\usetikzlibrary{decorations.pathreplacing,positioning}

\usepackage{rotating}

\usepackage[table]{xcolor}

\title{Benchmarking the expected performance of Bayesian parameter inference on Kilonovae}

\author[a]{Thomas Hussenot-Desenonges \orcidlink{0009-0009-2434-432X}}
\author[b]{Marion Pillas \orcidlink{0000-0003-3224-2146}}
\author[a,c]{Sarah Antier \orcidlink{0000-0002-7686-3334}}
\author[a]{Patrice Hello}
\author[d,e]{Peter T. H. Pang \orcidlink{0000-0001-7041-3239}}
\affiliation[a]{Universit\'e Paris-Saclay, CNRS/IN2P3, IJCLab, F-91405 Orsay, France}
\affiliation[b]{Institut d'Astrophysique de Paris, Sorbonne Université, 98 bd Arago, 75014, Paris, France}
\affiliation[c]{Observatoire de la Côte d'Azur, Nice, France}
\affiliation[d]{Nikhef, Science Park 105, 1098 XG Amsterdam, The Netherlands}
\affiliation[e]{Institute for Gravitational and Subatomic Physics (GRASP), Utrecht University, Princetonplein 1, 3584 CC Utrecht, The Netherlands}

\emailAdd{thomhus+astro@gmail.com}

\abstract{In the study of optical transients, parameter inference is a part of the process of extracting physical information, i.e. constraints on the source's characteristics, by comparing the observed lightcurves to the predictions of different astrophysical models and finding the model and parameter combination that make the closest match. 
In the developing field of the study of kilonovae (KNe), systematic uncertainties in modeling are still very large. Current-day models still struggle to fit adequately the whole multi-wavelength dataset of the AT2017gfo kilonova, associated to the Binary Neutron Star (BNS) merger GW170817, 
and thus obtain wide posterior distributions on the inferred source parameters.

After reviewing the many sources of uncertainty embedded in KN analyses and the process of Bayesian parameter inference, we discuss how to quantify the performance of parameter inference analyses using metrics on information gain from prior to posterior. We highlight the systematic error margin hyperparameter $\sigma_{\rm sys}$, which correlates to a model's goodness-of-fit, and exhibit its major influence on Bayesian inference performance metrics.
Using the example of two KN models from the literature with the Nuclear Multi-Messenger Astronomy (\texttt{NMMA}) Bayesian inference framework, we simulate the expected information gains that dedicated KN follow-ups with telescope networks could reasonably reach, for various observational cadences and error margins: we show that the wind ejecta's mass and electron fraction are the easiest KN parameters to constrain.

We finally apply our KN analysis to the dataset of AT2017gfo to show that our information gain predictions match results on real data and discuss the complementarity of multi-messenger approaches.}

\begin{document}
\maketitle
\flushbottom

\section{Introduction}
\label{sec:Intro}

Multi-messenger astronomy is the study of astrophysical phenomena via multiple observational channels: electromagnetic radiation (EM) at different wavelengths, gravitational wave signals (GW), detections of neutrinos or of very high-energy particles. The approach hinges on the redundancy and complementarity of the information extracted from different channels: analyzing astrophysical processes by joining multiple messengers at once will constrain a larger number of parameters than the constraints obtained by single-channel approaches, and some parameters will be constrained by multiple messengers. This for instance helps to degeneracies in highly-dimensional models.

One particular subject of interest in multi-messenger astronomy is the study of compact object mergers involving at least one neutron star, namely Binary Neutron Star (BNS) or Neutron Star-Black Hole (NSBH) mergers. The landmark example of this domain of research is the first coincident detection of GW and EM signals, on August 17th, 2017. Associated to the gravitational wave detection of the BNS GW170817 \citep{LIGOScientific:2017vwq}, were the detections of the Gamma-ray Burst (GRB) GRB 170817A \citep{Goldstein:2017mmi, LIGOScientific:2017zic} and the kilonova AT2017gfo \citep{LIGOScientific:2017ync, Smartt:2017fuw, Villar:2017wcc}; the location of the optical transient allowed the identification of the host galaxy NGC 4993 \citep{Coulter:2017wya} and estimations of its distance \citep{Hjorth:2017yza,Cantiello:2018ffy}.
Ensuing studies of this event have shown the potential of multi-messenger approaches to constrain the neutron star nuclear equation of state (EOS) \citep{Radice:2017lry, Coughlin:2018miv, Coughlin:2018fis, Dietrich:2020efo}, or the expansion rate of the Universe \citep{Hotokezaka:2018dfi, Coughlin_2020, Dietrich:2020efo}.
Other studies also probed the properties of the host galaxy and the environment in which the neutron star binary was formed \citep{Blanchard:2017csd}, and scenarios of the binary formation \citep{Palmese:2017yhz}.\\

In addition to the identification of the host galaxy, the kilonova (KN) \citep{Lattimer:1974slx, Li:1998bw, Metzger:2010sy, Kasen:2017sxr}, is crucial to understand the merger environment. These signatures are transients in the infrared, optical and ultraviolet bands, triggered by the radioactive decay of the products of r-process nucleosynthesis in the neutron-rich matter ejected during and after a BNS or NSBH compact object merger. Here, we restrict this work to the study of kilonovae from binary neutron star mergers.

Numerical relativity simulations (e.g. \cite{Hotokezaka:2012ze, Dietrich:2016fpt, Radice:2018pdn, Nedora:2020hxc}) indicate that the ejected matter in a neutron star merger is not perfectly spherical, but rather composed of different regions with anisotropic geometries.
KN modeling efforts, such as \cite{Metzger:2016pju, Kasen:2017sxr, Bulla:2019muo, Bulla:2022mwo, Nicholl:2021rcr, Breschi:2021tbm, Korobkin:2020spe, Wollaeger:2021qgf, Curtis:2021guz, Pognan:2023qhw, Rose:2026cnm},
generally consider at least two components to the ejecta. First, the matter ejected by tidal disruption of the neutron star outer layers before and during the merger, called the \textit{dynamical ejecta}. The lanthanides produced by r-process are typically concentrated around the neutron-rich equator, leading to a redder emission from equatorial regions compared to the bluer, neutron-poorer poles.
Then, following the merger, more matter is ejected by winds produced by the remnant object, e.g. due to neutrino emission or magnetic effects, the \textit{wind ejecta}, which tends to be lanthanide-free and more isotropic. 

Our example KN models in this work are the \textsc{Bu2019lm} model from \cite{Bulla:2019muo} and the \textsc{Bu2025MLP} from \cite{Koehn:2025zzb}. \textsc{Bu2019lm} predicts the (BNS) KN lightcurves for the following input parameters: the masses of the dynamical ejecta and the wind ejecta, $M_{\rm ej}^{\rm dyn}$ and $M_{\rm ej}^{\rm wind}$, the half-opening angle $\Phi$ of the equatorial lanthanide-rich component, the viewing angle $\iota$ ($\iota = 0$ when viewing the KN from its pole) and the luminosity distance $D_L$; whereas \textsc{Bu2025MLP} additionally considers the ejectas' velocities, $v_{\rm ej}^{\rm dyn}$ and $v_{\rm ej}^{\rm wind}$, and electron fractions, $Y_e^{\rm dyn}$ and $Y_e^{\rm wind}$. We label \textbf{A} and \textbf{A'} respectively each model's parameters' default prior ranges, given in appendix \S~\ref{app:priors}.\\

Several KN-candidate transients had been proposed before (and since) 2017, in association to GRBs\footnote{e.g. short GRBs 050709 \citep{Jin:2016pnm}, 130603B \citep{Berger:2013wna, Tanvir:2013pia} or long GRBs 211211A \citep{Rastinejad:2022zbg, Troja:2022yya, Kunert:2023vqd}, 230307A \citep{JWST:2023jqa} (see compilation and discussion of KN candidates in \cite{Gompertz:2017mbv, Rastinejad:2024zuk})}; but AT2017gfo was the first observation of a kilonova in coincidence with a GW detection, and had a consistent dataset in the full optical and near infrared ranges. 

Therefore we will use AT2017gfo (and AT2017gfo-like simulated kilonovae) as reference examples in this work. The AT2017gfo transient was observed many times over the span of two weeks in ultraviolet, optical and infrared energy bands. 
These multi-band observations can be used to extract physical information on the ejecta and its progenitor. To that end, the AT2017gfo lightcurve studied in this work is that of \cite{Coughlin:2018miv} based on the compiled dataset of \cite{Villar:2017wcc}, which we corrected from Milky Way extinction in \S~\ref{app:LC}. We refer to the studied dataset as \textbf{AT2017gfo}.\\

Given a model and a set of observed data, parameter inference is the process of evaluating which combinations of the model's parameters produce the lightcurves that match the dataset best. In Bayesian analysis frameworks such as the Nuclear Multi-Messenger Astronomy (\texttt{NMMA}) \citep{Pang:2022rzc, Rose:2026cnm}, \texttt{bajes} \citep{Breschi:2024qlc} or \texttt{REDBACK} \citep{Sarin:2024tja} pipelines, given a prior probability distribution on the model parameters across their allowed range, parameter inference analyses will return posterior distributions on the parameters, as well as a best-fit configuration (the combination of parameters that produced the modeled lightcurve with the highest likelihood, i.e. the smallest difference with the observed lightcurve). 

\texttt{NMMA}, \texttt{bajes} and \texttt{REDBACK} are able to perform joint analyses of multi-wavelength EM data and GW strains all at once. However, to highlight the current state of KN modeling and inference, as well as to avoid the heavy multi-messenger computing costs, we restrain the scope of this work to the information that can be gained on their own from KN-only analyses\footnote{We plan to extend our approach to multi-messenger inferences in future studies, thanks to recent speed-up optimizations of the \textsc{NMMA} framework~\citep{Rose:2026cnm}.}. It allows us to understand the extent of the information we can extract from this signature alone. In particular, our study will be useful for studying orphan KNe \citep{Zhu_2021}, whose number will increase in the coming years thanks to deep surveys \citep{Mochkovitch:2021prz}. We also ignore the potential optical contribution of a GRB afterglow, since we expect most BNS systems to be seen as off-axis GRB jets and thus particularly faint GRB afterglows, significantly fainter than their KN counterparts.

One goal of this work is to study what physical information can be extracted in KN-only parameter inference, quantifying the uncertainties of the obtained constraints, and showcasing what (hyper-)parameters improve or limit this performance. To that end, we focus in this work on the Bayesian inference framework \texttt{NMMA} \citep{Pang:2022rzc, Rose:2026cnm}. 
It uses the nested sampling approach~\citep{Skilling:2006gxv}, a variation of Monte Carlo sampling used for exploring high-dimensional likelihood landscapes, as implemented in \texttt{pymultinest}~\citep{Buchner:2014nha}. It estimates the Bayesian evidence and, in addition, returns the posterior samples as a by-product.

As an example, given the model's default priors, with a uniform prior on the logarithm of the dynamical ejecta mass, $\log_{10}(M_{\rm ej}^{\rm dyn}/M_\odot)\in [-3,-1]$, a KN-only \texttt{NMMA} analysis of AT2017gfo with \textsc{Bu2019lm} obtains a posterior distribution of the dynamical ejecta mass with a median value and 68\% credible interval of $\log_{10}(M_{\rm ej}^{\rm dyn}/M_\odot)= -2.23^{+0.11}_{-0.13}$, with a best-fit configuration at $\log_{10}(M_{\rm ej}^{\rm dyn}/M_\odot)= -2.24$.\footnote{See more parameter inference results of \textsc{Bu2019lm} AT2017gfo analyses in section~\ref{sub:AT2017gfo}.}
However, this result is only partially compatible with the $\log_{10}(M_{\rm ej}^{\rm dyn}/M_\odot)= -2.15^{+0.27}_{-0.15}$ 68\% credible interval obtained in \cite{Dietrich:2020efo} from GW-only studies of GW170817. 
Similar or worse discrepancies can be found for many KN models, as their AT2017gfo inference results sometimes clash with constraints from other messengers. For instance, the KN-only best-fit in \cite{Breschi:2021tbm} favors a luminosity distance $D_L>50$ Mpc whereas independent distance estimates\footnote{from Surface Brightness Fluctuations of infrared observations.} \citep{Jensen:2021ooi} place NGC 4993 at $40.2 \pm 2.2$ Mpc. 
Discrepancies also exist between KN models, as their predicted lightcurves for a given parameter combination can vary a lot depending on the model (see section~\ref{sub:ModelUnc}). 
An important question is therefore how to reconcile the results of parameter inference between KN models and other messengers when studied separately. 
Thus, we will also discuss in this work how to assess a model's compatibility with observational data and complementarity with other messengers' constraints.\\

In Section~\ref{sec:Uncertainties} we explore the inner workings of one implementation of a parameter inference framework (\texttt{NMMA}, focusing on the \textsc{Bu2019lm} model) and discuss the different sources of uncertainties involved. Then, in Section~\ref{sec:Performance} we study the typical performances of such frameworks to see what scientific output can be expected from future KN observations, and compare these predictions with a reanalysis of AT2017gfo, which we will also use to discuss the complementarity of KN inference with other messengers. 

\section{Exploration of uncertainties in parameter inference of KN properties} 
\label{sec:Uncertainties}
In this section, we discuss all the sources of uncertainty we identified in the process of KN analysis and parameter inference of the ejecta and source properties, from the data acquisition to KN modeling to the multi-physics framework for the Bayesian analysis of the observations. We aim to outline which of these effects have dominant contributions in terms of error uncertainties, and should be properly taken into account for KN analyses (with or without additional messengers) to produce robust and reliable results, 
and quantify coherence and tension with the current models.

\subsection{Sources of uncertainty in observational datasets and in models}

The first step of the study of any astronomical source is the acquisition and processing of observational data to produce a multi-band light curve, which will eventually be compared against the model's predictions. 

We refer to magnitude measurements in the absolute (AB) photometric system $\{m^{j}_{i}(t_i)\}$ and their associated statistical uncertainties $\sigma^{j}_{i}$ across different times $\{t_i\}$ and filters $\{j\}$. We denote as $\sigma_{meas}$ the typical uncertainty contributions in the estimation of observed magnitudes. 

In Appendix~\ref{sub:DataUnc} we discuss the error contributions of the various sources of uncertainties regarding optical data: extraction of the flux from the images, extinction along the line of sight and redshift effect. 

\begin{figure}
    \centering
    \includegraphics[width=0.8\columnwidth]{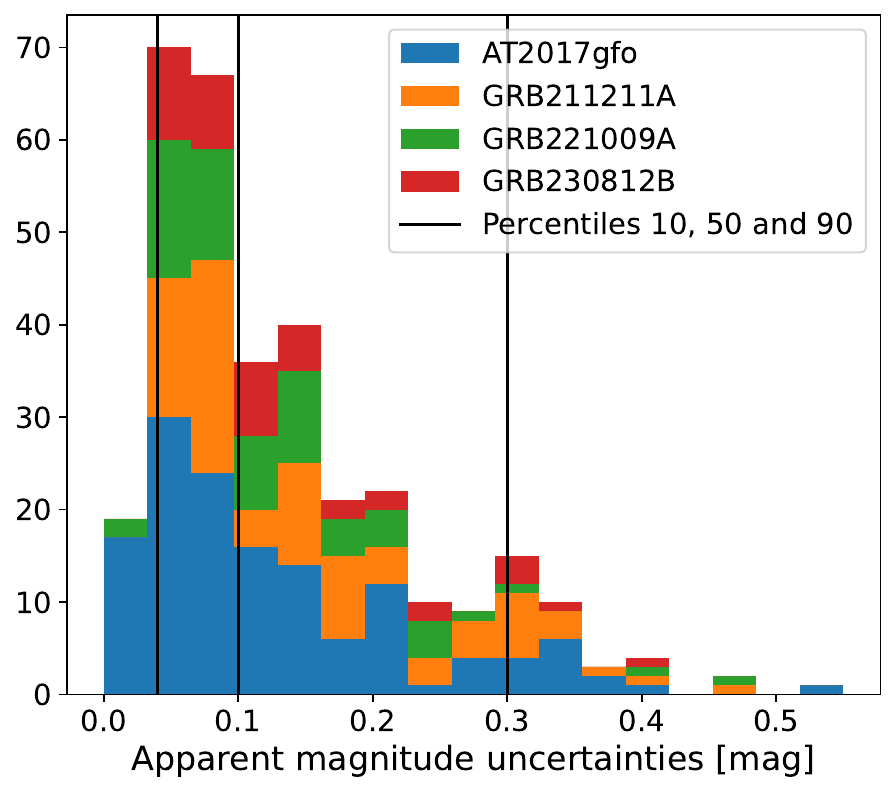}
    \caption{Distribution histogram of the observational magnitudes uncertainties for the \textbf{AT2017gfo} dataset of \cite{Coughlin:2018miv}(in blue), as well as datasets of GRBs 211211A~\citep{Kunert:2023vqd}(in orange), 221009A~\citep{Kann:2023ulv}(in green) and 230812B~\citep{Hussenot-Desenonges:2023oll}(in red). The overall median value is 0.10 mag.}
    \label{fig:histo_obs}
\end{figure}

When all of these effects are accounted for, for the typical range of magnitudes spanned by transients like GRB afterglows, KNe or SNe, the corresponding follow-up observations by multi-telescope collaborations are generally reported in the literature with total observational uncertainties on the order of $\sigma_{meas}\sim$ 0.04-0.3 mag \footnote{We highlight examples of a few well-observed events in Figure~\ref{fig:histo_obs}. The \cite{Rastinejad:2024zuk} catalogue of GRB+KN candidates exhibits a similar error range, with about 80\% of optical datapoints (excluding upper-limits) having $<0.3$ mag uncertainties.}. For example, the magnitude uncertainties of \textbf{AT2017gfo} span a [0.01-0.4] mag range, with most images with $1\sigma$ errors around $\sim 0.1$~mag, as shown in Figure~\ref{fig:histo_obs}.\\

The observed dataset will be compared to the lightcurves predicted by KN models, which come with their own sources of uncertainties, from modeling assumptions to numerical implementation.

We denote the typical uncertainty contributions of different modeling uncertainty factors as $\sigma_{mod}$.

As discussed in more detail in Appendix~\ref{sub:ModelUnc}, KN models are very diverse and subject to large systematic differences based on their construction (dimensionality, assumed ejecta structure, choice of analytical computations or photon transport simulations, etc.) and on their physical assumptions (thermalization and ionization rates of the ejecta due to decay of $r$-process products, description of the opacities of heavy elements, etc.), leading to variability on the order of $\sigma_{mod}\sim$~1-3~mag.
Some computationally expensive KN models are additionally handled via the implementation of surrogate models for quicker computation, adding typical surrogate errors of $\sigma_{mod}\lesssim 0.3$~mag.\\

Having compiled here the uncertainties inherited from upstream processes, specifically from the images and the models, we summarize in Figure~\ref{fig:Chart1} their different contributions.

\begin{figure}
    \hspace*{-0cm}
    \begin{tikzpicture}[
    level 1/.style = {sibling distance = 5cm},
    level 2/.style = {sibling distance = 2cm},
    level 3/.style = {sibling distance = 3cm}
    ]
    \node [draw] {Tightness of KN parameter inference posteriors}
        child {node [draw,align=center]{Total uncertainties\\involved (in mag)}
        child {node [draw,align=center]{Observational data\\uncertainties $\sigma_i$\\ \S \ref{sub:DataUnc}}
        child {node [draw,align=center, below=-4mm] {Photometry\\uncertainties\\\textcolor{blue}{ $\le$ 0.4 mag}}}
        child {node [draw,align=center, below=-4mm] {Extinction\\uncertainties\\\textcolor{blue}{ $\le$ 0.3 mag}}}}
        child [missing]
        child [missing]
        child {node [draw,align=center]{Modeling\\uncertainty $\sigma_{\rm sys}$}
        child [missing] 
        child {node [draw,align=center, below=-6mm] {Lightcurve\\prediction\\uncertainties\\ \S \ref{ssub:ModelIntrinsic}\\\textcolor{blue}{ $\le$ 3 mag}}}
        child {node [draw,align=center, below=-6mm] {Surrogate\\model\\uncertainties\\ \S \ref{ssub:Surrogate}\\\textcolor{blue}{ $\le$ 0.3 mag}}}
        child {node [draw,align=center, below=-6mm] {Model-data\\discrepancy\\ \S \ref{sub:err_sys}}}}
        }
        child {node [draw,align=center, sibling distance = 2cm] {Amount and type\\of data used\\ \S \ref{ssub:Cadence}}
        child [missing]
        child [missing]
        child {node [draw] {Cadence}}
        child {node [draw] {Filters}}};
    \end{tikzpicture}
    \caption{Breakdown of the sources of uncertainties and the sections in which they are treated in this work. In blue, we note the typical bounds for the uncertainties in KN datasets and models.}
    \label{fig:Chart1}
\end{figure}

\subsection{Sources of uncertainty in the parameter inference framework}
\label{sub:err_sys}

Now both observed and modeled lightcurves need to be compared within the parameter inference framework in order to extract information on the physical parameters. We focus on the \texttt{NMMA} framework \citep{Pang:2022rzc, Rose:2026cnm} to showcase this process.

\paragraph{Evaluating model-observation agreement and optimal error margins $\sigma_{\rm sys}$}

In Bayesian inference algorithms, such as \texttt{NMMA}'s nested sampling, for each sampled parameter combination $\vec{\theta}$ (ejecta masses, viewing angle, distance, etc.), the model computes the corresponding lightcurve magnitudes $m^{j, \rm{model}}_{i}(\vec{\theta})$, and then computes the likelihood as follows:

\begin{equation}
\begin{aligned}
    \log\mathcal{L}(\vec{\theta}) = \sum_{ij}- \frac{1}{2}\frac{\left(m^{j}_{i} - m^{j, {\rm model}}_{i}(\vec{\theta})\right)^2}{(\sigma^j_i)^2 + \sigma^2_{\rm sys}} -\frac{1}{2}\log2\pi ((\sigma^j_i)^2 + \sigma^2_{\rm sys}) 
    \label{eq:likelihood}
\end{aligned}
\end{equation}

This implements the systematic uncertainty $\sigma_{\rm sys}$, which is equivalent to allowing the "true" modeled magnitude to follow a Gaussian distribution of width $\sigma_{\rm sys}$ centered around $m^{j, \rm{model}}_{i}(\vec{\theta})$, instead of being exactly equal to $m^{j, \rm{model}}_{i}(\vec{\theta})$. This additional variance, added in quadrature to the observational uncertainties $\sigma_{i}^j$, is introduced in order to take into account the uncertainties of the models, and to consistently explain the dispersion between observed and modeled magnitudes.

Indeed, given an observation $\{m_i^j,\sigma_i^j\}$ and a model $m^{j, \rm{model}}_{i}(\vec{\theta})$ and varying only $\sigma_{\rm sys}$, the likelihood is maximal when $\sigma_{\rm sys}$ corresponds to a reduced chi-squared statistic $\chi^2_r$
\begin{equation}
    \chi^2_r = \frac{1}{N}\sum_{ij}\frac{(m_i^j-m_i^{j,\rm model}(\vec{\theta}))^2}{(\sigma^j_i)^2 + (\sigma_{\rm sys})^2}
\end{equation} of order 1, as would be expected for correctly sized Gaussian distributions.

The value traditionally chosen for previous \texttt{NMMA} KN studies was a conservative fixed $\sigma_{\rm sys}=1$ mag (discussed in \cite{Heinzel:2020qlt}, implemented in \cite{Pang:2022rzc, Kunert:2023vqd, Anand:2023jbz}), thus dominating the uncertainties from the data. We will discuss in the following how to choose well-adapted values of this $\sigma_{\rm sys}$ error margin.

\paragraph{Impact of $\sigma_{\rm sys}$ value}

\begin{figure}
    \centering
    \includegraphics[width=0.92\columnwidth]{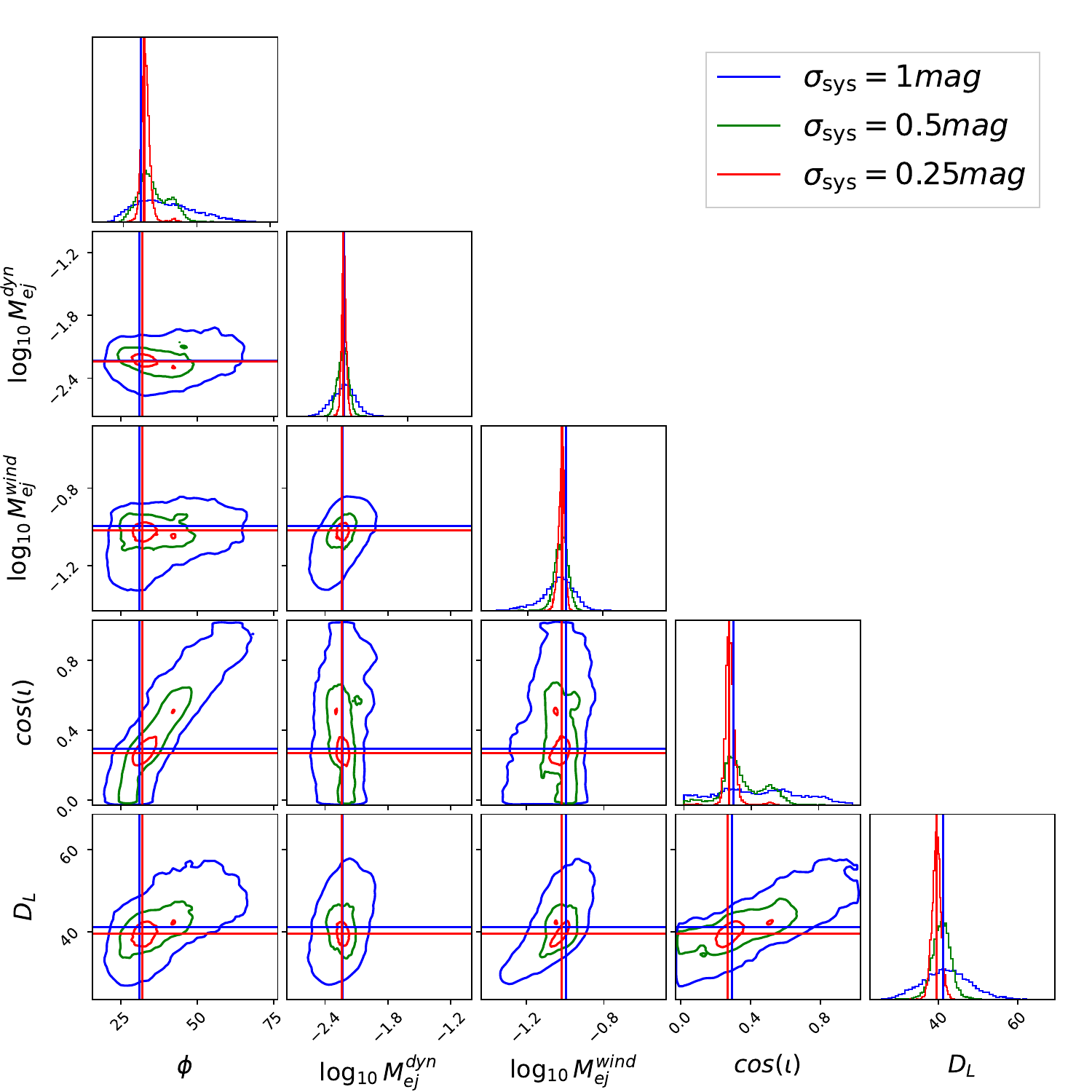}
    \caption{Posterior contours of KN-only (\textsc{Bu2019lm}) analysis of \textbf{AT2017gfo}, with default prior set \textbf{A} (from Table~\ref{tab:Priors}), with different choices of systematic modeling uncertainties. The overplot lines show the position of the best-fit parameters, in blue and red for $\sigma_{\rm sys}=1$ mag and $\sigma_{\rm sys}=0.25$ mag respectively. As a reminder, $\sigma_{\rm sys}$ is the dominant error contribution compared to the observational uncertainties, of order 0.1 mag for \textbf{AT2017gfo}.}
    \label{fig:sigma_sys}
\end{figure}

The bigger the $\sigma_{\rm sys}$ error margin gets, the more parameter combinations fall within its tolerance, so the wider are the posterior distributions. Conversely, smaller values of $\sigma_{\rm sys}$ will make the posterior distribution tighter around the best-fit configuration. For instance, Figure~\ref{fig:sigma_sys} shows the posterior distributions obtained from \textsc{Bu2019lm} analyses of \textbf{AT2017gfo}, only changing the value of $\sigma_{\rm sys}$. We observe that the posteriors have a large size for the biggest value of $\sigma_{\rm sys}$, and they get smaller and smaller as $\sigma_{\rm sys}$ decreases, making a tight peak around the parameter combination that maximizes likelihood (shown as overplot lines). 

Another effect of choosing a fixed modeling uncertainty is that the smaller the value of $\sigma_{\rm sys}$ is, the more constraining power is given to data points with small error bars relative to points with bigger photometric uncertainties, thus the best-fit parameters can vary slightly, as seen in Figure~\ref{fig:sigma_sys}.

Let us demonstrate these effects with a toy model, a simplified case where the measurement uncertainties are all equal (e.g $\sigma_i^j = \sigma_{meas}$)\footnote{One can show that the likelihood is then maximized when $\sigma_{\rm sys}$ verifies exactly: $\frac{1}{N}\sum_{ij}(m_i^j-m_i^{j,\rm model}(\vec{\theta}))^2 = \sigma_{meas}^2 + \sigma_{\rm sys}^2$, thus $\chi^2_r =1$.}.

We define a mock KN event by injecting the \textsc{Bu2019lm} parameters that best fit AT2017gfo\footnote{best-fit obtained in an NMMA analysis with default \textsc{Bu2019lm} prior set \textbf{A} from Table \ref{tab:Priors} and $\sigma_{\rm sys}=1$ mag.}: $\log_{10}(M_{\rm ej}^{\rm dyn}/M_\odot)=-2.24$, $\log_{10}(M_{\rm ej}^{\rm wind}/M_\odot)=-1.00$, $\phi = 31$ degrees, $\cos(\theta)=0.29$ and $D_L =41.2$ Mpc; that we will refer to as \textbf{mock17} parameters.
We simulate the corresponding AT2017gfo-like lightcurves with the \textsc{Bu2019lm} model, taking the datapoints with the same observational times and filters as the \textbf{AT2017gfo} dataset.
We then add a Gaussian noise of variance $\frac{1}{N}\sum_{ij}(m_i^j-m_i^{j,\rm model}(\vec{\theta}))^2 = (0.5~\text{mag})^2$ to all the datapoints, but declare the observational uncertainties to be smaller: $\sigma_{meas} = 0.3$ mag. The resulting \textbf{toymodel} dataset is then analysed to try to recover the injected parameters.

\begin{figure}
    \centering
    \includegraphics[width=0.81\columnwidth]{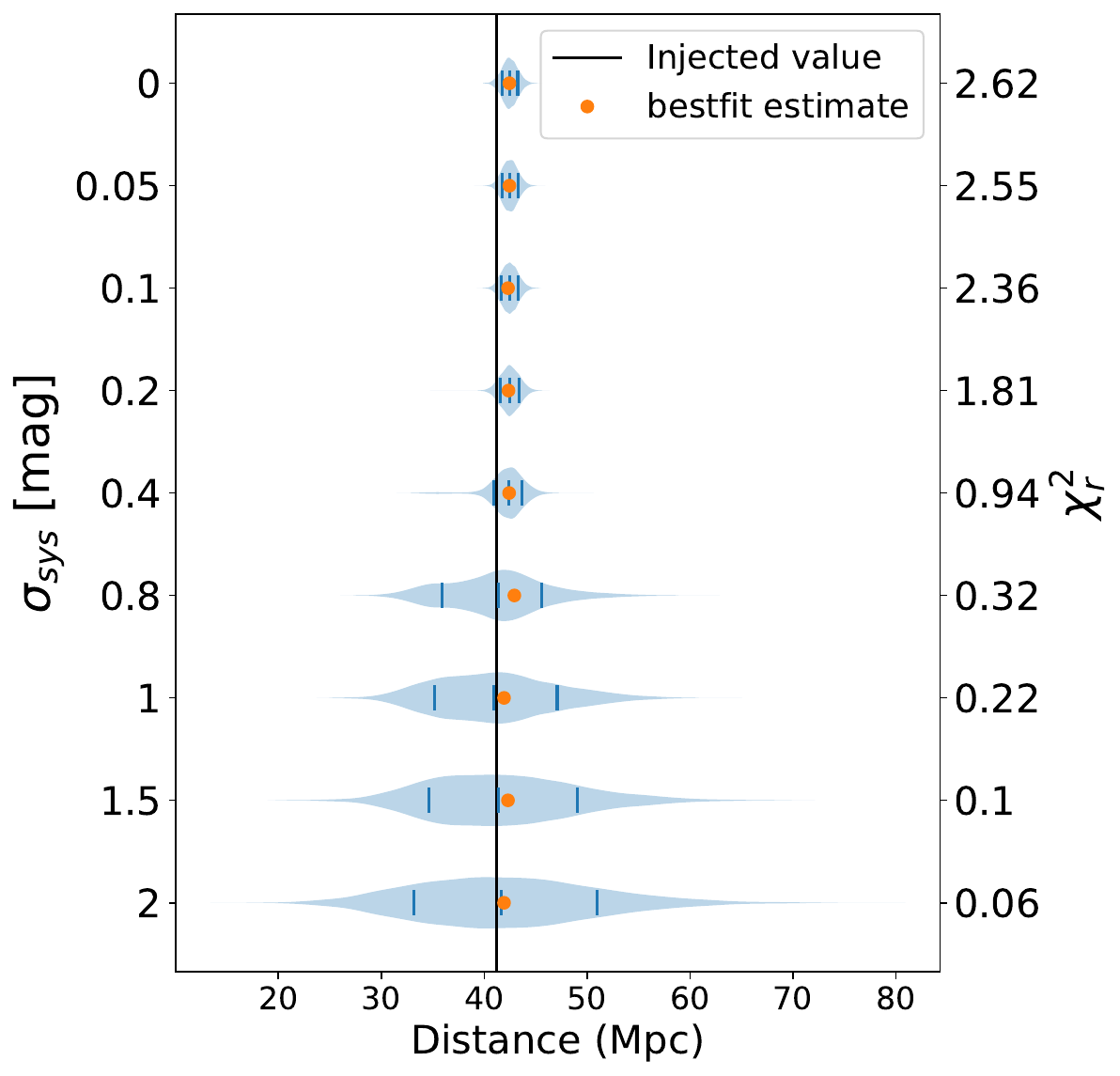}
    \includegraphics[width=0.17\columnwidth, trim = 480 0 0 0, clip]{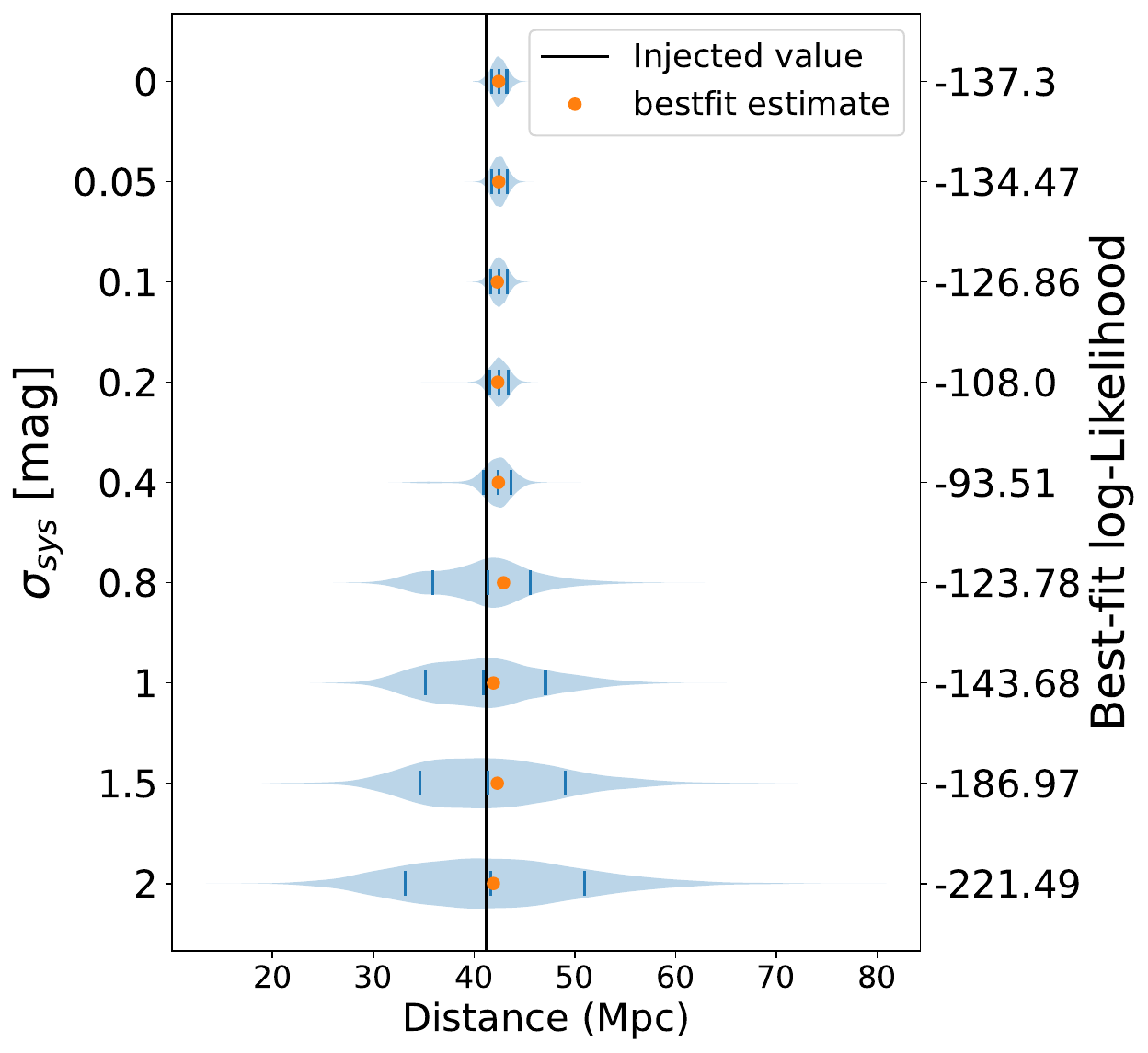}
    \caption{Posterior distribution of the distance, the best-fit log-likelihood $\log \mathcal{L}(\vec{\theta}_{\rm best})$ and $\chi^2_r$ for different fixed values of $\sigma_{\rm sys}$; resulting from \texttt{NMMA} \texttt{Bu2019lm} parameter inferences (default prior set \textbf{A} from Table~\ref{tab:Priors}) of the \textbf{toymodel} lightcurve (containing Gaussian noise). The injected \textbf{mock17} parameters are represented with a vertical black line: $D_L =$ 41.2~Mpc. The configuration with the highest log-likelihood is achieved for $\sigma_{\rm sys}=0.4$ mag, where $\chi^2_r \sim 1$.}
    \label{fig:sigma_sys_perf}
\end{figure}

Since we declared our observational uncertainties to be 0.3 mag, an added error margin $\sigma_{\rm sys} = 0.4$ mag is thus needed to explain the deviation ($0.3^2 + \sigma_{\rm sys}^2 = 0.5^2$). Indeed, as shown in Figure~\ref{fig:sigma_sys_perf}, when analyzing with different fixed values of $\sigma_{\rm sys}$, the best-fit log-likelihood is highest when $\sigma_{\rm sys}$ equals 0.4 mag, with a corresponding $\chi^2_r$ of order 1. More extensively, on one hand, if the $\sigma_{\rm sys}$ is $<$ 0.4 mag, it cannot explain the data-model dispersion (thus $\chi^2_r >1$) and the posterior distribution tightens around the best-fit value, disfavoring the true injected parameters. On the other hand, if $\sigma_{\rm sys} > 0.4$ mag the data fits within the margin ($\chi^2_r <1$) and is compatible with the injected value but the constraints are too loose (posteriors wider than they need to be). A correctly sized $\sigma_{\rm sys}$ error margin is thus paramount to a good balance between precision and accuracy of parameter inference.\\

\paragraph{Optimal $\sigma_{\rm sys}$ value as goodness-of-fit metric}
In the absence of prior knowledge on the optimal systematic uncertainty, $\sigma_{\rm sys}$ can be treated as a free parameter and sampled over during the nested sampling in Bayesian inferences like \texttt{NMMA}. In that case, the resulting posterior of $\sigma_{\rm sys}$ converges around values corresponding to a $\chi^2_r$ of order 1. This estimated $\sigma_{\rm sys}$ can then be interpreted as a goodness-of-fit indicator\footnote{It should be noted that \texttt{NMMA} can nowadays implement a time- and/or filter-dependent systematic error margin~\citep{Jhawar:2024ezm,Rose:2026cnm}, which too can be treated as free parameters and converge towards goodness-of-fit proxies restricted to desired time windows and filters.}: The lower the $\sigma_{\rm sys}$, the better the fit, and vice versa. 

When analyzing one event with different competing models, looking at each one's optimal $\sigma_{\rm sys}$ highlights which model fits the data the closest. For instance, AT2017gfo analyses in \cite{Villar:2017wcc} and \cite{Nicholl:2021rcr} determine their optimal error margins to be about 0.2 mag and 0.4 mag respectively, higher than the observational $\sigma_{\rm meas} \sim 0.1$ mag. 
Indeed, for KN parameter inferences in the literature, $\sigma_{\rm sys}$ usually is the dominant error contribution, and error margins on observational data are thus sub-leading.


Since the modeling biases (\S \ref{ssub:ModelIntrinsic}) and interpolation errors (\S \ref{ssub:Surrogate}) discussed in the appendix are possible error contributions encompassed in the $\sigma_{\rm sys}$ error margin, further studies will help to quantify these modeling uncertainties and try to diminish them, so that a better confidence in the model predictions can be achieved.

\subsection{Cadence, filter, and implementation effects}
\label{sub:Emergent}

After having explained the Bayesian framework and its handling of uncertainty contributions with the $\sigma_{\rm sys}$ hyperparameter, we now wish to quantify the performance of Bayesian inference analyses (both in terms of the width of the estimated parameter posteriors and of the offsets between posterior medians and the correct parameter values) and study the impact of hyperparameters over these metrics.
We will for instance study the impact of observational cadence and filter choice on the size of posterior distributions.\\
To quantify the performance of our parameter inference, we use in this work the Kullback-Leibler (K-~L) divergence \citep{Kullback:1951zyt}.~:
\begin{equation}
    \begin{split}
        D_{KL}(P||\pi) &= \int_{\vec{\theta} \in \text{Parameter Space}} d\vec{\theta} P(\vec{\theta}) \log\frac{P(\vec{\theta})}{\pi(\vec{\theta})}
    \end{split}  
    \label{eq:KL}
\end{equation}
which quantifies the information gain of the posterior $P$ as compared to the prior $\pi$.

We compute it using the approximation: 
\begin{equation}
\begin{aligned}
    D_{KL}(P||\pi) &\simeq  \frac{1}{N_{\rm samples}}\sum_{\rm posterior~samples} \log\frac{P(\vec{\theta})}{\pi(\vec{\theta})}\\
    &\simeq \frac{1}{N_{\rm samples}}\sum_{\rm posterior~samples} \log\mathcal{L}(\vec{\theta}) - \log\mathcal{Z}
\end{aligned}
\end{equation} where $\mathcal{Z}$ is the Bayesian evidence output by the analysis.

The information gain is quantified in nats, the base-$e$ equivalent to the base-2 bits: one nat of information is gained for instance when the posterior interval of one parameter has decreased in size by a factor $e$ compared to the prior interval.
Therefore, the tighter the posterior as compared to the prior, the higher the K-L divergence is, and vice-versa.

\subsubsection{Uncertainties arising from limited observational data}
\label{ssub:Cadence}

In this section, we use \textit{injection and recovery} tests to study the impact of the number of observational datapoints and their repartition on the tightness of the estimated posteriors. First, we generate a set of lightcurves called \textbf{sim17} by injecting the \textbf{mock17} parameters (see details in appendix \S~\ref{app:LC}) into the \textsc{Bu2019lm} model. We compute the corresponding lightcurves, producing a dense dataset with a point every 0.1 day from 0 to 15 days, in each filter ($g$-, $r$-, $i$-, $z$-, $y$-, $J$-, $H$- and $K$-band). We will denote this the \textbf{sim17} lightcurve, with the \textbf{10perday} cadence since it has 10 observations per day in each of the eight filters. 
This will be our nominal mock dataset that we will progressively downsample to quantify the effects of cadence and filter choice.

\paragraph{Cadence effect} 

We then downsample from the full \textbf{10perday} lightcurve to create lower cadence \textbf{sim17} datasets. For example, we define the \textbf{1perday} cadence lightcurve by randomly sampling from the \textbf{10perday} simulation 15 points in each filter across the [0,15]-day range, to get an average of one observation per filter per day. 
We conduct the same process to generate different observation cadences \textbf{2perday}, \textbf{1per2day}, \textbf{1per3day}, etc.

\begin{figure*}
    \centering
    \includegraphics[width=0.52\columnwidth]{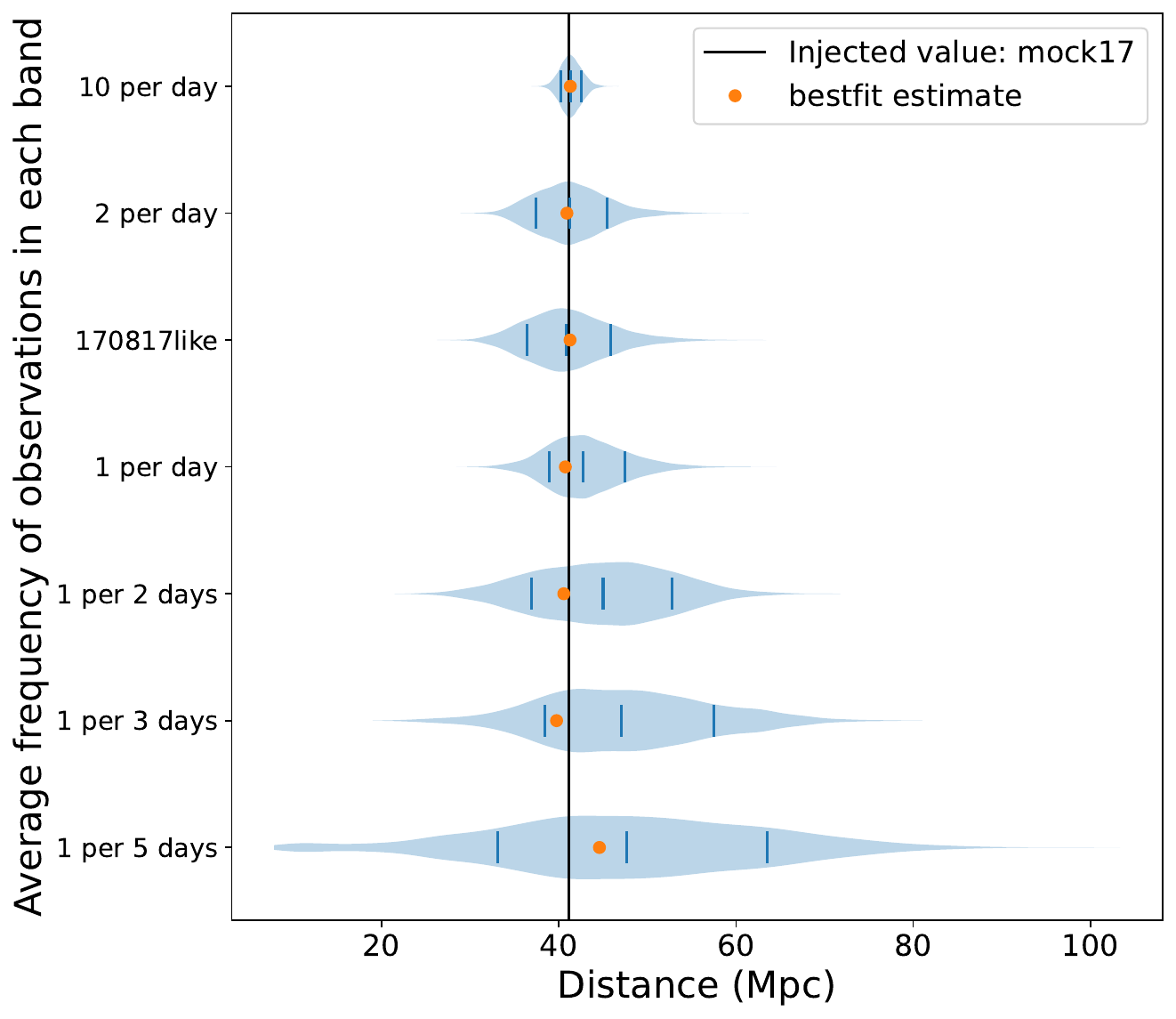}
    \includegraphics[width=0.47\columnwidth]{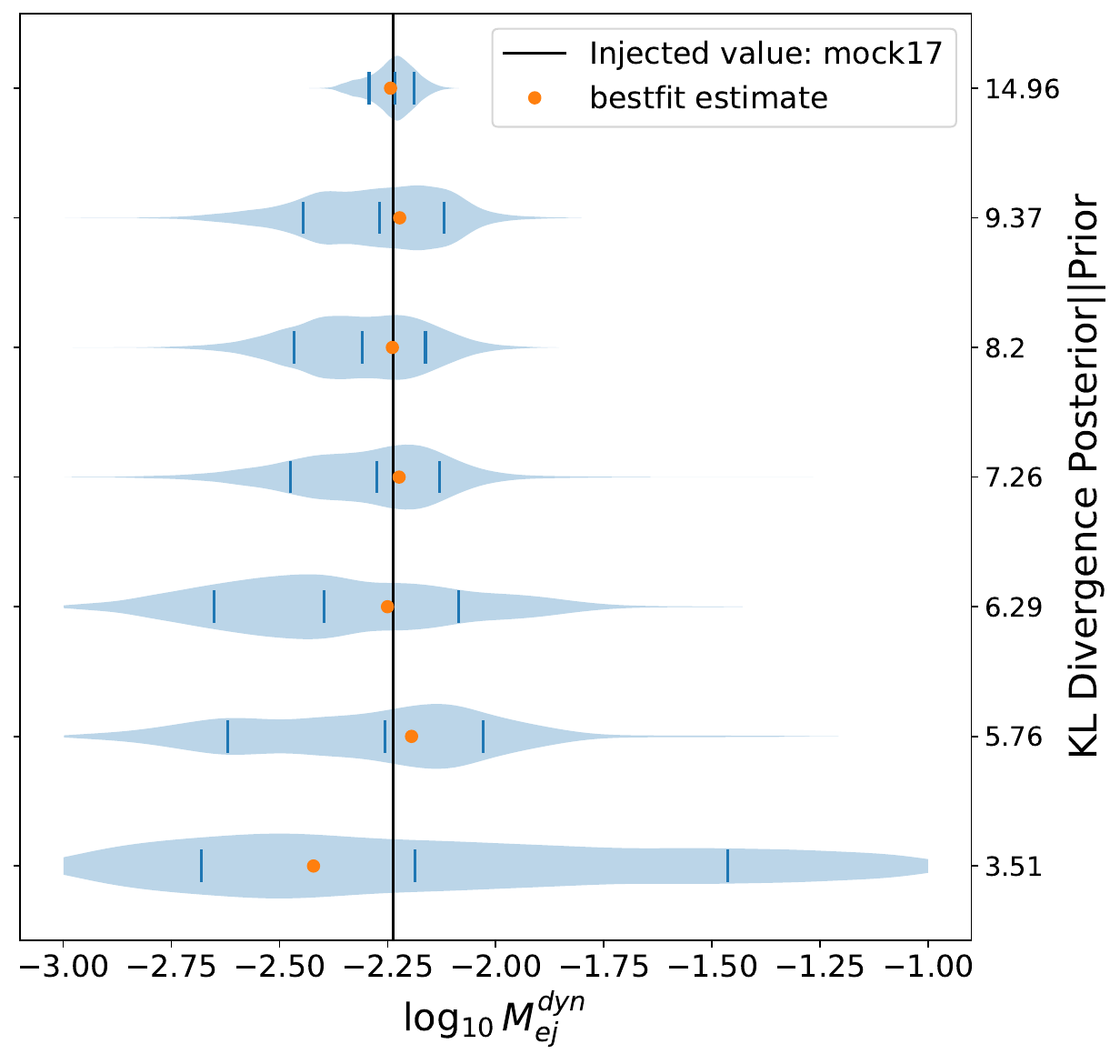}
    \caption{Posterior distributions (\textbf{Left:} Luminosity distance. \textbf{Right:} Dynamical ejecta mass) and K-L divergences resulting from \textsc{NMMA} \textsc{Bu2019lm} analyses of different \textbf{sim17} lightcurves ($grizyJHK$) with prior set \textbf{A1} (from Table~\ref{tab:Priors}); varying the cadence using observations more and more sparse in time for each filter.}
    \label{fig:KLcadence}
\end{figure*}

We then compare the posterior distributions that can be inferred using these different cadences, but all with $grizyJHK$ filters, using the default prior set \textbf{A1} from Table~\ref{tab:Priors}.  Figure~\ref{fig:KLcadence} shows two of them, e.g the luminosity and dynamical ejecta mass: each posterior distribution is represented as a violin plot, the median and the 16th and 84th percentiles (68\% credible interval) are highlighted in blue and the best-fit combination (the sampled combination of parameters with the highest likelihood) in orange. The expected value injected in the simulation is shown as a vertical black line. 

As expected, the more datapoints are used, the tighter is the posterior and the strongest is the constraint (highest K-L divergence). First, as reference, the full \textbf{10perday} dataset produces a gain $D_{KL}(P||\pi)\sim$ 15 nats, compared to about 7.3 nats for \textbf{1perday} cadence and 3.5 nats for \textbf{1per5days}. 
The \textbf{170817like} dataset (in which cadence is a little higher than 1/filter/day) finds $D_{KL}(P||\pi)\sim$ 8.2 nats. 
Moreover, it is interesting to notice that for the slowest cadence \textbf{1per5days}, the 3.5-nats information gain mostly comes from bounding the luminosity distance, whereas the other KN parameters return rather uniform posteriors, not gaining much physical information. 
In fact, computing the partial information gain of each parameter's posterior distribution (see method in Appendix~\ref{app:paramwisegains}), we see that all these other parameters have partial gains below 0.2 nats, our empirically defined threshold for significant constraints.

Therefore, a cadence of about one observation every three days in each of the $g, r, i, z, J, H,$ and $K$ bands appears to mark a threshold for \textsc{Bu2019lm} analyses (at $\sigma_{\rm sys}=1$ mag), under which one cannot expect to constrain the ejecta properties of a KN. This is consistent with recent studies for the LSST survey at the Vera Rubin Observatory, that show that their baseline cadences (below the threshold highlighted above) will be enough to detect a majority of close-by kilonovae \citep{Andrade:2025gme} but will be too sparse to obtain significant constraints in KN parameter inference on their own \citep{Ragosta:2024bfr}.

We perform the same \textit{injection and recovery} test on a simulated event with \textsc{Bu2025MLP}, and we similarly find that tighter constraints are obtained for denser cadences (see Fig~\ref{fig:KLcadence25}). Unlike for \textsc{Bu2019lm} however, the cadence threshold needed to start obtaining significant constraints varies from parameter to parameter: from \textbf{1per2days} for wind ejecta mass to \textbf{3perday} for dynamical ejecta velocity in the $\sigma_{\rm sys}=1$ mag case(see Figs \ref{fig:25Partial} and \ref{fig:25Partial2}). The only exception among ejecta parameters is the wind ejecta's electron fraction, for which significant constraints can be obtained even at a \textbf{1per5days} cadence. This means that analyses using \textsc{Bu2025MLP} might be able to constrain $Y_e^{\rm wind}$ for KN events observed by LSST, even with a large $\sigma_{\rm sys}=1$ mag.

\begin{figure}
    \centering
    \includegraphics[width=0.8\columnwidth]{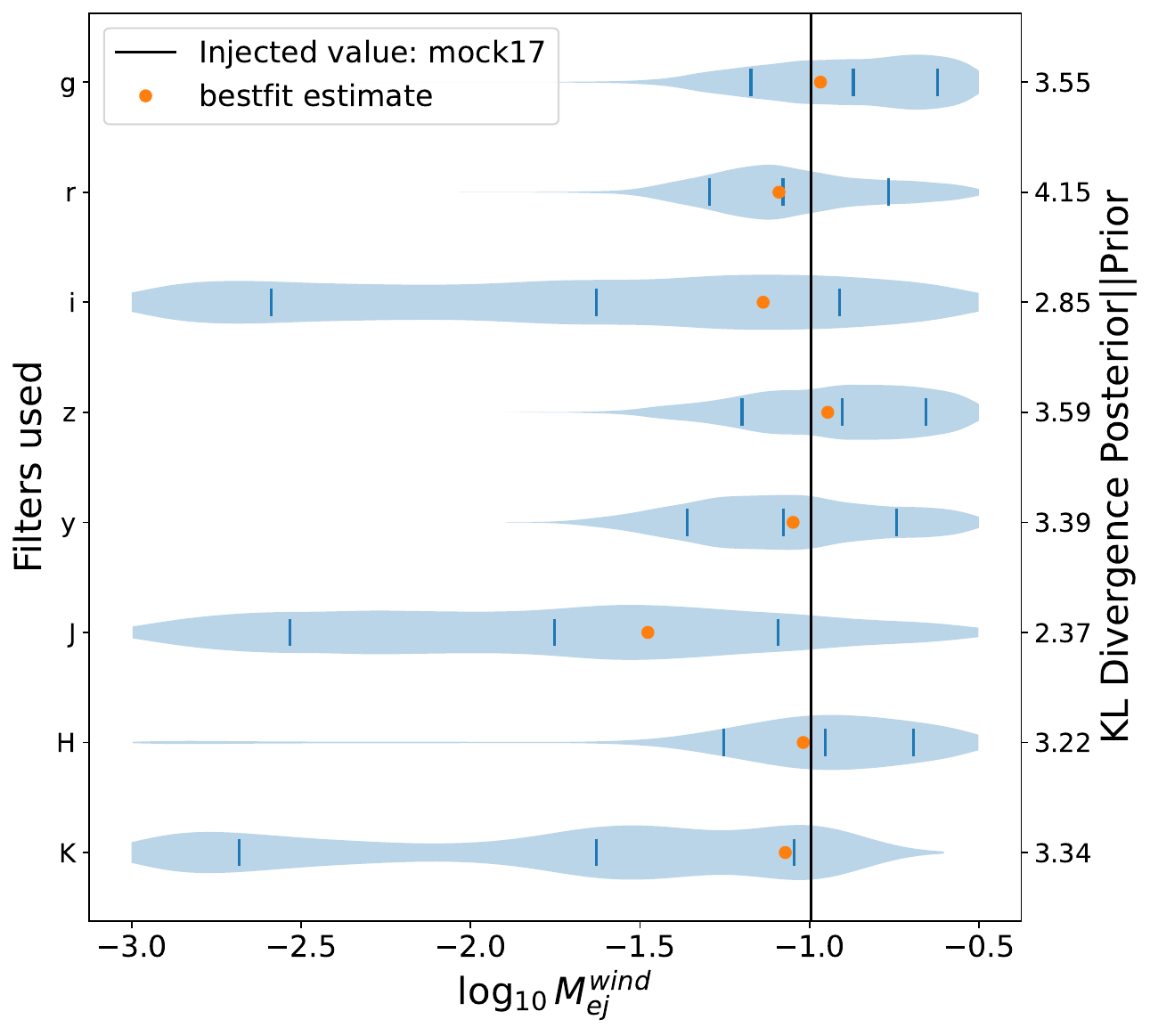}
    \caption{Posterior wind ejecta distributions and K-L divergences when trying to recover the \textbf{sim17} lightcurves with \textbf{170817like} cadence, analysed with prior set \textbf{A1} (from Table~\ref{tab:Priors}), using different filters on their own.}
    \label{fig:KL_individual_filts}
\end{figure}

\begin{figure}
    \centering
    \includegraphics[width=0.8\columnwidth]{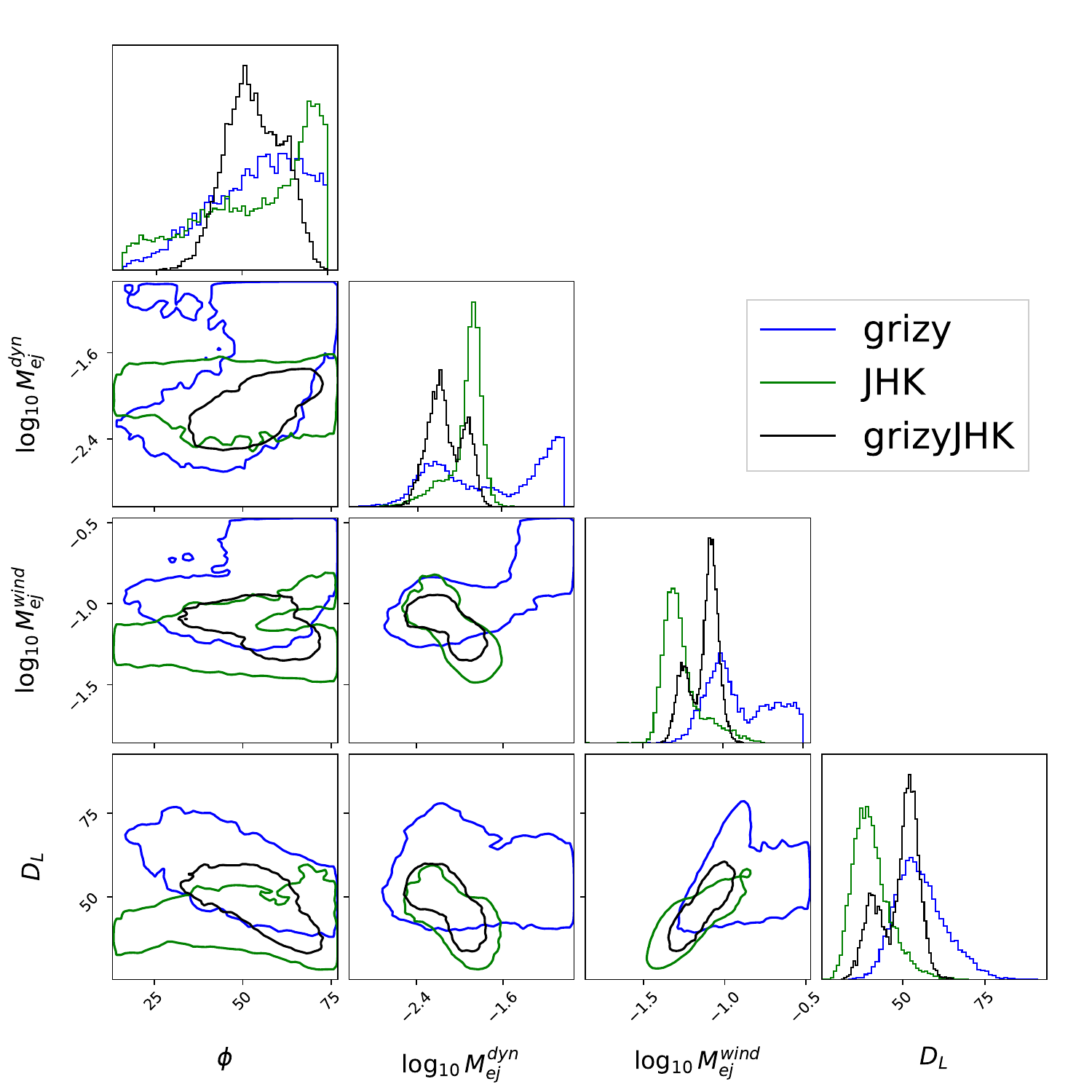}
    \caption{Posterior contours of KN-only (\textsc{Bu2019lm}) analysis of \textbf{AT2017gfo}, with prior set \textbf{B1} (from Table~\ref{tab:Priors}), with different choices of filters.}
    \label{fig:filterdiff}
\end{figure}

\paragraph{Filters} We now evaluate the effect of changing the number of filters included in the analysed dataset, the cadence being fixed for each filter. To do this, we produce \textbf{sim17} datasets, with a \textbf{170817like} cadence mimicking \textbf{AT2017gfo}'s (since its cadence is rather homogeneous across its eight filters, $grizyJHK$), using different filter subsets. 

Using the default prior set \textbf{A1} (from Table~\ref{tab:Priors}), we find that the optimal constraint when all filters are included ($D_{KL}(P||\pi) \sim 8.2$ nats) is slightly further degraded when using only near-infrared filters ($JHK$ bands, $D_{KL}(P||\pi) \sim 5.7$ nats) compared to only optical filters ($grizy$ bands, $D_{KL}(P||\pi) \sim 6.1$ nats). However, the distinction is small and within the range of variation possible due to the random sampling of points in time\footnote{Producing a lightcurve from the same injection but with a different choice of time points (even if the average cadence is the same) will result in slightly different results. The impact on resulting posterior distributions is however generally negligible~\citep{HussenotThesis}.}, so it is not universally applicable for future KN events.

In comparison, using individual filters alone results in information gains ranging only between 2 and 4 nats (see Figure~\ref{fig:KL_individual_filts}). 
More precisely, $r$-band provides the strongest constraint out of the individual filter analyses ($D_{KL}(P||\pi)\sim$ 4.2 nats), whereas other filters, e.g. $i$- $J$- or $K$-band, obtain looser constraints that barely change from the prior distributions. These poorer performances are likely due to more pronounced degeneracies in the \{model $\rightarrow$ lightcurve\} mapping for these filters. 

However, just because these filters extract poorer constraints when analysed on their own with only a handful of datapoints, does not mean that they cannot provide strong constraints when paired with other filters. We for instance note that by combining just the two filters $r$ and $K$, we improve the constraint up to $D_{KL}(P||\pi)\sim$ 5.5 nats, comparable to $grizy$'s 6.1 and $JHK$'s 5.7 nats; which shows the importance of complementarity between optical and infrared data to lift degeneracies.

These results demonstrate the contribution of different data subsets to the parameter inference process~: the likelihood used in Bayesian inference is naturally factorizable into a product of likelihoods for the lightcurve in each filter $\mathcal{L}(\vec{\theta}) = \prod_{j} \mathcal{L}_j(\vec{\theta})$. Thus, on their own, each filter's data constrain the model's parameters to match their lightcurve but data from different filters may favor different regions of the parameter space. However, analyzing multiple filters together theoretically gives more information than the products of its parts, since the model now also need to match the relative difference between filters in order to obtain the best likelihood match\footnote{This additional constraining power only disappears if the model assumes the source to be achromatic}. 
Another example of this phenomenon is the \textbf{AT2017gfo} analysis (real data) as shown in Figure~\ref{fig:filterdiff}: the posterior contours obtained from the filter subsets $grizy$ and $JHK$ highlight different sections of the parameter space, only partially overlapping. Then, when combining all filters together, not only does it converge to the bimodal region compatible with both previous analyses, but it produces an additional constraints by favoring one the two bimodality peaks.\\

\begin{figure}
    \centering
    \includegraphics[width=0.8\columnwidth]{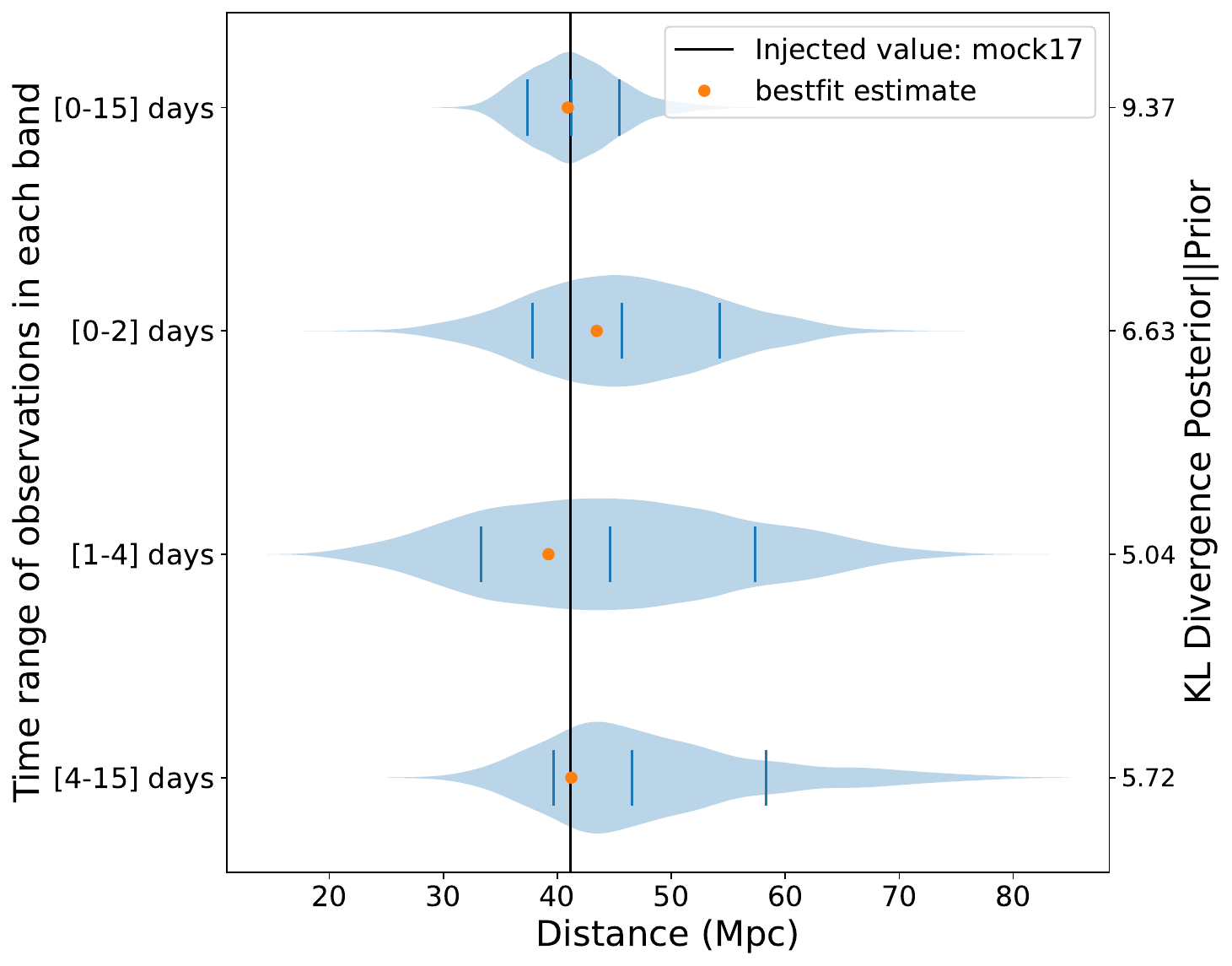}
    \caption{Posterior distance distributions and K-L divergences when trying to recover the \textbf{sim17} lightcurves with \textbf{2perday} cadence ($grizyJHK$) truncated to different time windows, analysed with prior set \textbf{A1} (from Table~\ref{tab:Priors}).}
    \label{fig:KL_timewindows}
\end{figure}

\paragraph{Observation time windows} We also explore the impact of different observation time windows. To do this, we truncated the \textbf{sim17 2perday} lightcurve to different time ranges, and analysed with the default prior set \textbf{A1} (from Table~\ref{tab:Priors}). Early-time points (0 to 2 days) yield $D_{KL}(P||\pi) \sim 6.4$ nats, while points around the peak (1 to 4 days) give 5.0 nats, and late-time points (4 to 15 days) result in 5.7 nats. In comparison, the full \textbf{2perday} light curve reaches 9.4 nats (see Figure \ref{fig:KL_timewindows}). This emphasizes the importance of early-time observations for parameter inference while also highlighting the need for late-time data to obtain a more complete picture. \\

\paragraph{Conclusion on these effects} 
We’ve exhibited complementarity of different data subsets (be it observing using different filters or during different time windows) in order to maximize information gain, and we have highlighted the overall trend that the more datapoints are considered, the more information is gained\footnote{Although we only presented these effects in the \textbf{sim17} example so far, we found they hold true all across the \textsc{Bu2019lm} parameter space, explored by the \textbf{ManySims} dataset constructed in \S~\ref{ssub:offsets}.}.
We have therefore established that a diminution of cadence will induce a relative information loss, but one should note that the information gains associated to any cadence are also highly dependent on other factors such as the systematic error margins and the prior parameter distributions used in the analysis. For instance, if the prior distribution narrowly constrained one of the parameters, it could potentially lift degeneracies and boost the information gain on the other parameters\footnote{For instance, we discuss the effects of external constraints on parameter inference information gains in the particular case of AT2017gfo, in \S~\ref{sub:AT2017gfo}.}. Relative losses from suboptimal cadences could thus be compensated by the presence of external multi-messenger information providing additional constraints on top of the KN model's default priors.

Another takeaway from simulating these effects is that, even if the strongest constraints are of course obtained when exploiting the most amount of data, knowing whether some filters and time windows gain in average more information than others means that one could adapt a telescope's observation strategy to optimize the information gained, in particular if observing time and resources are limited. 
Such studies on cadence optimization for KN follow-up have been explored (e.g., for the Vera Rubin Observatory \citep{Andrade:2025gme,Ragosta:2024bfr}), but the approach presented in this work can further enhance these efforts, for cadences reachable by larger telescope networks for instance.

\subsubsection{Potential systematic offsets}
\label{ssub:offsets}

In the inferences of \textbf{sim17} \textbf{1perday} lightcurves, the posterior distributions tend to be slightly biased toward recovering larger distances and smaller ejecta masses than were injected (as seen in Figure~\ref{fig:KLcadence}). 
Indeed, for these 170817-like parameters, the posterior medians for distance are in average  $\sim5\%$ larger than injected distance, median $M_{\rm ej}^{\rm wind}$ are $\sim4\%$ smaller and median $M_{\rm ej}^{\rm dyn}$ $\sim15\%$ smaller than injected. 

In order to study systematic offsets between the injected values and the posterior distribution medians in the general case, we construct a dataset (called \textbf{ManySims}) of \textbf{1perday} cadence lightcurves simulated for 22 different  parameter combinations sampled randomly all across the \textsc{Bu2019lm} parameter space\footnote{Detailed construction in Appendix~\ref{app:LC}}. Studying the results of the parameter inferences of these lightcurves (analysed with prior set \textbf{A1} from Table~\ref{tab:Priors}), we exhibit such offsets throughout the parameter space, although it should be noted the posterior distributions nevertheless remain compatible with the injected values.

These observed systematic offsets depend on the position of the injected parameters in the parameter space, leading to biases either over- or underestimating the recovered parameters. We summarize the median and extremal values of these offsets across the 22 parameter combinations of \textbf{ManySims} in Table~\ref{tab:offsets}.

When averaging this behavior across the whole parameter space though, the median offsets are close to zero, we do not identify strong biases common to all parameter combinations.

\begin{table}
\centering
\begin{tabular}{|c|c|c|c|c|c|}
\hline
& \multicolumn{5}{c|}{Parameters}\\
\hline
& $\phi$ & $\log_{10}M_{\rm ej}^{\rm dyn}$ & $\log_{10}M_{\rm ej}^{\rm wind}$ & $cos(\iota)$ & $D_L$ \\
& [degrees] & & & & [Mpc]\\
\hline
Min & -6.2 & -0.13 & -0.07 & -0.16 & -4.8\\
\hline
\textbf{Median} & +1.7 & -0.01 & 0.00 & +0.06 & +0.4\\
\hline
Max & +6.8 & +0.11 & +0.07 & +0.21 & +5.1\\
\hline
\end{tabular}
\caption{Distribution of median offsets (Posterior distribution median - Injected value) across the \textbf{ManySims} simulations analysed with prior set \textbf{A1} (from Table~\ref{tab:Priors}).}
\label{tab:offsets}
\end{table}

We similarly constructed and analysed a dataset (called \textbf{25MLPSims}, see \S~\ref{app:LC}) of \textsc{Bu2025MLP} lightcurves of \textbf{1perday} cadence, and obtained similar trends as the \textsc{Bu2019lm} results: recovery offsets identified in local regions of parameter space but an overall zero average bias (see Table~\ref{tab:offsets25}).

\begin{table*}
\hspace*{-1.5cm}
\begin{tabular}{|c|c|c|c|c|c|c|c|c|c|}
\hline
& & \multicolumn{8}{c|}{Parameters}\\
\hline
& & $\log_{10}M_{\rm ej}^{\rm dyn}$ & $v_{\rm ej}^{\rm dyn}$ & $Y_e^{\rm dyn}$ & $\log_{10}M_{\rm ej}^{\rm wind}$ & $v_{\rm ej}^{\rm wind}$ & $Y_e^{\rm wind}$ & $\iota$ & $D_L$ \\
& & & [c] & & & [c] & &  [radian] & [Mpc]\\
\hline
$\sigma_{\rm sys} = 1$ mag \\
\hline
& Min & -0.52 & -0.06 & -0.07 & -0.40 & -0.05 & -0.11 & -0.71 & -8.7\\
\hline
& \textbf{Median} & +0.03 & 0.00 & 0.00 & +0.02 & 0.00 & 0.00 & -0.06 & +2.2\\
\hline
& Max & +0.86 & +0.09 & +0.10 & +0.45 & +0.04 & +0.04 & +0.63 & +153\\
\hline
$\sigma_{\rm sys} = 0.5$ mag \\
\hline
& Min & -0.27 & -0.09 & -0.04 & -0.25 & -0.03 & -0.10 & -0.38 & -5.9\\
\hline
& \textbf{Median} & +0.01 & 0.00 & 0.00 & +0.02 & 0.00 & 0.00 & -0.02 & +0.3\\
\hline
& Max & +0.41 & +0.04 & +0.09 & +0.20 & +0.04 & +0.02 & +0.39 & +136\\
\hline
\end{tabular}
\caption{Distribution of median offsets (Posterior distribution median - Injected value) across the \textbf{25MLPSims} simulations analysed with \textsc{Bu2025MLP}'s default prior set \textbf{A'} (from Table~\ref{tab:Priors}) and varying error margins.}
\label{tab:offsets25}
\end{table*}

\section{Discussing the expected performance of Kilonova parameter inference}
\label{sec:Performance}

In previous sections, we have discussed the inner workings of parameter inference, highlighted the relevant sources of uncertainties, and presented tools to quantify the information gain of inference analyses. We now aim to quantify in more details the performance of KN parameter inference constraints that can be expected from KN lightcurves. Since we have shown the information gain depends on the quantity of data available, we assume in the following section the \textbf{1perday} cadence described earlier, which we deem to be reasonably achievable given the diversity of telescopes on Earth and in space, and study what the expected strengths of constraints on the different parameters are.

\subsection{Simulation study of expected performance with the \textsc{Bu2019lm} model}
\label{sub:perf}
\begin{figure}
    \centering
    \includegraphics[width=\columnwidth]{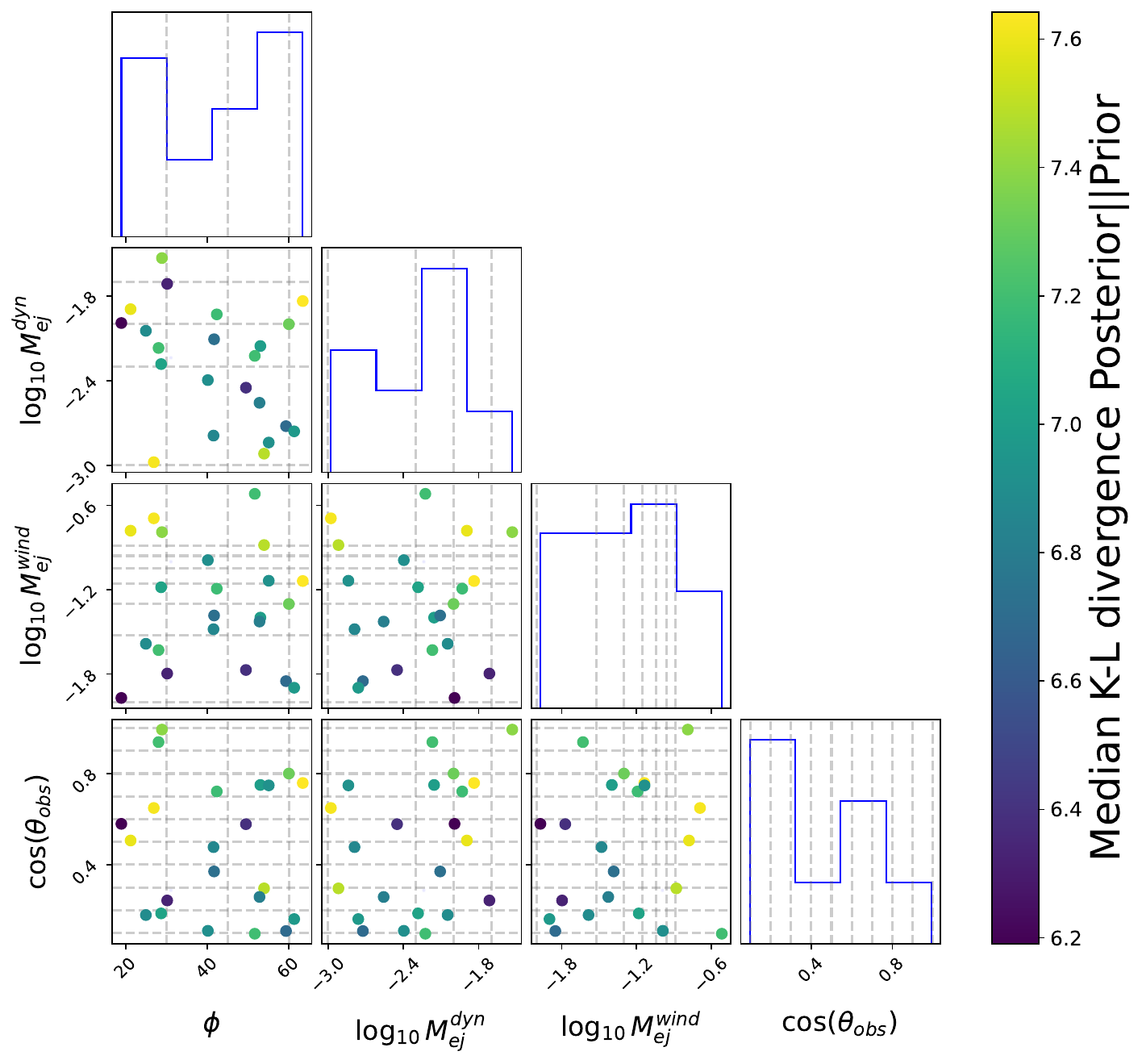}
    \caption{Median K-L divergence obtained with parameter inference (prior set \textbf{A1} from Table~\ref{tab:Priors}), for each of the 22 simulated parameter combinations in \textbf{ManySims}. Dashed grey lines indicate the values of the parameters of the grid of simulations from which the \textsc{Bu2019lm} is interpolated: We find no significant difference in performance between events simulated with parameters close to the gridpoints and those simulated with parameters further from the gridpoints (thus relying more on the interpolation).} 
    \label{fig:KLacrossgrid}
\end{figure}

\begin{figure}
    \centering
    \includegraphics[width=0.85\columnwidth]{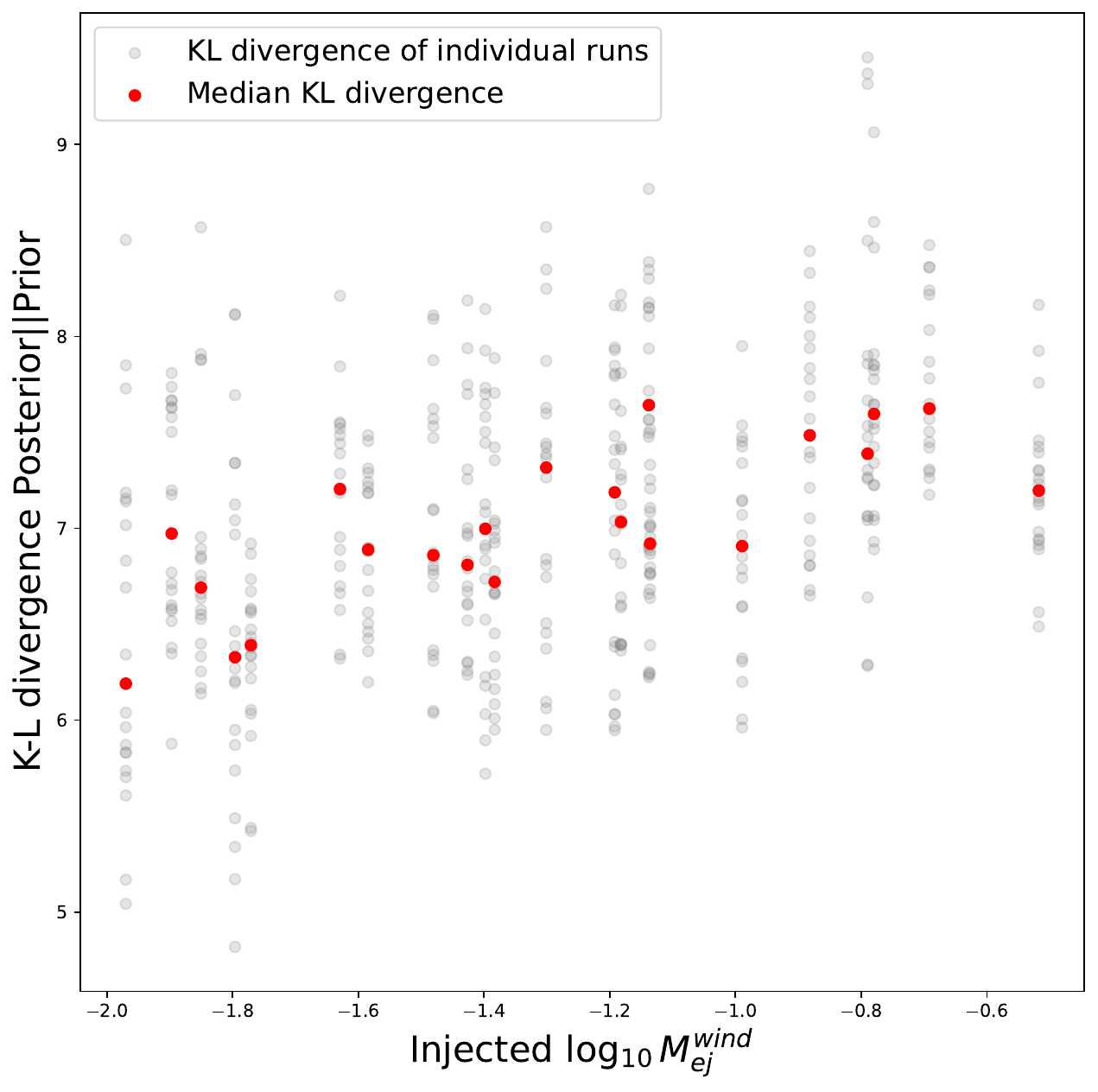}
    \caption{For each of the 22 simulated parameter combinations in \textbf{ManySims}, we show in grey the K-L divergences obtained with parameter inference (prior set \textbf{A1} from Table~\ref{tab:Priors}) for the 20 corresponding sampled lightcurves, and in red the median of these 20 K-L divergences.}
    \label{fig:KLwindcorr}
\end{figure}

\paragraph{Homogeneity of performance across the parameter space}
We want to probe whether some regions of the parameter space obtain tighter or looser parameter inference constraints than others.
We analyse, using the default prior set \textbf{A1} from Table~\ref{tab:Priors}, all the \textbf{1perday} lightcurves of the \textbf{ManySims} simulation set discussed in Section~\ref{ssub:offsets}. First of all, the performance obtained by inferences (measured as information gain) is quite homogeneous across the parameter space (see Figure~\ref{fig:KLacrossgrid})~: for all the parameter combinations (22 values across the parameter space), the median values\footnote{For each parameter combination, we simulate and analyse 20 different lightcurves, all with \textbf{1perday} cadence, and we report the median of the set of the obtained 20 performances.} of $D_{KL}(P||\pi)$ sit between 6.2 and 7.6 nats. This range of performance spanning 1.4 nats is smaller than the spread observed between the 20 realizations of each combination, which typically spans about 2 to 3 nats (see Figure~\ref{fig:KLwindcorr}). 

Therefore, given an input average cadence, the expected information gain can be estimated within a few nats, i.e. the size of the expected posterior volume can be estimated within an order of magnitude, no matter the values of the KN parameters. In more details, however, one statistically significant correlation we found is that injections where the wind ejecta mass are higher tend to be recovered with tighter constraints\footnote{This is likely because higher ejecta masses typically create brighter KNe, so more of the associated lightcurves will pass our $< 24$ mag threshold, and thus more datapoints end up analysed.}. Indeed, as shown in Figure~\ref{fig:KLwindcorr}, the median $D_{KL}(P||\pi)$ is about 1 nat greater on the high end of the wind mass range compared to the low end.

\paragraph{Properties of expected performance}
Given the relative homogeneity of parameter inference constraints, we now explore different properties of posterior distributions, and consider for each one the median of the 22x20 \textbf{ManySims} results as the representative performance to be expected from analyses with the default prior set \textbf{A}. 
Let us look at the breakdown of posterior distribution size for each model parameters, and study how it changes when varying error margins or cadences.

For each \textsc{Bu2019lm} parameter, the median performance in terms of reduction of prior size to posterior size, computed as the ratio of the 68\% credible interval to the prior interval, is presented in Table~\ref{tab:gains}. Given the same \textbf{1perday} cadence, we can see that when the systematic uncertainty $\sigma_{\rm sys}$ (dominating over the negligible observational uncertainties) changes from 1 mag to 0.5 mag, the typical posterior sizes diminish by about a factor 2 for each parameter. Consistently, the median K-L divergence improves from $\sim 7$ nats (i.e an effective reduction from prior volume to posterior volume of a factor $\sim 1100$) to $\sim 10$ nats (prior to posterior reduction of a factor $\sim 24000$). 

We also highlight the impact of the cadence as discussed in section~\ref{ssub:Cadence}: when simulating \textbf{ManySims} lightcurves with the \textbf{170817like} cadence (about 30\% more data points than the \textbf{1perday} cadence), we obtain slightly stronger constraints than for \textbf{1perday} (see Table~\ref{tab:gains2})~: $D_{KL}(P||\pi)\sim 8$ nats (posterior volume $\sim 3000$ times smaller than the prior.)\\

\begin{table}
\centering
\begin{tabular}{|c|c|c|c|c|c|c|}
\hline
& & \multicolumn{5}{c|}{Parameters}\\
\hline
Cadence & $\sigma_{\rm sys}$ & $\phi$ & $\log_{10}M_{\rm ej}^{\rm dyn}$ & $\log_{10}M_{\rm ej}^{\rm wind}$ & $cos(\iota)$ & $D_L$ \\
\hline
\textbf{1perday} & 1 mag & 0.35 & 0.18 & 0.08 & 0.49 & 0.04\\
\hline
\textbf{1perday} & 0.5 mag & 0.19 & 0.09 & 0.05 & 0.23 & 0.02\\
\hline
\end{tabular}
\caption{Ratio of Median posterior size (computed as the median of the lengths of posterior 68\% credible intervals obtained across the 22x20 \textbf{ManySims} analyses with prior set \textbf{A}) over the size of the prior interval (as defined in Table~\ref{tab:Priors}), for the different \textsc{Bu2019lm} parameters, for different values of $\sigma_{\rm sys}$.}
\label{tab:gains}
\end{table}

\begin{table}
\centering
\begin{tabular}{|c|c|c|c|c|c|c|}
\hline
& & \multicolumn{5}{c|}{Parameters}\\
\hline
Cadence & $\sigma_{\rm sys}$ & $\phi$ & $\log_{10}M_{\rm ej}^{\rm dyn}$ & $\log_{10}M_{\rm ej}^{\rm wind}$ & $cos(\iota)$ & $D_L$ \\
\hline
\textbf{1perday} & 1 mag & 0.35 & 0.18 & 0.08 & 0.49 & 0.04\\
\hline
\textbf{170817like} & 1 mag & 0.29 & 0.14 & 0.06 & 0.43 & 0.04\\
\hline
\end{tabular}
\caption{Ratio of Median posterior size (across the 22x20 \textbf{ManySims} analyses with prior set \textbf{A}) over the prior interval size, for the different \textsc{Bu2019lm} parameters, for different observational cadences.}
\label{tab:gains2}
\end{table}

The size of the posterior 68\% credible intervals being roughly proportional to the $\sigma_{\rm sys}$ error margin is expected. 
Indeed, the parameter region bound by the 68\% credible interval of each parameter roughly corresponds to the parameters that can produce a likelihood within a few log-units of the best-fit likelihood, as seen empirically in our analyses' posterior samples.
We consider the total error margin $\sigma_{\rm tot} = \sqrt{\sigma_{\rm meas}^2+\sigma_{\rm sys}^2}$ that will be considered equivalent to $\sigma_{\rm sys}$ in this example where observational uncertainties are negligible.
Looking at the likelihood equation \ref{eq:likelihood}, a likelihood margin of a few log-units correspond to a $\chi^2 = \sum_{ij}\frac{(m_i^j-m_i^{j,\rm model})^2}{\sigma_{\rm tot}^2} \lesssim 10$, and thus an average magnitude variation $\Delta m$ such that $N \frac{\Delta m ^2}{\sigma_{\rm tot}^2} \lesssim 10$, so $\Delta m \lesssim \sigma_{\rm tot}\sqrt{10/N}$, where $N$ is the number of datapoints in the analysed lightcurve.
In our \textbf{1perday} cadence examples ($N\sim110$) with $\sigma_{\rm tot} \simeq \sigma_{\rm sys} = 1$ mag, the characteristic magnitude deviation in the 68\% intervals is thus $\Delta m \lesssim \sigma_{\rm tot}\sqrt{10/N} \simeq 0.3$ mag.
This describes a "tolerance margin": if a parameter combination produces a lightcurve that deviates from the best-fit lightcurve by less than this magnitude margin, then it likely falls within the parameter region bound by each parameter's 68\% credible interval.

For example, the luminosity distance has a straightforward effect on the apparent magnitude, acting as a constant offset at all wavelengths: $m_i^j= M_i^j+5\times~\log_{10}(D_L/10 \text{ pc})$, where $M_i^j$ are the absolute magnitudes (i.e. at a reference distance of 10 pc). Therefore, a 0.3~mag tolerance margin in magnitude corresponds to a tolerance margin of 0.06 units in log-distance around the best-fit distance value. The distance posterior 68\% credible intervals reported in Table~\ref{tab:gains} are of the same order, typically twice as wide (this may be because degeneracies with other model parameters could partially cancel out the distance-caused variation, thus allowing a larger distance margin to get to the same magnitude deviation). A $\sigma_{\rm sys} = 0.5$ mag error margin would instead correspond to a 0.15 mag magnitude tolerance margin, and thus 0.03 units in log-distance, and we indeed see a reduction of posterior size by a factor 2 from $\sigma_{\rm sys} = 1$ mag to $\sigma_{\rm sys} = 0.5$ mag.

\begin{figure}
    \centering
    \includegraphics[width=0.9\columnwidth]{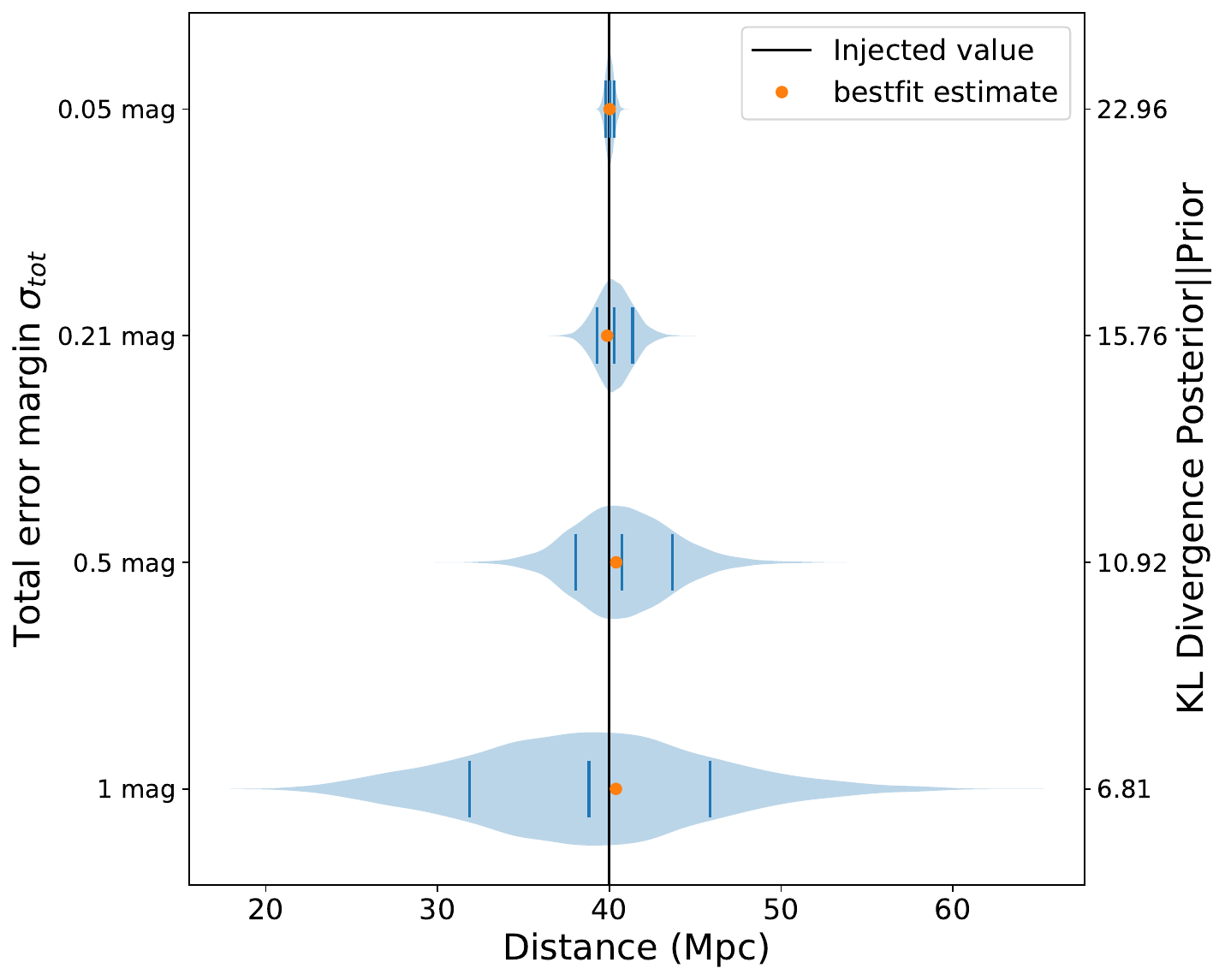}
    \caption{Posterior distance distributions and K-L divergences when trying to recover the first \textbf{ManySims} injection's first sampled lightcurve with \textbf{1perday} cadence ($grizyJHK$), analysed with prior set \textbf{A} (from Table~\ref{tab:Priors}) using different values of $\sigma_{\rm sys}$. We display each analysis' total error margin $\sigma_{\rm tot} = \sqrt{\sigma_{\rm meas}^2+\sigma_{\rm sys}^2}$.}
    \label{fig:KL_totalerr}
\end{figure}

The overall behavior of posterior credible interval sizes being roughly proportional to the total error margin is expected to hold for any value of $\sigma_{\rm tot}$ (see Figure \ref{fig:KL_totalerr} for example), as long as the corresponding parameter tolerance margins are small with respect to the prior size. For larger error margins, the parameter tolerance becomes as large as the prior interval, and the posterior distribution basically recovers the prior. 

\paragraph{Constraining power differentiated between parameters}
We now investigate which model parameters exhibit the strongest constraint, i.e. the strongest reduction from prior to posterior size. 
Comparing the recovered constraints on the different model parameters in Table~\ref{tab:gains}, and ignoring the luminosity distance whose strong reduction is linked to a very large and not physically motivated prior, we find that the wind ejecta mass is the parameter for which we get the strongest prior-to-posterior reduction. This could be explained by the fact that in the \{model parameters $\rightarrow$ lightcurves\} mapping, the magnitude gradient is highest along the wind ejecta mass axis.
Indeed, the higher the gradient, the smaller the tolerated parameter range to span the same magnitude tolerance margin\footnote{For example if the distance gradient was twice as large ($m= M+10\times~\log_{10}(D_L/10 \text{ pc})$), then a 0.3 mag margin in magnitude would correspond to 0.03 units of log-distance, twice as small as the 0.06 units in the default case.}.

\begin{figure}
    \centering
    \includegraphics[width=0.7\columnwidth]{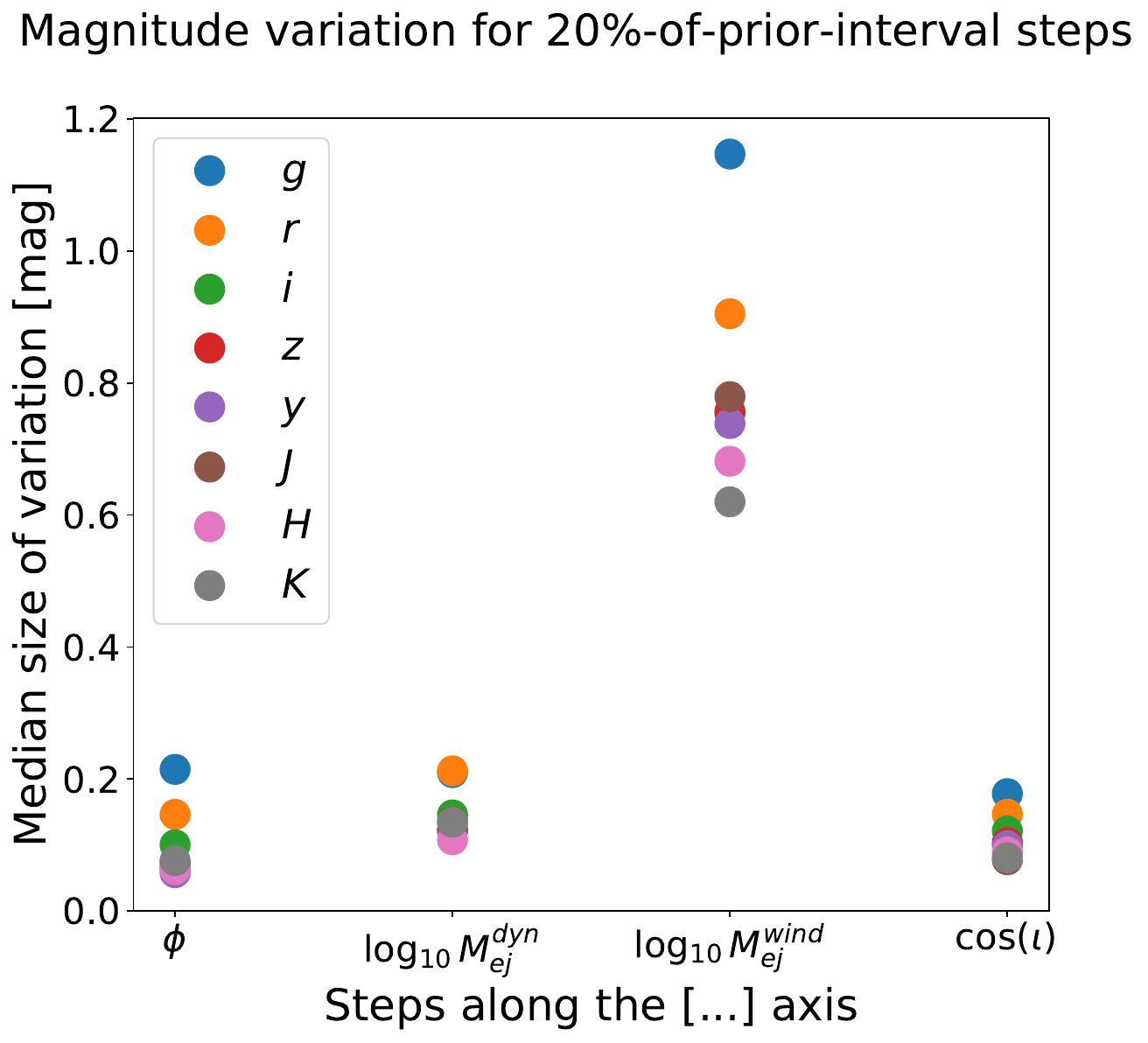}
   \caption{Size of lightcurve variations when moving along axes of the \textsc{Bu2019lm} parameter space in steps of 20\% of the prior interval, averaged over time and across the whole parameter space.}
    \label{fig:stepsBu19}
\end{figure}

In order to estimate the parameters-to-magnitude gradients, we divided the \textsc{Bu2019lm} parameter space into a grid, sampling 6 increments for each model parameter (the 0th to 5th quintiles of the default priors). For each parameter, we then computed the average variation in lightcurves obtained when moving by these "20\%-of-prior-range" steps along the given parameter, all other parameters being fixed. We then averaged the size of these steps (in magnitude) over all time epochs of the lightcurve and over all the possible configurations of the other parameters in the grid. 
We show the results in Fig~\ref{fig:stepsBu19} for many filters, and in almost all cases the biggest variations are indeed obtained when moving along the axis of wind ejecta mass.

Therefore, for a given total error margin $\sigma_{\rm tot}$, the tolerated parameter range (for which lightcurves fall within $\Delta m \lesssim \sigma_{\rm tot}\sqrt{10/N}$) is a smaller fraction of the prior interval for the wind ejecta mass than for the other parameters. Consequently, one can expect that in future observations of KNe, the physical parameter on which we have the strongest constraint is the wind ejecta mass, as simulated in Fig~\ref{fig:ParGainViolins}.

When comparing \texttt{NMMA}+\textsc{Bu2019lm} inference outputs with the results of other parameter inferences that use other messengers\footnote{For example, analyses of GW signals can indirectly constrain ejecta masses by using empirical fitting relations~\citep{Foucart:2018rjc,Kruger:2020gig,Dietrich:2020efo} involving the inferred progenitor masses, compactness and spins.} or alternative KN models, constraints on the wind ejecta mass are thus likely to be the strongest indicator of whether our results are compatible or in tension with the other analyses.\\

\subsection{Simulation study for higher-dimensional model \textsc{Bu2025MLP}}

\begin{table*}
\hspace*{-0.5cm}
\begin{tabular}{|c|c|c|c|c|c|c|c|c|c|}
\hline
& & \multicolumn{8}{c|}{Parameters}\\
\hline
Cadence & $\sigma_{\rm sys}$ & $\log_{10}M_{\rm ej}^{\rm dyn}$ & $v_{\rm ej}^{\rm dyn}$ & $Y_e^{\rm dyn}$ & $\log_{10}M_{\rm ej}^{\rm wind}$ & $v_{\rm ej}^{\rm wind}$ & $Y_e^{\rm wind}$ & $\iota$ & $D_L$ \\
\hline
\textbf{1perday} & 1 mag & 0.46 & 0.49 & 0.42 & 0.30 & 0.53 & 0.20 & 0.49 & 0.06\\
\hline
\textbf{1perday} & 0.5 mag & 0.25 & 0.30 & 0.24 & 0.20 & 0.30 & 0.12 & 0.28 & 0.03\\
\hline
\end{tabular}
\caption{Ratio of Median posterior size (computed as the median of the lengths of posterior 68\% credible intervals obtained across the 100 \textbf{25MLPSims} simulations) over the prior interval size, for the different \textsc{Bu2025MLP} parameters.}
\label{tab:gainsBu25}
\end{table*}

We apply the same \textit{injection-recovery} approach to the \textbf{25MLPSims} dataset in order to study the expected performance of \textsc{Bu2025MLP} inferences. We find that when analyzing \textbf{1perday}-cadence datasets with default priors and $\sigma_{\rm sys}=1$ mag, the obtained K-L divergences range between $\sim[5,10]$ nats, wider but similar range to \textsc{Bu2019lm}'s $\sim[5,9]$ nats range.

Consistently with our \textsc{Bu2019lm} findings, we find here too that changing $\sigma_{\rm sys}$ from 1 mag to 0.5 mag corresponds to roughly halving the size of typical 68\% posterior credible intervals in all parameters, and equivalently increasing $D_{KL}$ by a few nats, the range now being $\sim[8,15]$ nats.

\begin{figure}
    \centering
    \includegraphics[width=0.7\columnwidth]{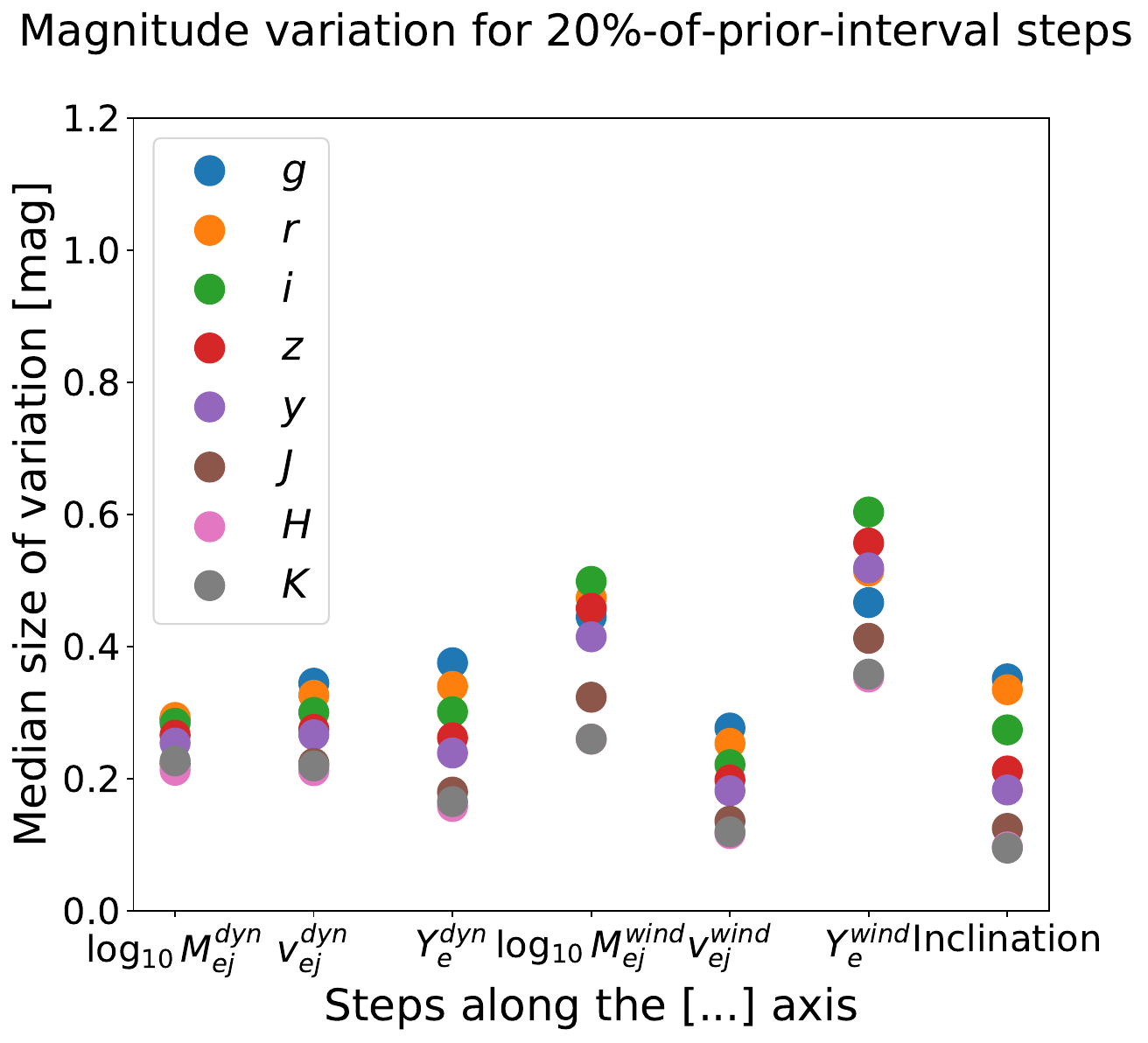}
   \caption{Size of lightcurve variations when moving along axes of the \textsc{Bu2025MLP} parameter space in steps of 20\% of the prior interval, averaged over time and across the whole parameter space.}
    \label{fig:stepsBu25}
\end{figure}

However, when looking at the breakdown of constraining power between different parameters (see Table~\ref{tab:gainsBu25}), the wind ejecta mass is no longer the parameter associated to the tightest constraints: the wind ejecta's electron fraction is. This fact, also visible in Fig~\ref{fig:ParGainViolins}, is consistent with the strong magnitude gradient along the electron fraction axis hinted at in Figure~\ref{fig:stepsBu25}.

Finally, we note that, cadence and $\sigma_{\rm sys}$ being equal, the typical posterior intervals for ejecta masses and distance are larger for \textsc{Bu2025MLP} analyses than for \textsc{Bu2019lm}. This can likely be attributed to \textsc{Bu2025MLP} being a higher-dimensional model, with potentially more degeneracies between parameters.

\subsection{Application to real data~: AT2017gfo}
\label{sub:AT2017gfo}

\begin{sidewaystable}
\hspace{-1.5cm}
\scalebox{0.9}{
\begin{tabular}{|l|r|lllll|l|l|l|l|}
\hline
\multicolumn{2}{|l|}{Prior set} & \multicolumn{5}{l|}{Parameters} &  & \multicolumn{3}{l|}{K-L divergences}\\
\hline
\multicolumn{2}{|l|}{} & $\phi$ & $\log_{10}M_{\rm ej}^{\rm dyn}$ & $\log_{10}M_{\rm ej}^{\rm wind}$ & $cos(\iota)$ & $D_L$ & best-fit & $D_{KL}(\pi||\textbf{A})$ & $D_{KL}(P_{170817}||\pi)$ & $D_{KL}(P_{170817}||\textbf{A})$ \\
\multicolumn{2}{|l|}{} & [deg] & [$M_\odot$] & [$M_\odot$] & & [Mpc] & $\chi^2_r$ & & & \\
\hline
\textbf{A}: Default Prior Range & Prior & [15,75] & [-3,-1] & [-3,-0.5] & [0,1] & Log$\mathcal{U}(1,500)$& & 0 &  &  \\
\multicolumn{2}{|r|}{\textbf{A1}: $\sigma_{\rm sys} = 1$ mag Analysis Posterior} & $38.03^{+23.48}_{-14.36}$ & $-2.23^{+0.25}_{-0.24}$ & $-1.03^{+0.15}_{-0.22}$ & $0.44^{+0.47}_{-0.41}$ & $41.95^{+12.36}_{-11.27}$ & 0.26 & & 8.13 & 8.13\\
\multicolumn{2}{|r|}{\textbf{A0.6}: $\sigma_{\rm sys} = 0.6$ mag Analysis Posterior} & $48.68^{+11.67}_{-7.26}$ & $-1.97^{+0.09}_{-0.23}$ & $-1.13^{+0.07}_{-0.08}$ & $0.64^{+0.26}_{-0.11}$ & $41.82^{+9.34}_{-4.93}$ & 0.90 & & 11.79 & 11.79\\
\hline
\textbf{B}: Angle constraint & Prior & [15,75] & [-3,-1] & [-3,-0.5] & [0.91,0.95] & Log$\mathcal{U}(1,500)$& & 3.41 & & \\
\multicolumn{2}{|r|}{\textbf{B1}: $\sigma_{\rm sys} = 1$ mag Analysis Posterior} & $53.08^{+15.35}_{-15.27}$ & $-2.15^{+0.33}_{-0.28}$ & $-1.09^{+0.11}_{-0.21}$ & $0.93^{+0.02}_{-0.02}$ & $50.53^{+7.23}_{-13.86}$ & 0.36 & & 7.75 & 11.16\\
\multicolumn{2}{|r|}{\textbf{B0.6}: $\sigma_{\rm sys} = 0.6$ mag Analysis Posterior} & $52.04^{+13.96}_{-9.41}$ & $-2.18^{+0.32}_{-0.16}$ & $-1.09^{+0.06}_{-0.21}$ & $0.93^{+0.02}_{-0.01}$ & $51.07^{+4.28}_{-13.26}$ & 0.96 & & 9.83 & 13.25\\
\hline
\textbf{C}: Angle \& GW distance & Prior & [15,75] & [-3,-1] & [-3,-0.5] & [0.91,0.95] & $\mathcal{N}(43.4, 2^2)$ & & 6.90 & & \\
\multicolumn{2}{|r|}{\textbf{C1}: $\sigma_{\rm sys} = 1$ mag Analysis Posterior} & $60.44^{+7.91}_{-12.68}$ & $-1.93^{+0.14}_{-0.31}$ & $-1.22^{+0.12}_{-0.09}$ & $0.93^{+0.02}_{-0.02}$ & $43.07^{+3.97}_{-3.28}$ & 0.37 & & 5.49 & 12.38\\
\multicolumn{2}{|r|}{\textbf{C0.6}: $\sigma_{\rm sys} = 0.6$ mag Analysis Posterior} & $61.30^{+4.45}_{-6.91}$ & $-1.92^{+0.09}_{-0.12}$ & $-1.24^{+0.09}_{-0.05}$ & $0.93^{+0.02}_{-0.01}$ & $42.10^{+4.01}_{-2.68}$ & 0.97 & & 7.63 & 14.53\\
\hline
\textbf{D}: Angle \& GW distance+masses & Prior & [15,75] & [-2.6,-1.5] & [-1.9,-1.2] & [0.91,0.95] & $\mathcal{N}(43.4, 2^2)$ & & 8.77 & & \\
\multicolumn{2}{|r|}{\textbf{D1}: $\sigma_{\rm sys} = 1$ mag Analysis Posterior} & $59.38^{+6.86}_{-9.53}$ & $-1.94^{+0.16}_{-0.15}$ & $-1.25^{+0.04}_{-0.07}$ & $0.93^{+0.01}_{-0.02}$ & $42.47^{+3.42}_{-3.01}$ & 0.37 & & 4.74 & 13.50\\
\multicolumn{2}{|r|}{\textbf{D0.6}: $\sigma_{\rm sys} = 0.6$ mag Analysis Posterior} & $61.28^{+4.61}_{-5.25}$ & $-1.93^{+0.08}_{-0.09}$ & $-1.24^{+0.04}_{-0.05}$ & $0.93^{+0.02}_{-0.02}$ & $41.78^{+2.47}_{-2.53}$ & 0.97 & & 6.25 & 15.01\\
\hline
\end{tabular}}
\caption{Prior ranges for different combinations of constraints, and the corresponding performances of \textsc{Bu2019lm} parameter inferences on \textbf{AT2017gfo}: Posterior 95\% credible intervals, best-fit $\chi^2_r$ and K-L divergences between each prior $\pi$, the default prior $\textbf{A}$, and each analysis' posterior $P_{170817}$. Log$\mathcal{U}(a,b)$ describes a log-uniform distribution from $a$ to $b$, and $\mathcal{N}(\mu, \sigma^2)$ is a Gaussian distribution with mean $\mu$ and variance $\sigma^2$; all other prior intervals obey a uniform distribution over their given range.}
\label{tab:PriorsKL}
\end{sidewaystable}

We now perform a parameter inference analysis of the AT2017gfo kilonova, with \textsc{Bu2019lm} model, using the \textbf{AT2017gfo} dataset\citep{Villar:2017wcc, Coughlin:2018miv}.  We would like to check if the performances estimated with our simulation set \textbf{ManySims} are consistent with the inference performances obtained on real data. The results of the following analyses are compiled in Table~\ref{tab:PriorsKL}.

Let us first fix $\sigma_{\rm sys} = 1$ mag in order to compare with the simulations of the previous section.

Running a KN-only parameter inference with the default prior set \textbf{A1} yields constraints of similar strength to the section~\ref{sub:perf} results of \textbf{ManySims} simulations with \textbf{170817like} cadence: we obtain $D_{KL}(P||\pi)= 8.1$ nats, which is consistent with the simulated analyses' $D_{KL}(P||\pi)\sim 8$ nats.

However, looking at the associated best-fit, we see that $\chi^2_r = 0.26$, which indicates that the $\sigma_{\rm sys} = 1$ mag error margin was unnecessarily large (see discussion in section~\ref{sub:ModelUnc}).

Performing another analysis where we leave the error margin $\sigma_{sys}$ as a free parameter, the maximum likelihood is found for an error margin of order $\sigma_{\rm sys} = 0.57$ mag. 
This means that even with using its widest uniform prior, the \textsc{Bu2029lm} model has no parameter combination able to fit the \textbf{AT2017gfo} dataset within the observational uncertainties alone, it needs the additional systematic error to be compatible with data. This can mean than the \textsc{Bu2019lm} is limited by its model assumptions (e.g. simplified approximation of lanthanide opacities, assuming local thermal equilibrium at all times, etc.), and that changing/relaxing its hypotheses could allow to produce different lightcurves that better fit AT2017gfo's. For comparison, AT2017gfo analyses with other KN models in \cite{Villar:2017wcc} and \cite{Nicholl:2021rcr} determine their optimal error margins to be $\sigma_{\rm sys} \sim 0.23$ mag and $\sigma_{\rm sys} \sim 0.36$ mag respectively. This means that other KN models also struggle to reproduce AT2017gfo without an added systematic error, although they obtain lower values of $\sigma_{\rm sys}$ and thus a closer match than \textsc{Bu2019lm}\footnote{Their better fits may be partially explained by their models' higher number of free parameters, 9 and 8 respectively, compared to \textsc{Bu2019lm}'s 4.}.

We will thus fix $\sigma_{\rm sys} = 0.6$ mag in the following analyses, whose results are compiled in Table~\ref{tab:PriorsKL}. 

\begin{figure*}
    \centering
    \includegraphics[width=0.7\columnwidth]{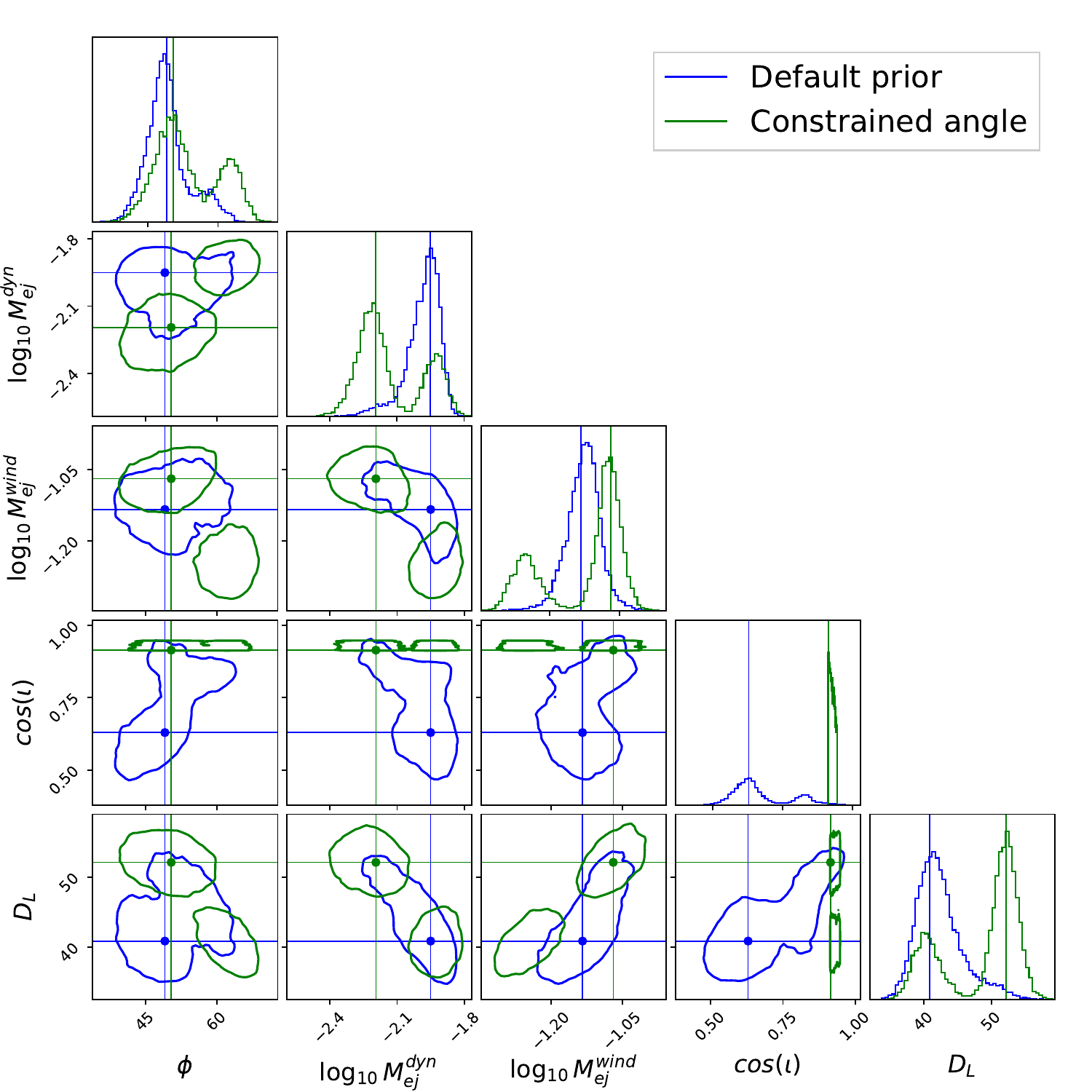}
    \includegraphics[width=0.7\columnwidth]{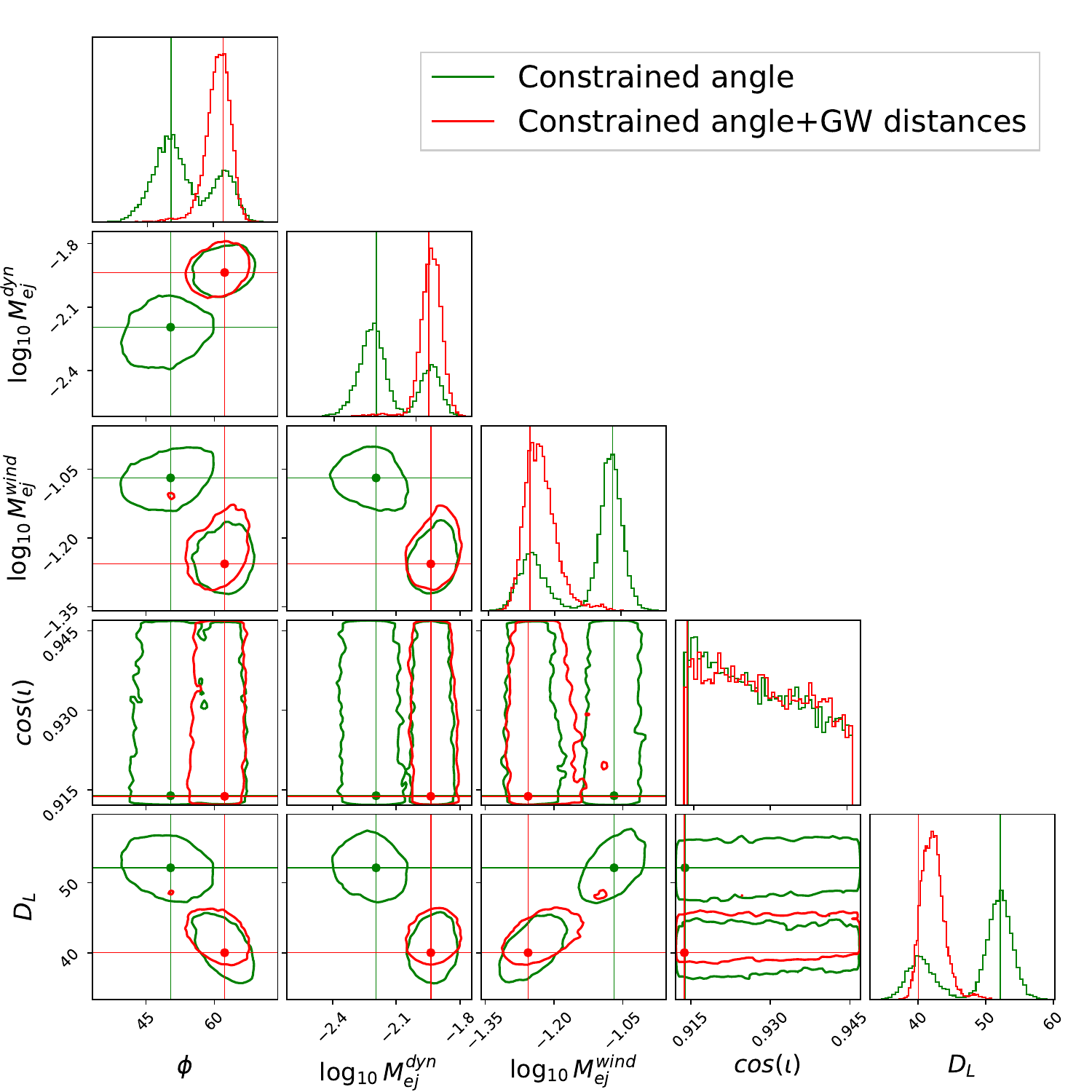}
    \caption{Parameter inference of \textbf{AT2017gfo} using \textsc{Bu2019lm} with different constraints enforced in the priors: Prior set \textbf{A0.6} in blue, \textbf{B0.6} in green, \textbf{C0.6} in red. For each analysis, the overlaid point and solid lines show the position of the best-fit (i.e highest-likelihood) parameter combination. }
    \label{fig:localmaxima}
\end{figure*}

\paragraph{KN-only uninformed analysis (default priors)}
With the prior \textbf{A0.6} (i.e. default prior set \textbf{A} from Table~\ref{tab:Priors} and $\sigma_{\rm sys} = 0.6$ mag), we now obtain a stronger constraint ($D_{KL}(P||\pi)= 11.8$ nats)\footnote{The gain of a few nats is consistent with our \S\ref{sub:perf} findings: moving from $\sigma_{\rm sys} = 1$ mag to $\sigma_{\rm sys} = 0.6$ mag roughly halves the size of posterior intervals, as shown in Table~\ref{tab:PriorsKL}.} with a $\chi^2_r = 0.90$ (see Table~\ref{tab:PriorsKL}). However, when looking at the estimated posteriors, we notice that the analysis favors a parameter region with incidence angles with $cos(\iota) \in  [0.5,0.9]$ (i.e. $\iota \in [25,60]$ degrees). This is unfortunately incompatible with independent estimates of the incidence angle using the study of superluminal motion of the relativistic jet with radio very-long-baseline interferometry \citep{Mooley:2018qfh} and optical astrometry \citep{Mooley:2022uqa}, which place the angle between $cos(\iota) \in [0.91,0.95]$.\\

\paragraph{KN-only analysis with angle constraints from GRB afterglow observations} To ensure that our results are compatible with this angle estimate, we produce another prior set where we enforce the \cite{Mooley:2022uqa} angle constraint: Prior set \textbf{B} is constructed from the default set \textbf{A} by restricting the incidence angle to $cos(\iota) \in [0.91,0.95]$.
Performing a parameter inference with this prior set \textbf{B0.6}, we now obtain a new set of posterior distributions, which displays a bimodality (see the green distribution in Figure~\ref{fig:localmaxima}). Unfortunately, the favored peak has luminosity distance $D_L = 52.1^{+3.5}_{-3.4}$ Mpc (95\% credible interval), contradicting the distance estimates from Surface Brightness Fluctuations \citep{Cantiello:2018ffy} or from GW analyses \citep{LIGOScientific:2017vwq}. Nevertheless, the second peak, although having a slightly worse likelihood than the first peak (best-fit $\chi^2_r$ of 0.97 compared to 0.96), has a luminosity distance of $40.4^{+4.6}_{-3.6}$ (95\% credible interval), compatible with the SBF estimate of $40.7 \pm 2.4$ Mpc.

\begin{figure*}
    \centering
    \includegraphics[width=0.49\columnwidth]{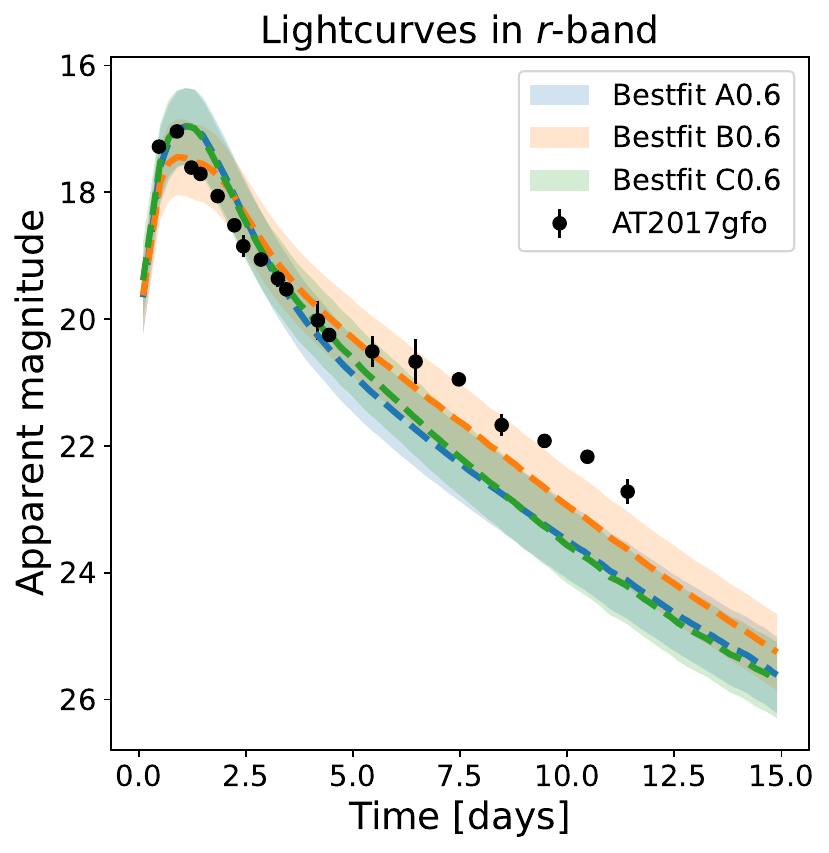}
    \includegraphics[width=0.49\columnwidth]{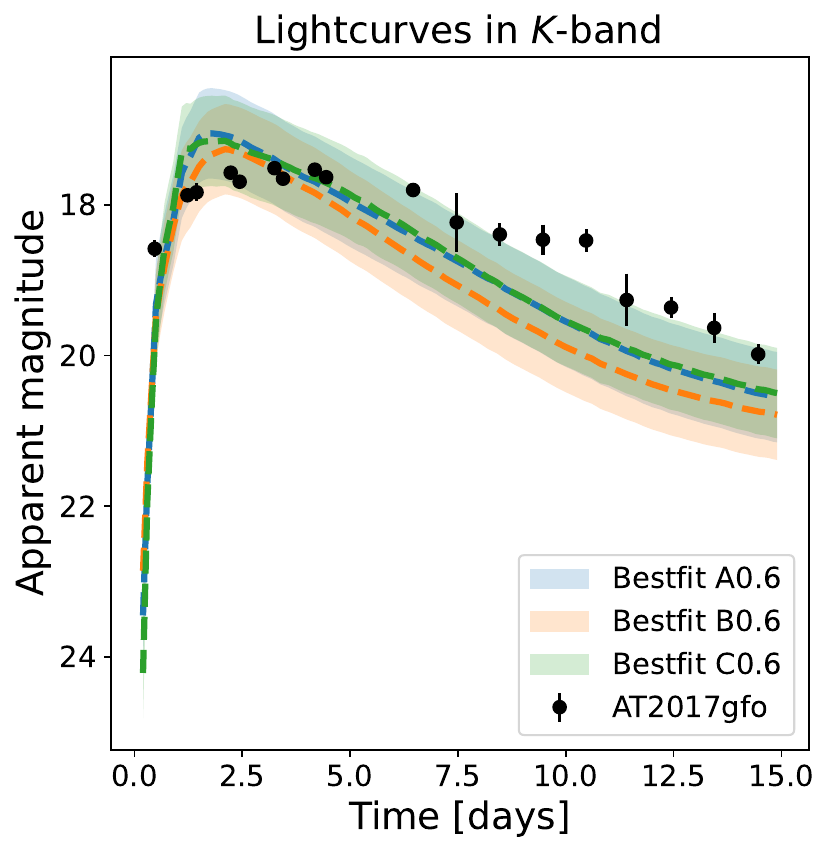}
    \caption{Lightcurves of best-fit parameter configurations obtained from KN-only analyses using prior sets \textbf{A0.6}, \textbf{B0.6} and \textbf{C0.6}, compared with the \textbf{AT2017gfo} datapoints, for a selection of filters.}
    \label{fig:170817fits}
\end{figure*}

\paragraph{KN-only analysis with GRB angle constraints and GW constraints on distance and masses} Rejecting the parameters incompatible with distance constraints, we thus produce additional prior sets \textbf{C} and \textbf{D} where, on top of the angle constraint, we constrain the distance and the ejecta masses within the bounds of the distributions of the GW-only posteriors of \cite{Dietrich:2020efo} (keeping only the subset of samples with $cos(\iota) \in [0.91,0.95]$). Prior set \textbf{C} adds only the GW distance constraint to prior \textbf{B}, then prior set \textbf{D} adds the ejecta masses constraints on top of that, as seen in Table~\ref{tab:Priors}.
Compared to the GW-only ejecta masses estimates of \cite{Dietrich:2020efo} ($\log_{10}(M_{\rm ej}^{\rm dyn}/M_\odot)= -2.15^{+0.27}_{-0.15}$, $\log_{10}(M_{\rm ej}^{\rm wind}/M_\odot)= -1.58^{+0.18}_{-0.07}$, 68\% credible intervals), our KN results with the prior set \textbf{C0.6} (angle- and distance-informed) are tighter, with 95\% credible intervals $\log_{10}(M_{\rm ej}^{\rm dyn}/M_\odot)= -1.92^{+0.09}_{-0.12}$, $\log_{10}(M_{\rm ej}^{\rm wind}/M_\odot)= -1.24^{+0.09}_{-0.05}$. Our estimated masses are compatible within, though pushing towards the high-mass end of, the bounds of the \cite{Dietrich:2020efo} posteriors.
We also note a sizable difference with the estimates of \cite{Nicholl:2021rcr} who find lower ejecta masses $\log_{10}(M_{\rm ej}^{\rm dyn}/M_\odot)\sim -2.2$ and  $\log_{10}(M_{\rm ej}^{\rm wind}/M_\odot)\sim-1.7$, that may be due to a difference of physical assumptions between KN models. 

Although the recovered incidence angle and luminosity distance basically overlapped their enforced priors, the analyses with prior sets \textbf{C} and \textbf{D} still obtain sizable prior-to-posterior reductions in ejecta masses and half-opening angles (see Table~\ref{tab:PriorsKL}); thus highlighting what constraints the KN parameter inference can produce in addition to the external constraints.\\

Every time we add a constraint in the priors, we notice that the best-fit $\chi^2_r$ gets a bit larger, indicating a slightly worse fit\footnote{Interestingly, it seems that the added constraints can make the model fit certain energy bands better, but only to the detriment of other bands. For instance, we can see in Figure~\ref{fig:170817fits} that compared to \textbf{A0.6}, the \textbf{B0.6} best-fit matches more closely the $r$-band data of \textbf{AT2017gfo} but is worse at recovering the $K$-band.}.
This is consistent with the fact that implementing the prior constraints effectively rejects regions of the parameter space which may have otherwise been favored, and thus the parameter inference sampling converges towards lower-likelihood regions: If the global maximum is rejected then we converge towards the next compatible local maximum. This is made apparent in Figure~\ref{fig:localmaxima}: the displayed best-fit parameter configurations settle on the next-best peak when constraints are added.\\

\paragraph{Complementarity of information between messengers} Looking at the K-L divergences (Table~\ref{tab:PriorsKL}) obtained with the different prior choices, we notice that as the prior gets tighter, the information gain of the KN analysis relative to that prior $D_{KL}(P_{170817}||\pi)$ decreases, but the overall constraint $D_{KL}(P_{170817}||\textbf{A})$ increases: the external constraints and the KN parameter inference are complementary. It should however be noted that adding many constraints eventually has diminishing returns: the overall divergence $D_{KL}(P_{170817}||\textbf{A})$ improves by the smallest amount when adding the ejecta mass constraints (prior set \textbf{D}) to the Angle+GWdistance prior set \textbf{C}. This means that these mass constraints are almost redundant with what the Angle+GWdistance-informed KN analysis could already obtain.

\paragraph{\textsc{Bu2025MLP} Analysis of AT2017gfo}

We now perform a parameter inference of the same \textbf{AT2017gfo} dataset but using the \textsc{Bu2025MLP} model. 
Analyses with the default prior set \textbf{A'1} yields constraints a $D_{KL}(P||\pi)= 15.4$ nats, which is much higher than the $\sim[6-12]$ nats range that would have been expected for \textbf{170817like} given the performances of \textbf{25MLP} injections. Indeed, when comparing each parameter's credible posterior intervals to their empirical typical size (Table~\ref{tab:gainsBu25}), we found that the posterior intervals for the dynamical ejecta's mass and electron fraction and the wind ejecta velocity are at least twice as small as the simulated average, whereas other parameter's intervals fall within expected ranges. Interestingly, these three parameters are also the ones exhibiting posterior distributions squished against one edge of their prior interval, which hints at the fact that posteriors concentrated at a prior edge may often have smaller-than-average characteristic interval sizes, explaining the stronger information gain.

\begin{figure*}
    \centering
    \includegraphics[width=0.49\columnwidth]{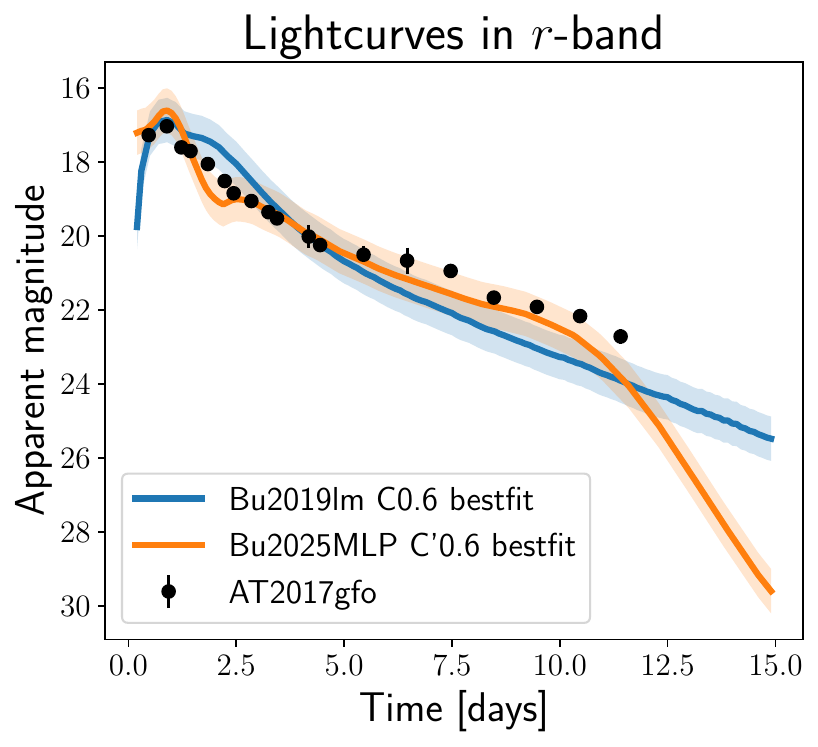}
    \includegraphics[width=0.49\columnwidth]{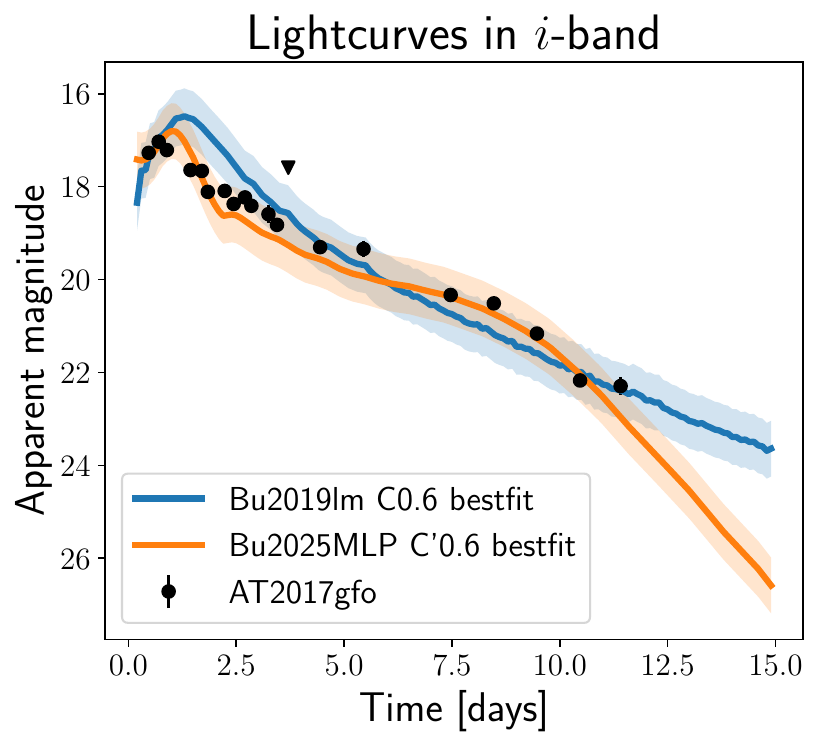}
    \includegraphics[width=0.49\columnwidth]{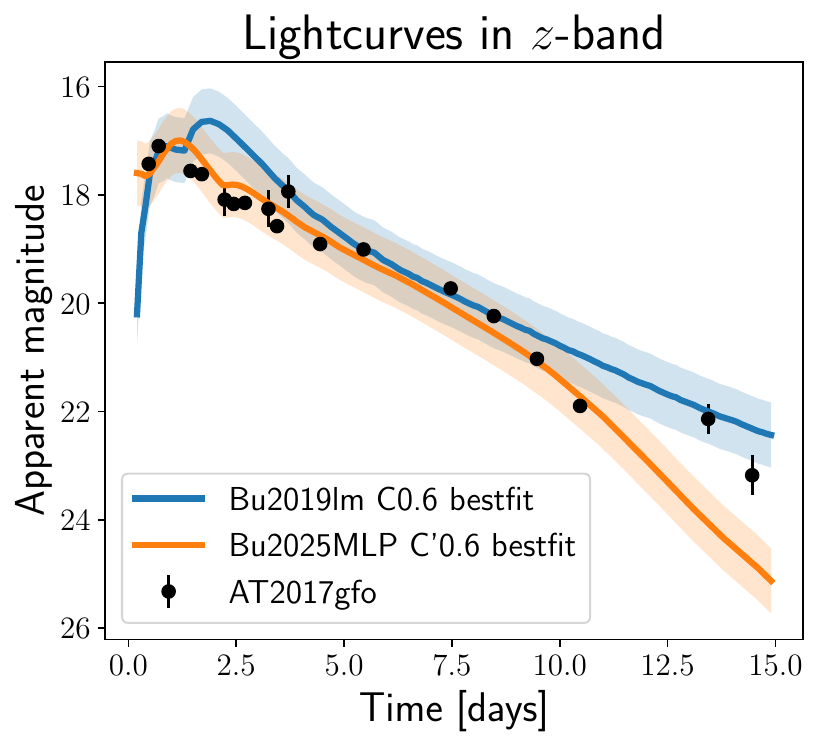}
    \includegraphics[width=0.49\columnwidth]{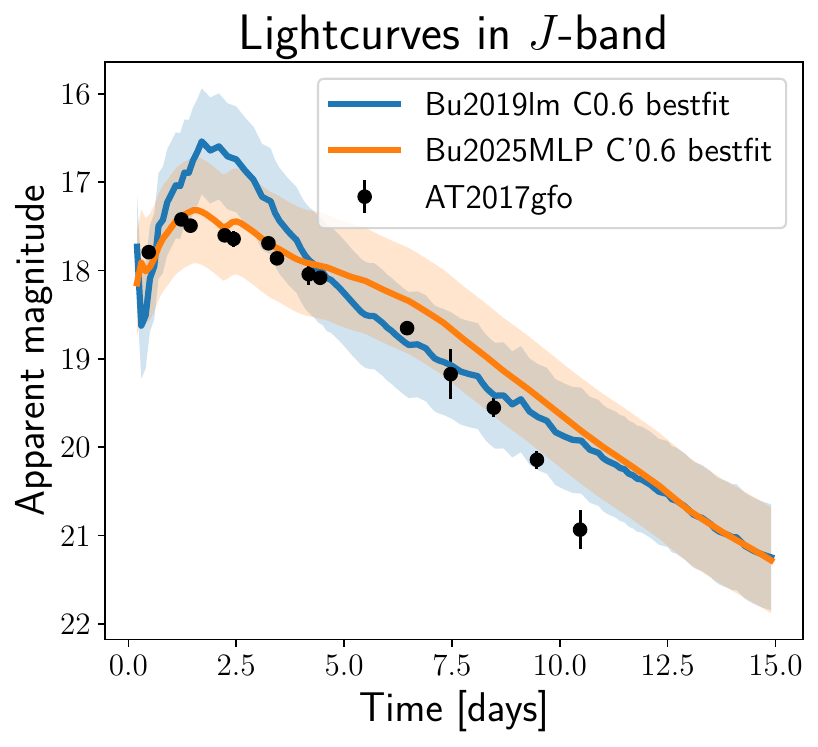}
    \includegraphics[width=0.49\columnwidth]{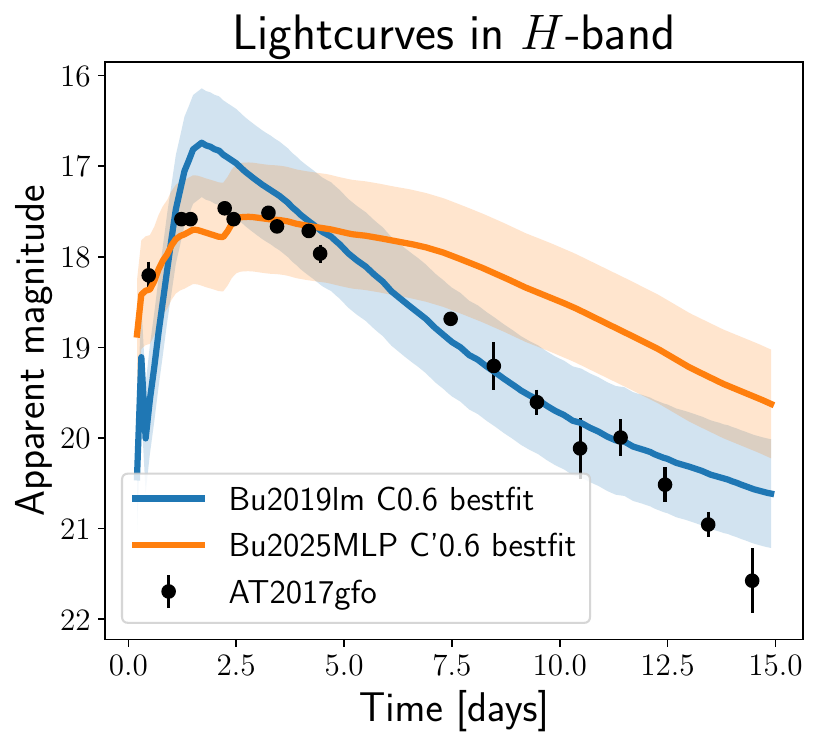}
    \includegraphics[width=0.49\columnwidth]{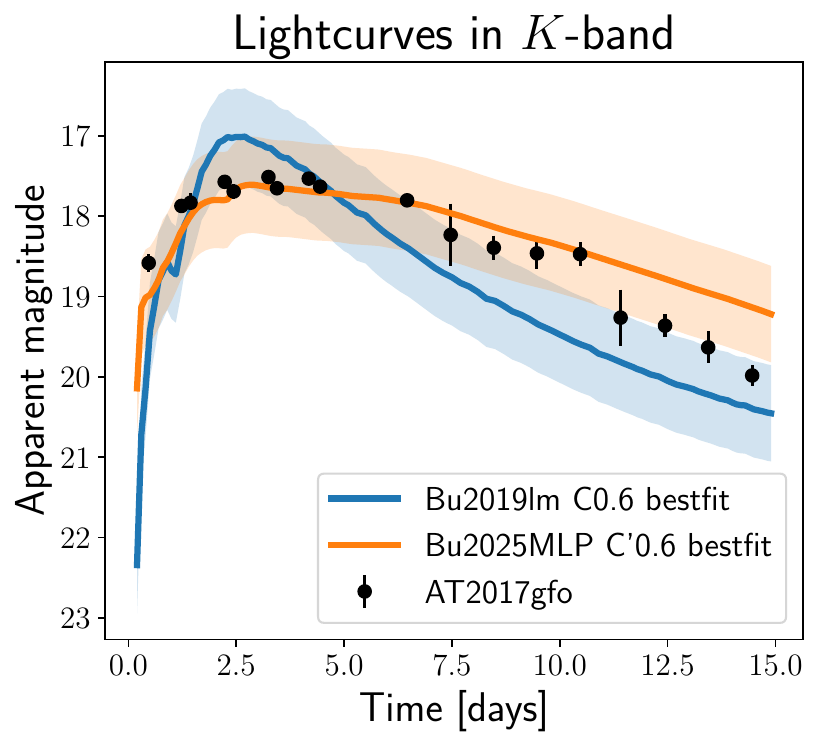}
    \caption{Lightcurves of best-fit parameter configurations obtained from AT2017gfo KN-only analyses using \textsc{Bu2019lm} with \textbf{C0.6} priors and \textsc{Bu2025MLP} with \textbf{C'0.6} priors, compared with the \textbf{AT2017gfo} datapoints, for a selection of filters.}
    \label{fig:170817fitsBu19Bu25}
\end{figure*}

When leaving the error margin $\sigma_{sys}$ as a free parameter, the maximum likelihood is found for an error margin of order $\sigma_{\rm sys} = 0.54$ mag, so \textsc{Bu2025MLP} struggles about as much as \textsc{Bu2019lm} at fitting AT2017gfo\footnote{In particular, as illustrated in Figure~\ref{fig:170817fitsBu19Bu25}, the \textsc{Bu2025MLP} bestfits tend to better match datapoints out to ten days compared to \textsc{Bu2019lm} bestfits, but then exhibit poorer fits on late-time infrared points.}. We will thus fix $\sigma_{\rm sys}=0.6$ mag in the following too.\\

Just like the \textsc{Bu2019lm} case, the inference with prior \textbf{A'0.6} obtains incidence angle posteriors hardly compatible with the \cite{Mooley:2022uqa} constraint (although this time the preferred angle values are smaller than the VLBI ones). We thus similarly construct prior set \textbf{B'} to enforce this angle constraint. Both \textbf{A'06} and \textbf{B'0.6} inferences prefers luminosity distances around 70 Mpc, incompatible with GW distances.

Unlike \textsc{Bu2019lm} however, enforcing the \cite{Dietrich:2020efo} distance and masses constraints in priors \textbf{C'} and \textbf{D'} does not solve the distance discrepancy: the \textbf{C'0.6} distance posterior sits around 51 Mpc, at the $3-\sigma$ high-end tail of the enforced Gaussian $\mathcal{N}(43.4, 2^2)$ Mpc prior.

Compared to \textsc{Bu2019lm}, our \textbf{C'0.6} inference obtains a much higher dynamical ejecta mass $\log_{10}(M_{\rm ej}^{\rm dyn}/M_\odot)= -1.57^{+0.09}_{-0.08}$, but a lower wind ejecta mass $\log_{10}(M_{\rm ej}^{\rm wind}/M_\odot)= -1.24^{+0.09}_{-0.05}$ (both 95\% credible intervals) in compensation; both models thus agree on a total ejecta mass around 0.07 $M_\odot$ \footnote{As a point of comparison, recent multi-messenger joint analyses of GW170817+GRB170817A+AT2017gfo in \cite{Rose:2026cnm} favor a higher total ejecta mass around 0.09 $M_\odot$ with very little dynamical ejecta ($\log_{10}(M_{\rm ej}^{\rm dyn}/M_\odot)= -2.47^{+0.18}_{-0.12}$, $\log_{10}(M_{\rm ej}^{\rm wind}/M_\odot)= -1.06^{+0.06}_{-0.07}$, 68\% credible intervals), but with a higher inclination angle around 0.44 radians, outside the angle-constrained prior we restricted ourselves to.}.\\

The overall behavior of losing in fit accuracy (increasing $\chi^2_r$) when adding prior constraints, while still producing complementary information (increasing the overall $D_{KL}$ information gain), holds for \textsc{Bu2025MLP} just like it was for \textsc{Bu2019lm}.

\begin{sidewaystable}
\hspace*{-1.5cm}
\scalebox{0.84}{
\begin{tabular}{|l|r|llllllll|l|l|l|l|}
\hline
\multicolumn{2}{|l|}{Prior set} & \multicolumn{8}{l|}{Parameters} &  & \multicolumn{3}{l|}{K-L divergences}\\
\hline
\multicolumn{2}{|l|}{} & $\log_{10}M_{\rm ej}^{\rm dyn}$ & $v_{\rm ej}^{\rm dyn}$ & $Y_e^{\rm dyn}$ & $\log_{10}M_{\rm ej}^{\rm wind}$ & $v_{\rm ej}^{\rm wind}$ & $Y_e^{\rm wind}$ & $\iota$ & $D_L$ & best-fit & $D_{KL}(\pi||\textbf{A'})$ & $D_{KL}(P_{170817}||\pi)$ & $D_{KL}(P_{170817}||\textbf{A'})$ \\
\multicolumn{2}{|l|}{} & [$M_\odot$] & [c] & & [$M_\odot$] & [c] & &  [radian] & [Mpc] & $\chi^2_r$ & & & \\
\hline
\textbf{A'}: & Prior & [-3,-1.3] & [0.12,0.28] & [0.15,0.35] & [-2,-0.89] & [0.05,0.15] & [0.2,0.4] & [0,$\pi/2$] & Log$\mathcal{U}(1,500)$& & 0 &  &  \\
\multicolumn{2}{|r|}{Results: \textbf{A'1}} & $-1.34^{+0.04}_{-0.14}$ & $0.24^{+0.03}_{-0.05}$ & $0.33^{+0.02}_{-0.02}$ & $-1.15^{+0.25}_{-0.12}$ & $0.051^{+0.003}_{-0.001}$ & $0.24^{+0.08}_{-0.01}$ & $0.23^{+0.62}_{-0.22}$ & $71.17^{+12.15}_{-9.72}$ & 0.32 & & 15.40 & 15.40\\
\multicolumn{2}{|r|}{\textbf{A'0.6}} & $-1.33^{+0.02}_{-0.08}$ & $0.25^{+0.02}_{-0.04}$ & $0.33^{+0.02}_{-0.01}$ & $-1.20^{+0.12}_{-0.05}$ & $0.050^{+0.001}_{-0.000}$ & $0.24^{+0.01}_{-0.01}$ & $0.21^{+0.27}_{-0.20}$ & $70.66^{+6.04}_{-5.82}$ & 0.83 & & 19.71 & 19.71\\
\hline
\textbf{B'}: & Prior & [-3,-1.3] & [0.12,0.28] & [0.15,0.35] & [-2,-0.89] & [0.05,0.15] & [0.2,0.4] & [0.33,0.42] & Log$\mathcal{U}(1,500)$& & 2.86 & & \\
\multicolumn{2}{|r|}{Results: \textbf{B'1}} & $-1.34^{+0.03}_{-0.12}$ & $0.24^{+0.04}_{-0.04}$ & $0.33^{+0.02}_{-0.02}$ & $-1.15^{+0.23}_{-0.11}$ & $0.051^{+0.003}_{-0.001}$ & $0.24^{+0.07}_{-0.01}$ & $0.37^{+0.05}_{-0.04}$ & $69.90^{+12.58}_{-8.42}$ & 0.32 & & 14.66 & 17.52\\
\multicolumn{2}{|r|}{\textbf{B'0.6}} & $-1.32^{+0.02}_{-0.06}$ & $0.25^{+0.02}_{-0.03}$ & $0.33^{+0.02}_{-0.01}$ & $-1.19^{+0.10}_{-0.05}$ & $0.050^{+0.001}_{-0.000}$ & $0.24^{+0.01}_{-0.01}$ & $0.36^{+0.05}_{-0.03}$ & $69.21^{+5.53}_{-5.25}$ & 0.82 & & 18.67 & 21.53\\
\hline
\textbf{C'} & Prior & [-3,-1.3] & [0.12,0.28] & [0.15,0.35] & [-2,-0.89] & [0.05,0.15] & [0.2,0.4] & [0.33,0.42] & $\mathcal{N}(43.4, 2^2)$ & & 6.35 & & \\
\multicolumn{2}{|r|}{Results: \textbf{C'1}} & $-1.61^{+0.12}_{-0.15}$ & $0.24^{+0.03}_{-0.02}$ & $0.345^{+0.004}_{-0.016}$ & $-1.31^{+0.06}_{-0.06}$ & $0.050^{+0.002}_{-0.000}$ & $0.24^{+0.01}_{-0.01}$ & $0.38^{+0.04}_{-0.05}$ & $47.69^{+2.90}_{-3.09}$ & 0.42 & & 15.56 & 21.91\\
\multicolumn{2}{|r|}{\textbf{C'0.6}} & $-1.57^{+0.09}_{-0.08}$ & $0.24^{+0.02}_{-0.01}$ & $0.348^{+0.002}_{-0.011}$ & $-1.30^{+0.03}_{-0.03}$ & $0.050^{+0.001}_{-0.000}$ & $0.243^{+0.004}_{-0.005}$ & $0.37^{+0.04}_{-0.04}$ & $50.73^{+2.88}_{-2.5}$ & 1.00 & & 23.71 & 30.06\\
\hline
\textbf{D'} & Prior & [-2.6,-1.5] & [0.12,0.28] & [0.15,0.35] & [-1.9,-1.2] & [0.05,0.15] & [0.2,0.4] & [0.33,0.42] & $\mathcal{N}(43.4, 2^2)$ & & 7.24 & & \\
\multicolumn{2}{|r|}{Results: \textbf{D'1}} & $-1.61^{+0.09}_{-0.14}$ & $0.24^{+0.03}_{-0.02}$ & $0.345^{+0.005}_{-0.016}$ & $-1.31^{+0.05}_{-0.06}$ & $0.050^{+0.002}_{-0.000}$ & $0.24^{+0.01}_{-0.01}$ & $0.38^{+0.04}_{-0.05}$ & $47.60^{+2.96}_{-2.90}$ & 0.43 & & 15.58 & 22.83\\
\multicolumn{2}{|r|}{\textbf{D'0.6}} & $-1.58^{+0.07}_{-0.08}$ & $0.24^{+0.02}_{-0.01}$ & $0.347^{+0.002}_{-0.008}$ & $-1.30^{+0.03}_{-0.03}$ & $0.050^{+0.001}_{-0.000}$ & $0.244^{+0.003}_{-0.004}$ & $0.37^{+0.04}_{-0.04}$ & $50.61^{+2.63}_{-2.43}$ & 1.02 & & 23.25 & 30.50\\
\hline
\end{tabular}}
\caption{Prior ranges for different combinations of constraints, and the corresponding performances of \textsc{Bu2025MLP} parameter inferences on \textbf{AT2017gfo}: Posterior 95\% credible intervals, best-fit $\chi^2_r$ and K-L divergences between each prior $\pi$, the default prior $\textbf{A'}$, and each analysis' posterior $P_{170817}$. Log$\mathcal{U}(a,b)$ describes a log-uniform distribution from $a$ to $b$, and $\mathcal{N}(\mu, \sigma^2)$ is a Gaussian distribution with mean $\mu$ and variance $\sigma^2$; all other prior intervals obey a uniform distribution over their given range.}
\label{tab:PriorsKLBu25}
\end{sidewaystable}

\section{Conclusion and Discussion}

In this work, we have identified many sources of uncertainties involved in the observation and modeling of KN lightcurves and the associated parameter inference framework. We first listed the effects that should be taken into account when computing the uncertainty of observational datapoints, noting that these effects combine into observational errors that are typically of order 0.1-0.2~mag. Then, on the modeling side, we enumerated systematic uncertainties due to the model's physical assumptions (variation of 1-3 mag between different models) or to their surrogate model implementations (interpolation errors of order 0.1-0.3 mag). We discussed how these effects combine under the umbrella of the model error margin $\sigma_{sys}$, which is shown to dominate over the observational error contribution in the literature of KN parameter inferences. Further studies are needed on these modeling biases in order to understand their error contributions better and put stricter bounds on them, so that modeling errors can be lowered.\\

We also propose our simulation recovery tests as diagnostic tests to hunt for systematic biases and assess average performances when benchmarking model implementations. This approach can be useful for comparing the performance of observing strategies and identifying which parts of the lightcurves (such as filters, or time windows) would bring the biggest information gain, so that observatory networks could optimize their strategy given their available observing resources. For example, using the metrics of K-L divergence and prior-to-posterior reduction, we showcased the performance of a KN parameter inference with a multi-band ($grizyJHK$) observational cadence averaging one observation per filter per night, in the example of the \textsc{Bu2019lm} model: The expected performance is an information gain of $D_{KL}(P||\pi)\sim 8$ nats (for an error margin $\sigma_{sys}$=1~mag), and we show that the strongest constraints are obtained for the distance and the wind ejecta mass, whereas the poorest constraints are on the incidence angle; with the actual size of posterior distributions strongly correlated with the $\sigma_{sys}$ error margin. We similarly study the \textsc{Bu2025MLP} model, for which we find that the strongest constraints are obtained for the wind ejecta electron fraction.

We quantify the overall trend that the more datapoints are considered, the more information is gained; but we also highlight the complementarity of using both optical and infrared data, as well as using both early- and late-time points, which is crucial to design observation strategies that maximize information gain.
Finally, for various values of $\sigma_{sys}$, we identify cadence thresholds that are needed to obtain significant constrains on a parameter's posterior distribution, and we show these thresholds depends on the considered model and parameter.\\

To check the validity of our simulated predictions, we performed in section \ref{sub:AT2017gfo} an analysis of the \textbf{AT2017gfo} dataset with the \textsc{Bu2019lm} model. For the $\sigma_{sys}$=1~mag analysis with default priors, we indeed obtained constraints that were quantitatively compatible with the performances estimated from simulations in section \ref{sub:perf}.
Then, searching for the optimal error margin, we find a maximum likelihood for $\sigma_{sys}\sim0.6$~mag. The corresponding analysis with default priors unfortunately favors incidence angles incompatible with the $cos(\iota) \in [0.91,0.95]$ constraint from \cite{Mooley:2022uqa}. By enforcing in the \textsc{Bu2019lm} priors this angle constraint as well as GW distance constraints \citep{Dietrich:2020efo}, we eventually obtain the results of Table~\ref{tab:PriorsKL}: we estimate ejecta masses $\log_{10}(M_{\rm ej}^{\rm dyn}/M_\odot)= -1.92^{+0.09}_{-0.12}$ and $\log_{10}(M_{\rm ej}^{\rm wind}/M_\odot)= -1.24^{+0.09}_{-0.05}$ (95\% credible intervals), compatible with GW results \citep{Dietrich:2020efo} but with higher values and tighter constraints.

Repeating the analysis with the \textsc{Bu2025MLP} model, we also find an optimal $\sigma_{sys}\sim0.6$~mag and similar behavior (full results in Table~\ref{tab:PriorsKLBu25}), albeit with a higher $\log_{10}(M_{\rm ej}^{\rm dyn}/M_\odot)= -1.57^{+0.09}_{-0.08}$ and a lower $\log_{10}(M_{\rm ej}^{\rm wind}/M_\odot)= -1.24^{+0.09}_{-0.05}$ (95\% credible intervals).\\

Finally, we wish to address the question of compatibility between models and observations. One difficulty of Bayesian parameter inference is that there is no clear-cut, binary criterion to say whether a model is compatible or not with the observations of a given event. We highlight in this work the metrics of $\sigma_{sys}$ and $\chi^2_r$ as the relevant quantities to discuss these questions, and we recommend the value of $\sigma_{sys}$ maximizing the likelihood (i.e for which $\chi^2_r \sim 1$) as a reasonable goodness-of-fit metric. Since this metric has a continuous spectrum of possible values, we do not identify a particular threshold over which a model is deemed incompatible with data, but we prefer to compare the performance of different models to see which one best fits the data: the lower this optimal $\sigma_{sys}$, the better the match. 

Moreover, the performance of a model strongly depends on the priors imposed on the parameter space, as seen in the example of our AT2017gfo analyses. In our opinion, if a model's KN-only analysis of an event returns posterior distributions whose highest-likelihood regions do not match the preferred parameter region obtained by independent studies of the event (e.g using other messengers), that KN model should not be rejected altogether for that event, but can instead still provide inference constraints in the (lower-likelihood) parameter regions compatible with the other studies' results. This is rather an indication that separate messengers each hold incomplete information and that the best constraint would come from joining all the datasets in a multi-messenger approach.

Then, by enforcing constraints from other messengers in the KN priors like in section~\ref{sub:AT2017gfo}, one can look at the $D_{KL}(P||\pi)$ divergence to get an idea, not of model-data compatibility, but of the usefulness of the model to get constraints from KN analysis in addition to the applied external constraints.
As an example, if the prior is too tight and/or the $\sigma_{sys}$ error margin is too large, one can imagine a case in which all of the allowed KN parameter combinations fit within the error margin, and thus no significant gain of information from prior to posterior would be obtained; which would show the KN analysis to be redundant with the enforced constraints. 

In a general case, in order to assess the information gain of a KN parameter inference and its complementarity to external analyses from other messengers, one can therefore run analyses with multiple sets of prior constraints in order to see which ones are compatible, redundant or contradictory to the KN results. We note that such an approach is not limited to KN analyses, but could be applied to many kinds of multi-messenger studies.

\section*{Acknowledgements}

T.H.D. thanks Tim Dietrich, Michael Coughlin, Brendan King, Thibeau Wouters, Hauke Koehn and the rest of the NMMA collaborators for interesting discussions during weekly NMMA calls.
M.P. acknowledges support from FNRS and IISN 4.4503.
This work used computing resources from Expanse at the San Diego Supercomputer Cluster, supported by the NSF. P.T.H.P. is supported by the research program of the Netherlands Organization for Scientific Research (NWO) under grant number VI.Veni.232.021. (\url{https://doi.org/10.61686/WEYCA68506})  

\section*{Data Availability}

The datasets used in this work (either simulations or \textbf{AT2017gfo}) and their corresponding parameter estimation results are available in \url{https://github.com/Thomhus/kilonovae-H0}.

\appendix

\section{Defining the datasets and inference parameters used in this work}

\subsection{Lightcurves used}
\label{app:LC}

\paragraph{\textbf{AT2017gfo}}
The AT2017gfo lightcurve studied in this work is that of \cite{Coughlin:2018miv} based on the compiled dataset of \cite{Villar:2017wcc}. Since these studies reported only reported the apparent magnitudes, we corrected this dataset from Milky Way extinction.

The absorption by Milky Way dust is computed at a given wavelength $\lambda$ as:
\begin{equation}
    A(\lambda) = r_{\lambda/V} * R_V * E(B-V)
\end{equation}
where $E(B-V) = A(B)-A(V)$ is the reddening, catalogued over the whole sky, $R_V=A(V) / E(B-V)$ is the relative visibility, estimated at $R_V=3.1$ for most of the Milky Way, and $r_{\lambda/V} =A(\lambda) / A(V)$ is the relative absorption of different wavelengths (of order unity, catalogued for $R_V=3.1$, has different values if $R_V$ changes).
We applied to the \cite{Coughlin_2020} dataset an extinction correction computed with $E(B-V) \sim 0.123$ mag \citep{Schlafly2011}.

The resulting dataset is referred to as \textbf{AT2017gfo} in this work.

\paragraph{\textbf{sim17}} We define a mock lightcurve simulated by the \textsc{Bu2019lm} model. We inject the \textsc{Bu2019lm} parameters that best fit AT2017gfo under prior set \textbf{A1} ($\log_{10}(M_{\rm ej}^{\rm dyn}/M_\odot)=-2.24$, $\log_{10}(M_{\rm ej}^{\rm wind}/M_\odot)=-1.00$, $\phi = 31$ degrees, $\cos(\theta)=0.29$ and $D_L =41.2$ Mpc), that we will refer to as \textbf{mock17} parameters. We compute the corresponding lightcurves, producing a dense dataset with a point every 0.1 day from 0 to 15 days, in each filter ($g$-, $r$-, $i$-, $z$-, $y$-, $J$-, $H$- and $K$-band). We will call this the \textbf{sim17} lightcurve, with the \textbf{10perday} cadence since it has 10 observations per day in each of the eight filters. 

We then sample from the full \textbf{10perday} lightcurve to create lower cadence \textbf{sim17} datasets. For instance, we define the \textbf{1perday} cadence by randomly sampling 15 points in each filter across the [0,15]-day range, to get an average of one observation per filter per day.
We can also sample to keep only the same observation times and filters as \textbf{AT2017gfo}'s (thus an average of a bit more than 1 observation per day in each filter), which we will call the \textbf{170817like} cadence.

Although no noise is added to the magnitudes simulated by the \textsc{Bu2019lm} surrogate model, non-zero uncertainties are needed for \texttt{NMMA} to function. The observational uncertainties of \textbf{sim17} datapoints are thus declared to be $\sigma_i^j = 0.05$ mag, negligible before the other uncertainty contributions involved in the analyses.

\paragraph{\textbf{toymodel}} For the toy model study of section \ref{sub:err_sys}, we produced the \textbf{toymodel} as follows:
We used the simulated \textbf{sim17} lightcurve with the \textbf{170817like} cadence, and added a Gaussian noise of variance $\frac{1}{N}\sum_{ij}(m_i^j-m_i^{j,\rm model}(\vec{\theta}))^2 = (0.5~\text{mag})^2$ mag to all the datapoints, but declared the observational uncertainties to be smaller: $\sigma_{meas} = 0.3$ mag.

\paragraph{\textbf{ManySims}} We sampled 22 parameter combinations sampled randomly all across the \textsc{Bu2019lm} parameter space, and simulated for each the corresponding \textbf{10perday} lightcurves (a point every 0.1 day from 0 to 15 days, in each filter $grizyJHK$). For each simulation we then drew 20 subsets of datapoints to construct sibling lightcurves with the \textbf{1perday} cadence. To emulate a detectability similar to AT2017gfo, we place the simulated events at $D_L=40$ Mpc and remove from the lightcurves any point fainter than magnitude 24, as few instruments can consistently reach these depths\footnote{This magnitude 24 threshold only removes a few percent of datapoints for the infrared filters that fade slowly, but it removes 11\% of $i$-band points, 28\% of $r$-band points and 49\% of $g$-band
points (when averaged across the 22 injected lightcurves), mainly at the late-time end of the lightcurves.}. 
The observational uncertainties are declared to be $\sigma_i^j = 0.05$ mag.

\paragraph{\textbf{25MLPSims}} We sampled 100 parameter combinations sampled randomly all across the \textsc{Bu2025MLP} parameter space, and simulated for each the corresponding \textbf{10perday} $grizyJHK$ lightcurves, which were then randomly downsampled to a \textbf{1perday} cadence. We place the simulated events at $D_L=40$ Mpc and remove from the lightcurves any point fainter than the magnitude 24 threshold\footnote{In the \textsc{Bu2025MLP} case, lightcurves tend to be slightly fainter than the \textsc{Bu2019lm} predictions, and so more than half of optical datapoints (up to 80\% for $g$-band) are removed, effectively only keeping points around the lightcurve's peak in the analyses.}. The observational uncertainties are declared to be $\sigma_i^j = 0.05$ mag.

\paragraph{\textbf{Injection5}} Out of the 100 \textbf{25MLPSims} injections, we arbitrarily chose the fifth injection and produced its lightcurves at various cadences to be analysed with many $\sigma_{\rm sys}$ values. Its injected parameters are $\log_{10}M_{\rm ej}^{\rm dyn}\simeq-2.789$, $v_{\rm ej}^{\rm dyn}\simeq 0.210~c$, $Y_e^{\rm dyn}\simeq 0.267$, $\log_{10}M_{\rm ej}^{\rm wind}\simeq-1.476$, $v_{\rm ej}^{\rm wind}\simeq0.077~c$, $Y_e^{\rm wind}\simeq0.269$, $\iota\simeq0.281$ rad and $D_L=40$ Mpc.

\subsection{\textsc{Bu2019lm} and \textsc{Bu2025MLP} priors used}
\label{app:priors}

The default \textsc{Bu2019lm} priors, denoted in this work as prior set \textbf{A}, are uniform distributions in $\phi$, $\log_{10}M_{\rm ej}^{\rm dyn}$n $\log_{10}M_{\rm ej}^{\rm wind}$, $cos(\iota)$ and $\log(D_L)$, as described in Table~\ref{tab:Priors}.

When running the analysis with a fixed error margin $\sigma_{\rm sys}$=1 mag, we denote it as \textbf{A1}; for $\sigma_{\rm sys}$=0.6 mag we note \textbf{A0.6}.

Prior set \textbf{B} is constructed from the default set \textbf{A} by enforcing the incidence angle constraint from \cite{Mooley:2022uqa}: $cos(\iota) \in [0.91,0.95]$.

To add the constraints of GW-only analyses, we iteratively produce additional prior sets where, on top of the angle constraint, we constrain the distance and the ejecta masses within the bounds of the distributions of the GW-only posteriors of \cite{Dietrich:2020efo} (keeping only the subset of samples with $cos(\iota) \in [0.91,0.95]$).
Prior set \textbf{C} adds the distance constraint to prior \textbf{B}, then prior set \textbf{D} adds the ejecta masses constraints on top of that, as seen in Table~\ref{tab:Priors}.\\

We similarly denote the default \textsc{Bu2025MLP} priors as prior set \textbf{A'}, and construct from it iteratively the prior sets \textbf{B'}, \textbf{C'} and \textbf{D'} by enforcing the same angle, distance and ejecta masses constraints (see Table~\ref{tab:PriorsBu25}).

\begin{table*}
\centering
\begin{tabular}{|c|ccccc|}
\hline
Prior set & \multicolumn{5}{c|}{Parameters} \\
\hline
 & $\phi$ & $\log_{10}M_{\rm ej}^{\rm dyn}$ & $\log_{10}M_{\rm ej}^{\rm wind}$ & $cos(\iota)$ & $D_L$ \\
& [deg] & [$M_\odot$] & [$M_\odot$] & & [Mpc] \\
\hline
\textbf{A}: Default Prior Range & [15,75] & [-3,-1] & [-3,-0.5] & [0,1] & Log$\mathcal{U}(1,500)$ \\
\hline
\textbf{B}: Angle constraint & [15,75] & [-3,-1] & [-3,-0.5] & [0.91,0.95] & Log$\mathcal{U}(1,500)$ \\
\hline
\textbf{C}: Angle \& GW distance & [15,75] & [-3,-1] & [-3,-0.5] & [0.91,0.95] & $\mathcal{N}(43.4, 2^2)$ \\
\hline
\textbf{D}: Angle \& GW distance+masses & [15,75] & [-2.6,-1.5] & [-1.9,-1.2] & [0.91,0.95] & $\mathcal{N}(43.4, 2^2)$ \\
\hline
\end{tabular}
\caption{\textsc{Bu2019lm} Prior ranges for different combinations of constraints. Log$\mathcal{U}(a,b)$ describes a log-uniform distribution from $a$ to $b$, and $\mathcal{N}(\mu, \sigma^2)$ is a Gaussian distribution with mean $\mu$ and variance $\sigma^2$; all other prior intervals obey a uniform distribution over their given range.}
\label{tab:Priors}
\end{table*}

\begin{table*}
\hspace*{-2.8cm}
\scalebox{0.85}{\begin{tabular}{|c|cccccccc|}
\hline
Prior set & \multicolumn{8}{c|}{Parameters} \\
\hline
 &  $\log_{10}M_{\rm ej}^{\rm dyn}$ & $v_{\rm ej}^{\rm dyn}$ & $Y_e^{\rm dyn}$ & $\log_{10}M_{\rm ej}^{\rm wind}$ & $v_{\rm ej}^{\rm wind}$ & $Y_e^{\rm wind}$ & $\iota$ & $D_L$  \\
& [$M_\odot$] & [c] & & [$M_\odot$] & [c] & &  [radian] & [Mpc] \\
\hline
\textbf{A'}: Default Prior Range & [-3,-1.3] & [0.12,0.28] & [0.15,0.35] & [-2,-0.89] & [0.05,0.15] & [0.2,0.4] & [0,$\pi/2$] & Log$\mathcal{U}(1,500)$ \\
\hline
\textbf{B'}: Angle constraint & [-3,-1.3] & [0.12,0.28] & [0.15,0.35] & [-2,-0.89] & [0.05,0.15] & [0.2,0.4] & [0.33,0.42] & Log$\mathcal{U}(1,500)$ \\
\hline
\textbf{C'}: Angle \& GW distance & [-3,-1.3] & [0.12,0.28] & [0.15,0.35] & [-2,-0.89] & [0.05,0.15] & [0.2,0.4] & [0.33,0.42] & $\mathcal{N}(43.4, 2^2)$  \\
\hline
\textbf{D'}: Angle \& GW distance+masses & [-2.6,-1.5] & [0.12,0.28] & [0.15,0.35] & [-1.9,-1.2] & [0.05,0.15] & [0.2,0.4] & [0.33,0.42] & $\mathcal{N}(43.4, 2^2)$ \\
\hline
\end{tabular}}
\caption{Same table as Table~\ref{tab:Priors}, applied to the \textsc{Bu2025MLP} model.}
\label{tab:PriorsBu25}
\end{table*}

\section{Details on sources of uncertainty}
\subsection{Sources of uncertainty in the estimation of observed magnitudes}
\label{sub:DataUnc}

The process from observation images to analysis-ready dataset contains many steps which can introduce uncertainty in the estimated magnitude, which we discuss in the following.

\subsubsection{Data analysis uncertainties}

Many steps involved in the photometric analysis of telescope images contribute to the magnitude uncertainties reported in observational datasets. We summarize them below:

\paragraph{Separating the transient signal from the images' background} 

In order to quantify the flux corresponding to the transient, one must first estimate the local mean and root-mean-square deviation (RMS) of the background noise in a region of the image around the source. Once the mean background flux is subtracted, the transient flux in computed by summing the pixel values within a given aperture\footnote{a subset of pixels centered the transient coordinates, covering a small region corresponding roughly to the telescope+detector's point spread function characteristic.}. The signal-to-noise ratio (SNR $= \frac{\text{ transient flux}}{\text{background flux RMS}}$) quantifies the significance of the detection, and the instrumental magnitude uncertainties ($\sim \frac{2.5}{\ln (10)\times SNR}$) are large when the SNR is small. 
Lower SNRs happen as sources get fainter or as the noise level get bigger, where the latter highly depends on the sensitivity of the instrument and the exposure time, and can be further polluted by bad weather conditions or the presence of bright sources (e.g. the Moon) in the field of view.

One must also be cautious to subtract other overlapping sources that might contribute to flux in the integration aperture. If the transient is within a host galaxy, then the galaxy flux in the aperture should be estimated and accounted for. Another difficulty can be crowding, when separating the transient from neighbouring sources is difficult, which can happen when the image has poor pixel resolution or when imaging a crowded field such as the galactic plane.

\paragraph{Photometric calibration}
The observed instrumental magnitude must then be converted to the transient's apparent magnitude in order to correct for absorption by Earth's atmosphere. For many telescopes, this can be done by crossmatching reference catalogues with the image's sources: identifying reference stars whose magnitudes are catalogued and building a calibration scale. 
Such calibration step adds a magnitude uncertainty contribution, that typically increases when few catalogued stars are detected in the image or when the telescope's filters poorly match the catalogue's bandpasses. \\

\subsubsection{Correction uncertainties}

The estimated apparent magnitudes then need to be corrected for the attenuation and redshifting of the radiation along the line of sight, in order to recover the intrinsic magnitude emitted at the source. The obtained lightcurve will then be the one compared with model predictions.

\paragraph{Correction of the extinction along the line of sight, $\sigma_{meas} <$ 0.5 mag}
First, we want to correct for the absorption of light by dust and gas along the line of sight, mainly occurring in the Milky Way and host galaxy. We note that photometric calibration using reference stars already compensated the absorption by Earth's atmosphere.

The absorption by Milky Way dust can be computed from dust reddening maps (see \cite{Schlafly2011} as a reference) using catalogued filter-dependent absorption factors (see \S~\ref{app:LC}).
For example, these corrections applied for the AT2017gfo coordinates are of the size of a few tenths of mag and should not be neglected. 
A comparison of various reddening catalogues~\citep{HussenotThesis} exhibits a typical between-maps systematic uncertainty of $\sim 10 \%$ relative error in reddening, going up to $\sim 50 \%$ relative error in some sky locations. The corresponding systematic uncertainty on the extinction correction is therefore usually on the order of hundredths of mag, but can reach tenths of mag in high-reddening regions.

Additionally, since the absorption depends on the wavelength, instruments using slightly different passbands but calibrated with the same catalogue filter will have small discrepancies in extinction correction, leading to (usually negligible) uncertainties around hundredths of mag.\\

Moreover, if the emission spectrum of the host galaxy has been observed, one can then infer the expected absorption from host galaxy dust, although it should be noted that the estimated average extinction may differ from the exact extinction at the location of the transient. For example, there is no significant dust extinction from NGC4993, the host galaxy of AT2017gfo \citep{Blanchard:2017csd}, 
so this correction is not included further in our study.\\ 

\paragraph{Redshift effect} The light collected in our observer frame has been redshifted with respect to the emitter frame. In this sense, the flux integrated in the observer-frame energy ranges should be compared to the model fluxes in corresponding blue-shifted ranges.
For KN models that predict the spectral energy distribution of a kilonova~\citep{Koehn:2025zzb}, one can do this accurately by applying the redshift to the SED and integrating the result over the observer filter's bandpass.
\subsection{Sources of uncertainty in simulating KN lightcurves}
\label{sub:ModelUnc}

The prediction of lightcurves by KN models is also subject to multiple error contributions arising from model assumptions or implementations, which we discuss in the following. 

\subsubsection{Model intrinsic uncertainties}
\label{ssub:ModelIntrinsic}

\paragraph{Diversity of KN modeling approaches, $\sigma_{mod} <$ 3 mag}
Many different KN models (predicting lightcurves of either NSBH or BNS mergers) have been developed in the past decade, but their choices of physical approximations and computing architecture vary significantly. Some models rely on (semi-)analytical computations \cite{Metzger:2016pju, Breschi:2021tbm, Nicholl:2021rcr}, while others perform full simulations of photon transport through the ejecta \cite{Kasen:2017sxr,Bulla:2022mwo, Korobkin:2020spe, Wollaeger:2021qgf, Curtis:2021guz, Pognan:2023qhw} which are much more computationally expensive. The dimensionality of the modeling also differs: some assume an isotropic distribution for 1D modeling, whereas others adopt more complex 2D or 3D descriptions, implementing angular variations in the ejecta structures. Additionally, models may either assume a single ejecta component with a given lanthanide fraction and associated opacities or incorporate multiple components (interacting or not) that differ in their lanthanide fraction and/or geometry. As a consequence of these different assumptions, models predict vastly different theoretical lightcurves, with differences between predicted magnitudes that can easily reach 1-3 mag~\citep{Heinzel:2020qlt}.

\paragraph{Physical assumption of r-process and heating rate, $\sigma_{mod}~<$~1~mag}

When choosing a single KN model architecture, its underlying physical assumptions come with their own systematic uncertainties.
Indeed, in order to model KN evolution with proper matter-radiation interactions, one needs to accurately describe the behavior of nuclear matter containing many r-process products, namely lanthanides, because these heavy elements highly contribute to the ejecta opacity in the optical and infrared energy ranges. The opacity is indeed a critical parameter to model the transition of the ejecta from a mostly opaque to a mostly transparent medium, which strongly correlates to the time of peak magnitude in the lightcurves. 

However, the abundance of these heavy elements as produced by r-process is not precisely known yet, and their exact atomic structure is not well constrained \citep{2011PrPNP..66..346T, Metzger:2016pju, Cowan_2021}.
Competing atomic models \citep{Tanaka:2019iqp,Fontes:2019tlk,Kasen:2017sxr} therefore differ in their assumptions of matter-radiation interaction and thus in their predictions of the wavelength-dependent opacities of the ejecta matter, and of the heating rate of r-process production and the thermalization efficiency of the ejecta, which are also crucial in KN color effects. 
Using different lanthanide atomic datasets within the same kilonova modeling scheme, \cite{Brethauer:2024zxg} show sizable variations in the output lightcurves with~$\sim 1$~mag discrepancies that grow even bigger after a week. Moreover, different heating rate prescriptions can change the magnitude at peak by up to 1 mag, but also impact the whole lightcurves' temporal evolution \citep{Sarin:2024tja}.

\subsubsection{Lightcurves predictions using surrogate models, associated uncertainty}
\label{ssub:Surrogate}

Some KN models as discussed in previous section can be extensively computationally expensive. For example, in photon-transport schemes, such as \texttt{POSSIS} \citep{Bulla:2019muo, Bulla:2022mwo}, each simulation composed with a fixed combination of ejecta properties and geometry can take hundreds of CPU hours or more. However, Bayesian inference sampling algorithms requires computing lightcurves for thousands of parameter combinations in order to converge to posterior distributions. 

Therefore, in order to reduce computing resources, a grid of already-simulated lightcurves\footnote{Or flux densities for models not restricted to a few filter bandpasses.} is generated, regularly spaced in the model parameters hypervolume, and interpolation between them is performed to quickly approximate the lightcurves for any combination of model parameters in their allowed range. This interpolated approximation is called the surrogate model.

\paragraph{Mitigate Poisson noise in photon transport} In the case of photon-transport models, the first step to build the grid is to smooth out the modeled lightcurves. Indeed, the output of photon-transport simulations is affected by statistical noise, mainly Poisson noise from counting the photon packets collected in each time- and energy- bin. More specifically, the Poisson noise is usually more pronounced at early and late times when the kilonova is fainter and thus photons are more sparse. A Principal Component Analysis (PCA) of the matrix of magnitudes as a function of time step is performed in order to average out these fluctuations, as well as reduce dimensionality of the data to optimize computing cost. This decomposition is performed for each filter that the surrogate model needs to be trained on.

\begin{figure}
    \centering
    \includegraphics[width=0.9\columnwidth]{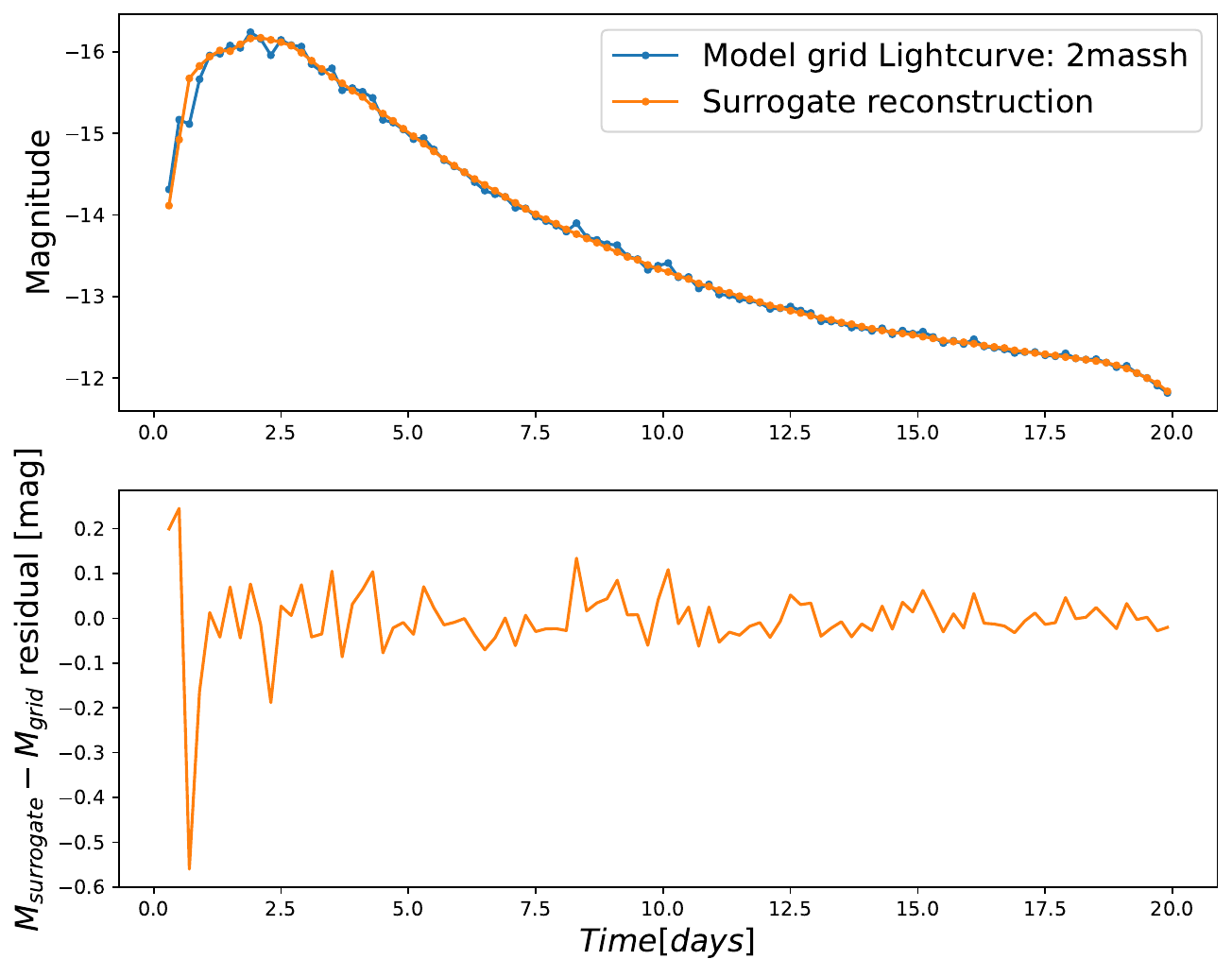}
    \caption{\textbf{Top:} Example of a representative \textsc{Bu2019lm} lightcurve in $H$-band simulated by \texttt{POSSIS} (corresponding to ejecta parameters $M_{\rm ej}^{\rm dyn} = 0.005 M_\odot$, $M_{\rm ej}^{\rm wind} = 0.09 M_\odot$, $\phi = 45$ degrees and $\cos(\theta)=0.6$) in absolute magnitude, in blue, and its corresponding reconstruction from the first 10 PCA coefficients, in orange. \textbf{Bottom:} Residual difference between the two curves. This PCA reduction seems to smooth out a Poisson noise of order 0.1 mag, and this behavior is consistently found across the whole \textsc{Bu2019lm} model grid.}
    \label{fig:Poisson_noise_SVD_smooting}
\end{figure}

The number of PCA components used is chosen to compromise between computing resources and accuracy: the more components are kept, the more features of the lightcurves are embedded in the surrogate, but the heavier is the memory footprint, slowing down the computing times when applying the surrogate.
For the \textsc{Bu2019lm} grid of \texttt{POSSIS} lightcurves, we found that keeping the first ten PCA components (for each filter) is enough to reproduce the lightcurves faithfully (near-zero average deviation in each 1-day time-bin, see Figure~\ref{fig:Poisson_noise_SVD_smooting}). More PCA components may be needed for KN model grids with a number of input model parameters higher than the four of \textsc{Bu2019lm}'s model. 

\paragraph{Interpolation and ensuing errors, $\sigma_{mod}$ $<$ 0.3 mag }
We now have a \{model parameters $\rightarrow$ PCA coefficients\} mapping defined on the simulated grid points, that we need to extend to the full parameter space.
We can either perform a direct analytical interpolation but it becomes very expensive in computing time and memory when the model has many parameters; or train a neural network to serve as a surrogate model.
For instance, the implementation of the four-parameter \textsc{Bu2019lm} model uses a Gaussian Process Regression interpolation, which needs gigabytes of memory to store and to run\footnote{This is because the size of the model roughly scales with $N_{sim}\times N_{filts}\times N_{coef}$ with $N_{sim}$ the number of simulations in the training grid, $N_{filt}$ the number of filters to be trained, and $N_{coef}$ the number of PCA coefficients used. For \textsc{Bu2019lm} this is $1586\times26\times10$}. Higher-dimensional model grids like the seven-parameter \textsc{Bu2025MLP} \citep{Koehn:2025zzb} would need much larger memory footprints with a similar implementation\footnote{Both because the number of simulations in the training grid would scale exponentially with the dimensionality and because the number of PCA coefficients needed to reproduce the nuances of a more complex model would also increase.}, which would be computationally impractical and expensive; which is why neural network surrogates are used instead to cut down on computing resources\footnote{Incidentally, carefully trained neural networks additionally seem to achieve more faithful reconstructions of their training models, with smaller surrogacy errors than Gaussian Process Regression for high-dimensional models.}.

\begin{figure}
    \centering
    \includegraphics[width=0.9\columnwidth]{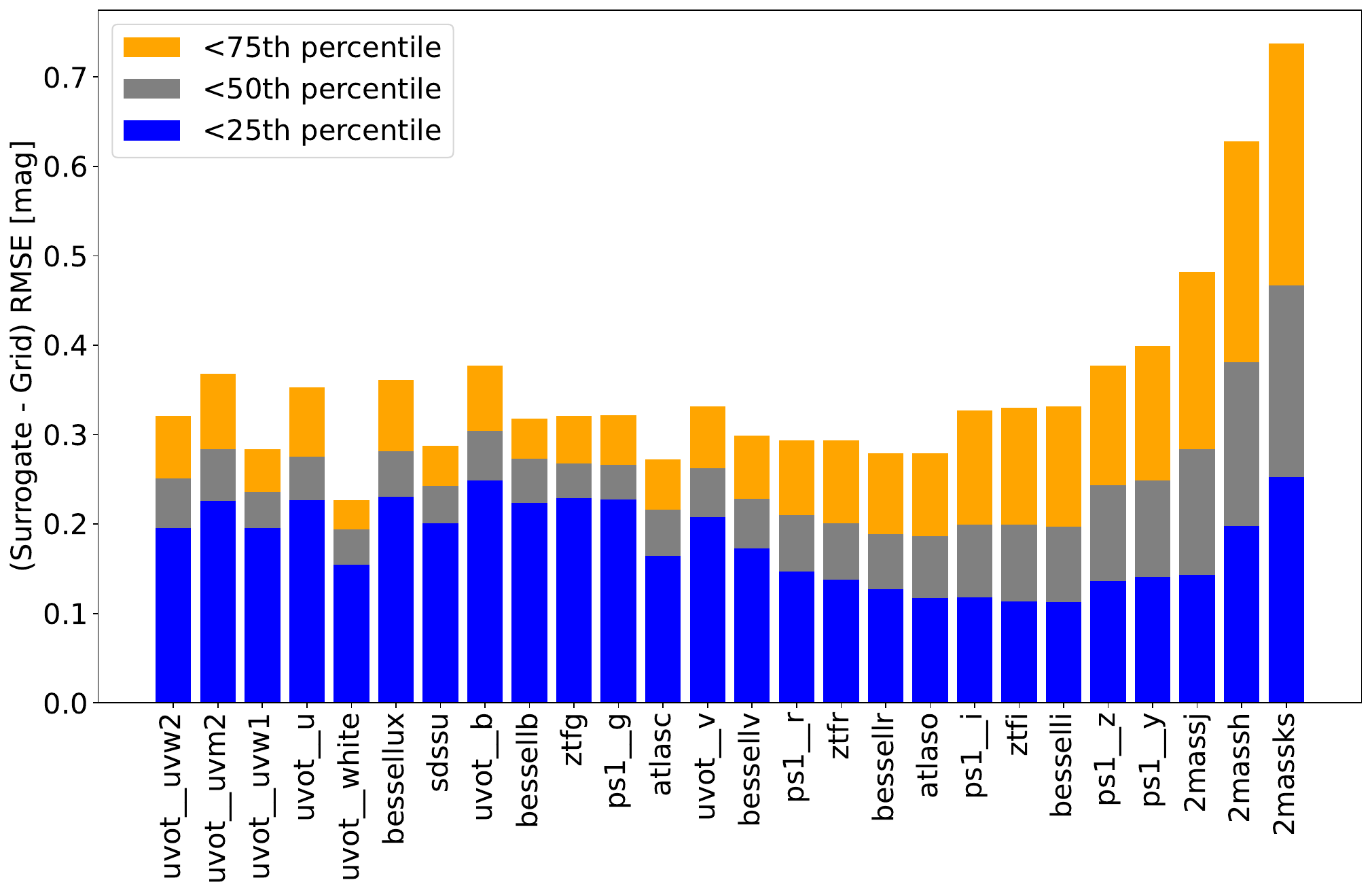}
    \caption{Quartiles of the distribution of the RMS error (in $\rm mag$) between the simulated model and its reconstruction using a Gaussian Process Regression interpolator, for all the \textsc{Bu2019lm} grid lightcurves, and across different filters (in ascending order of effective wavelength $\lambda_{eff}$).}
    \label{fig:surrogate_inaccuracy}
\end{figure}

Depending on the model's dimensionality and the size and structure of the interpolation network, the surrogate model can exhibit limited accuracy, introducing errors of up to tenths of mag. 
For instance, in the \textsc{Bu2019lm} implementation used in this work, we quantify the average deviation between model grid lightcurve and surrogate reconstruction with the Root Mean Square Error: $RMSE = \sqrt{\frac{1}{N} \sum_{i=1}^N (m_i - m_i^\star)^2}$; where $m_i$ are the grid magnitudes, $m_i^\star$ the magnitudes predicted by the surrogate, and $N$ the total number of time points. For most filters, we find a $RMSE<0.3$ mag for the majority of the grid simulations, as shown in Figure~\ref{fig:surrogate_inaccuracy}. 
The exception to this are the 2massh and 2massks filters, for which the discrepancies are almost twice as large, due to a very high level of numerical noise in the \texttt{POSSIS} lightcurves at early times for these filters. Across all filters, the median\footnote{Median taken from the set of $RMSE$ values computed for all the lightcurves in the training grid of \textsc{Bu2019lm}} $RMSE$ obtained here is 0.25 mag.

Recent implementations of \textsc{Bu2019lm} with neural network surrogates \citep{Jhawar:2024ezm} obtain lower interpolation errors, they quantify their median $RMSE$ at 0.1 mag. 
For higher-dimensional models however, recent surrogates still exhibit median $RMSE$ values of 0.2-0.3 mag~\citep{Koehn:2025zzb}.

\section{Parameter-wise partial information gains}
\label{app:paramwisegains}

To further breakdown the performance of parameter inferences, we wish to quantify the resulting information gain for each individual parameter's distribution. We recall the definition of the Kullblack-Leibler divergence, that measured the overall information gain:

\begin{equation}
    D_{KL}(P||\pi) = \int_{\vec{\theta} \in \text{Parameter Space}} d\vec{\theta} P(\vec{\theta}) \log\frac{P(\vec{\theta})}{\pi(\vec{\theta})}
\end{equation}

Let us define the marginalized posterior distribution for each parameter:
\begin{equation}
    \tilde{P}_i (x_i) = \int dx_1 \dots dx_{i-1}dx_{i+1}\dots dx_n P(x_1\dots x_n)
\end{equation}

In our study, we approximate marginalized distributions as the density histogram of parameter values in our posterior sample sets. These sets typically contain thousands of samples, we arbitrarily choose to consider histograms with 20 equal-width bins.\\

In the case where all parameters $\vec{\theta}=\{p_1,\dots , p_n\}$ have independent priors: $\pi(\vec{\theta}) = \prod_{i=1}^n \pi_i(x_i)$, the K-L divergence can then be rewritten as:

\begin{equation}
\begin{aligned}
    &D_{KL}(P||\pi) = \int_{\vec{\theta}} d\vec{\theta} P(\vec{\theta}) \log(\frac{P(\vec{\theta})}{\prod_i \tilde{P}_i} \frac{\prod_i \tilde{P}_i}{\prod_i \pi_i})\\
    &= \int_{\vec{\theta}} d\vec{\theta} P(\vec{\theta}) \log\frac{P(\vec{\theta})}{\prod_i \tilde{P}_i} +\sum_i \log \frac{\tilde{P}_i(x_i)}{\pi_i(x_i)}
    = M.I. + \sum_i D_i
\end{aligned}
\end{equation}

where $M.I. = \int_{\vec{\theta}} d\vec{\theta} P(\vec{\theta}) \log\frac{P(\vec{\theta})}{\prod_i \tilde{P}_i}$ is called the \textit{mutual information} and 

\begin{equation}
\begin{aligned}
    D_i = \int_{\vec{\theta}} P(\vec{\theta}) \log \frac{\tilde{P}_i(x_i)}{\pi_i(x_i)} = \int_{x_i} \tilde{P}_i(x_i) \log \frac{\tilde{P}_i(x_i)}{\pi_i(x_i)} \simeq \frac{1}{N_{\rm samples}}\sum_{\rm posterior~samples} \log\frac{\tilde{P}_i(x_i)}{\pi_i(x_i)}
\end{aligned}
\end{equation}
is each parameter's \textit{partial information}, analogue of the K-L divergence on the marginalized distribution.\\

\begin{figure*}
    \centering
    \includegraphics[width=0.56\columnwidth]{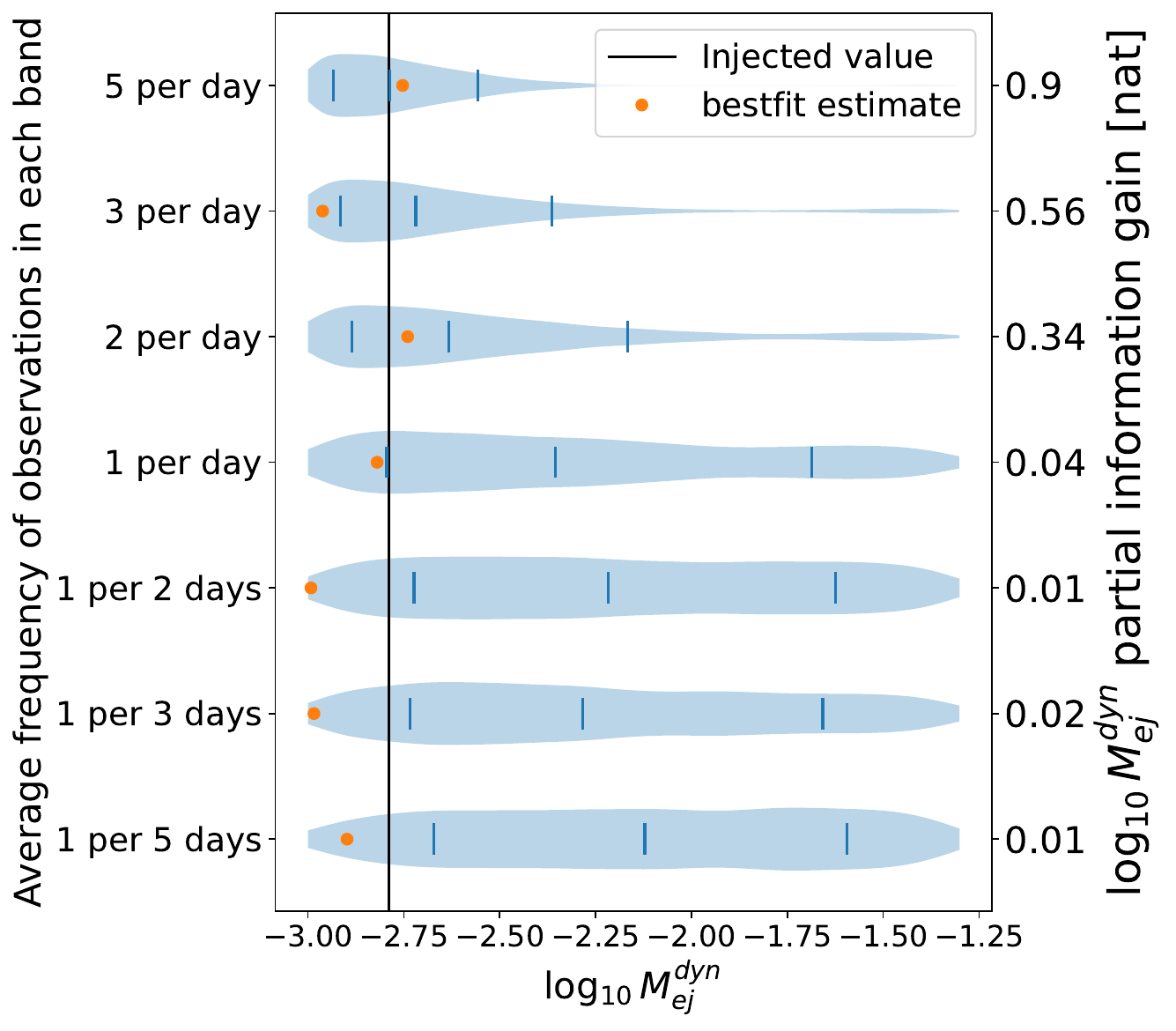}
    \includegraphics[width=0.43\columnwidth,trim = 150 0 0 0, clip]{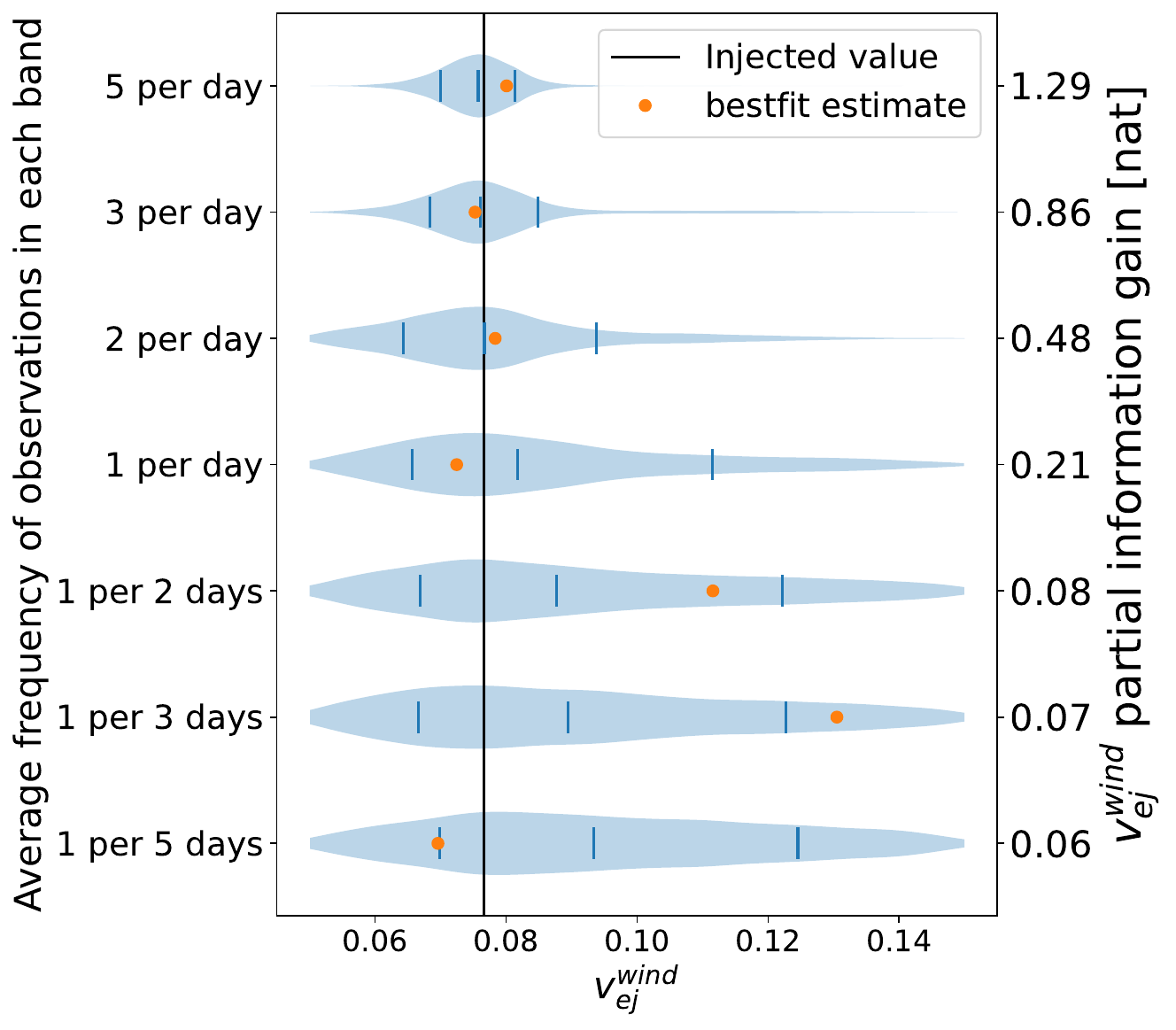}
    \includegraphics[width=0.56\columnwidth]{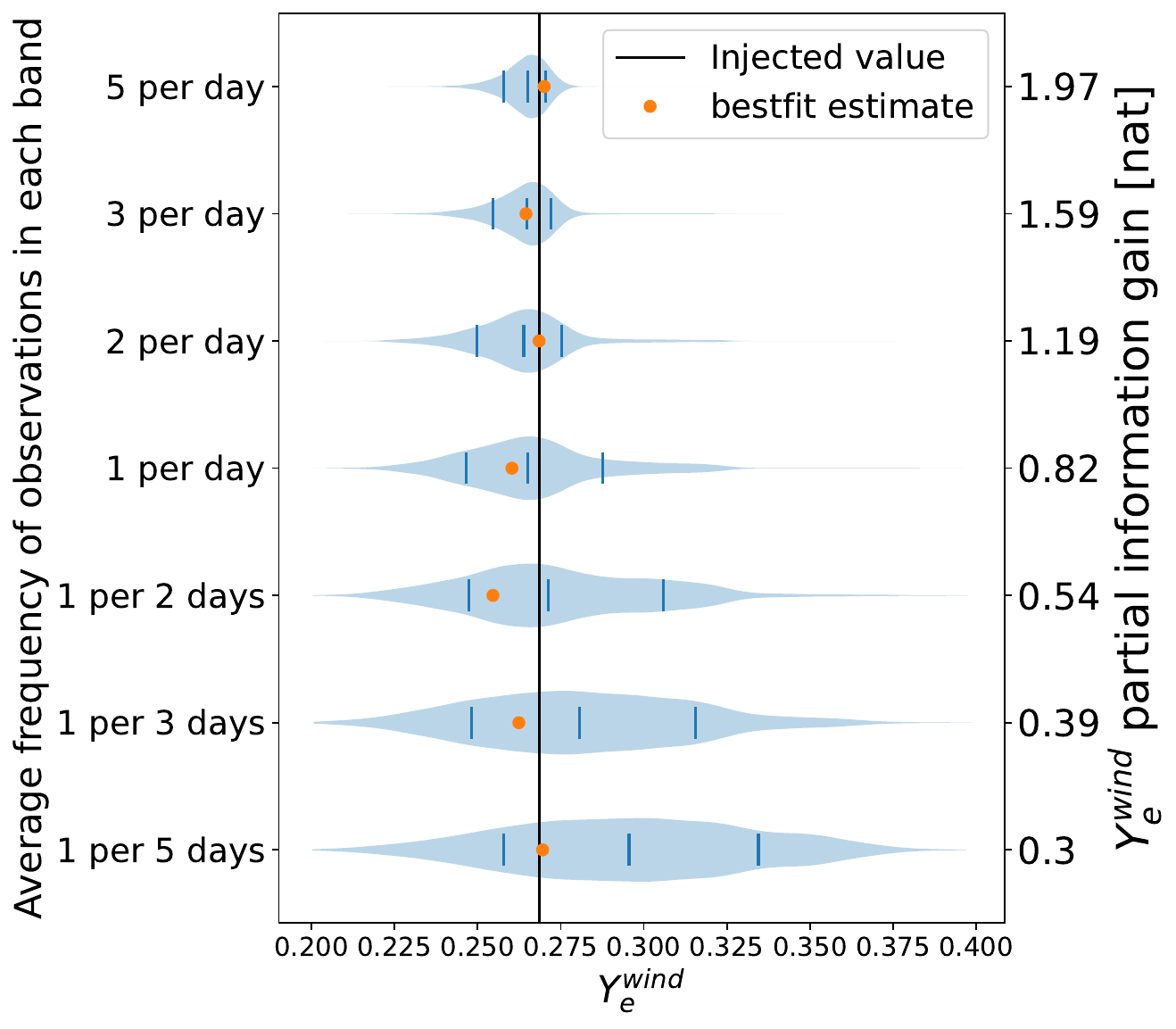}
    \includegraphics[width=0.43\columnwidth,trim = 150 0 0 0, clip]{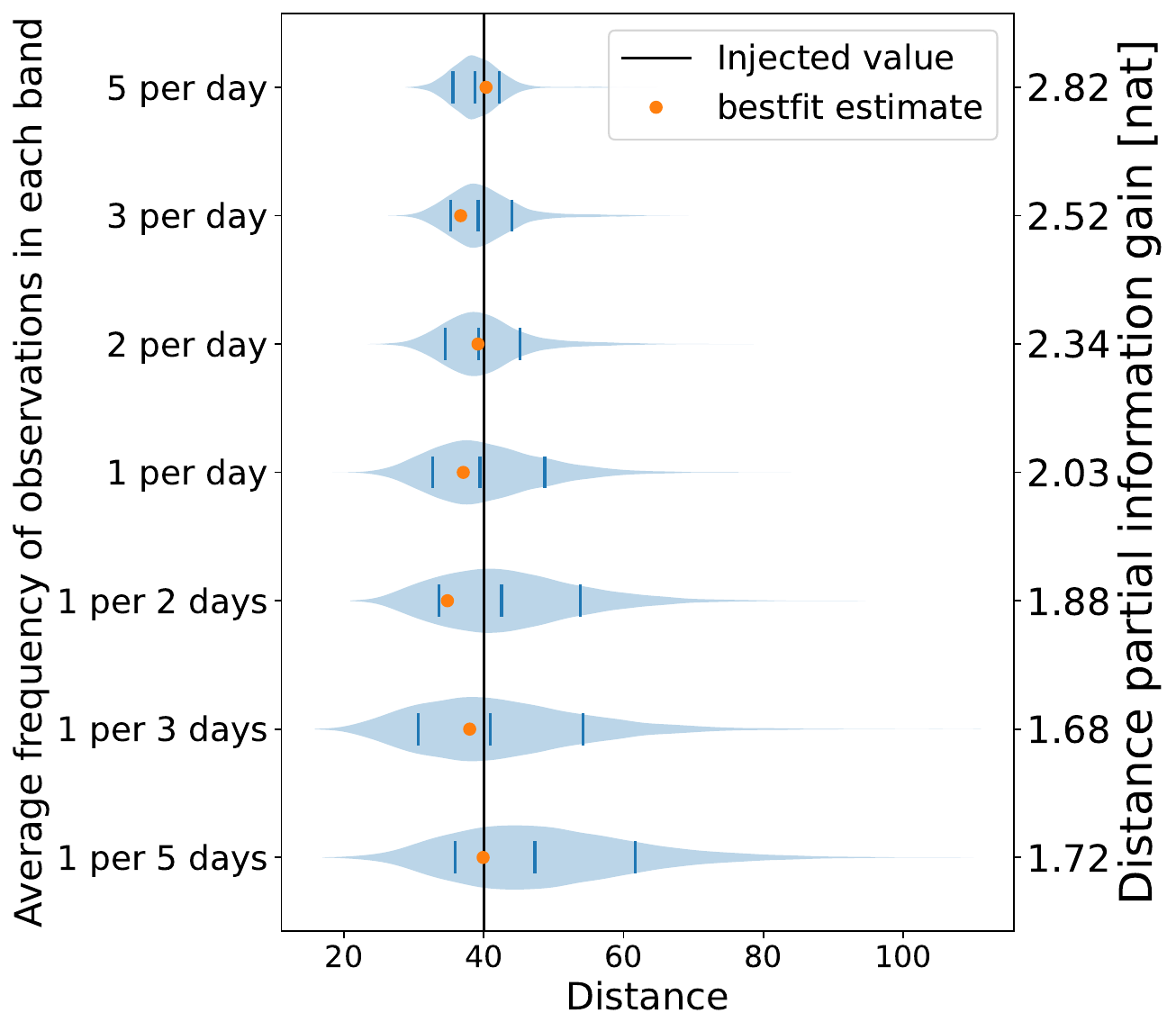}
    \caption{Posterior distributions (\textbf{Top Left:} Dynamical ejecta mass. \textbf{Top Right:} Wind ejecta velocity. \textbf{Bottom Left:} Wind ejecta electron fraction. \textbf{Bottom Right:} Luminosity distance) and corresponding partial information gains, obtained when analyzing \textbf{Injection5} lightcurves ($grizyJHK$) with \textsc{Bu2025MLP} and prior set \textbf{A'1} (from Table~\ref{tab:PriorsBu25}); varying the cadence using observations more and more sparse in time. For each distribution, the median and the 16th and 84th percentiles (68\% credible interval) are highlighted in blue.}
    \label{fig:KLcadence25}
\end{figure*}

We showcase in Figure~\ref{fig:KLcadence25} the posterior distributions and partial information gain for different parameters and different simulated cadences analysed with \textsc{Bu2025MLP}. We notice a transition between low-gain configurations, where the posterior roughly recovers the uniform prior, and higher-gain configurations where posteriors are noticeably tighter. We empirically identify a threshold around partial gains of 0.2 nats, above which posterior distribution become significantly constrained (in particular the posterior 68\% confidence interval size falls below $\sim 50$\% of the prior range). \\

\begin{figure*}
    \centering
    \includegraphics[width=0.435\columnwidth]{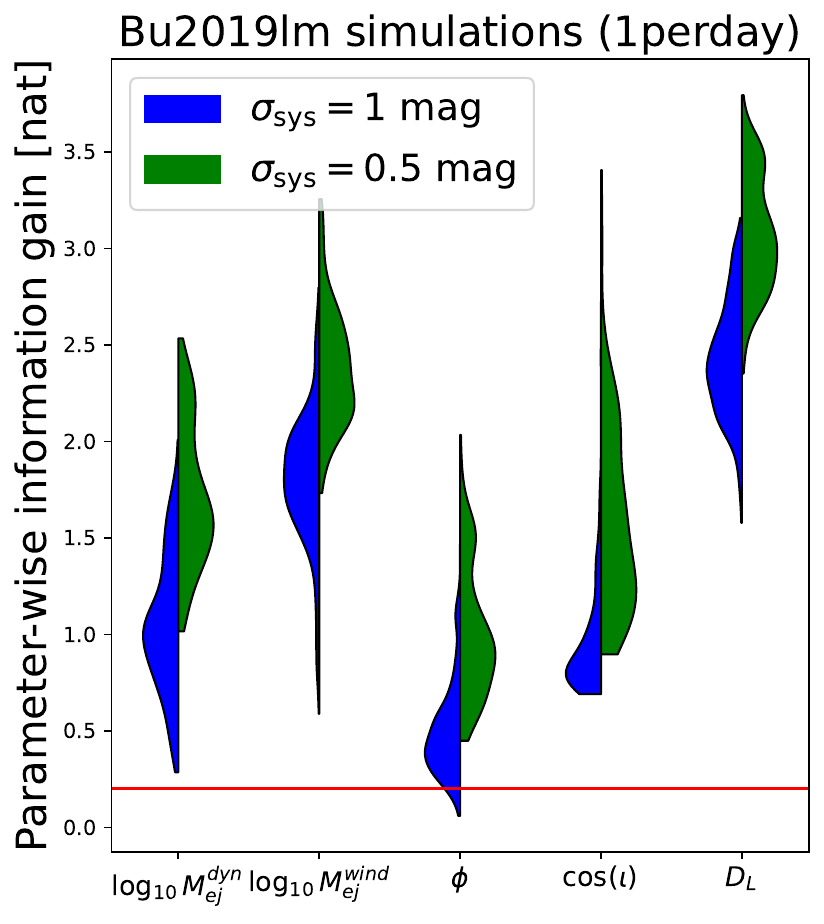}
    \includegraphics[width=0.55\columnwidth]{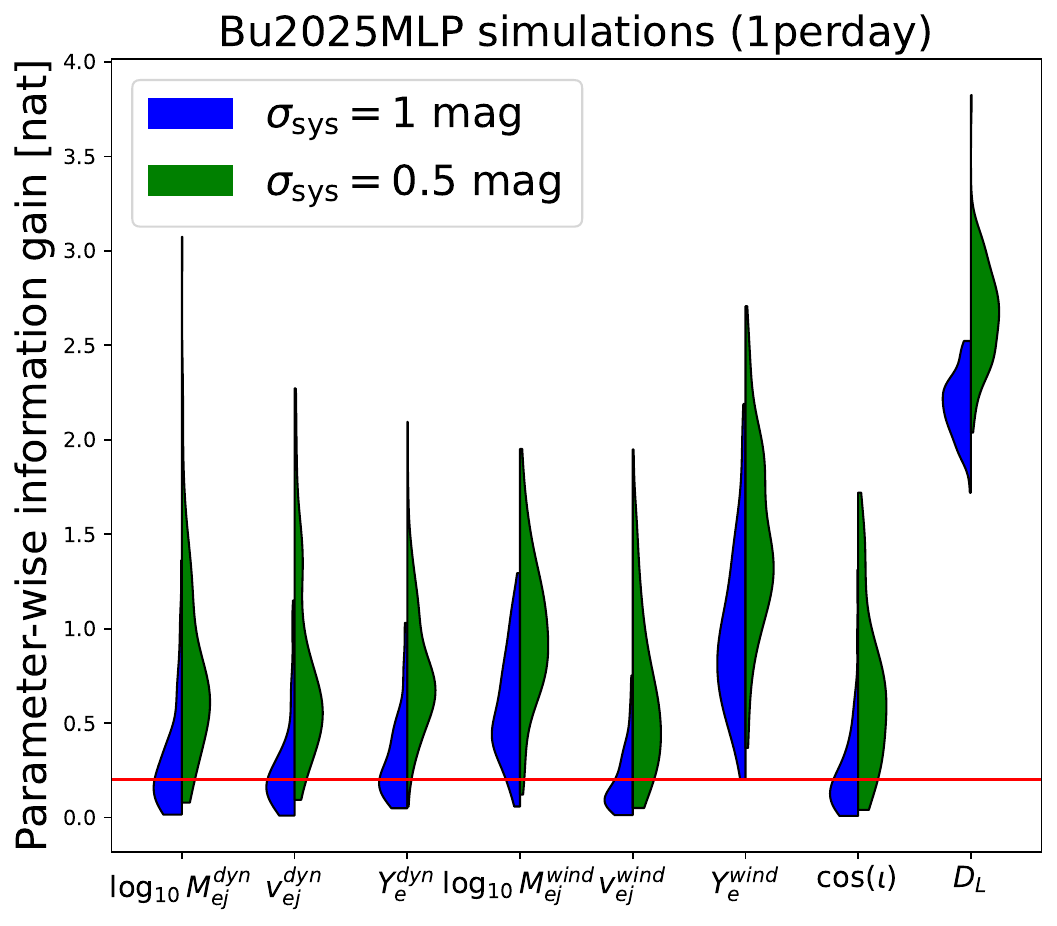}
    \caption{\textbf{Left:} Distribution of each \textsc{Bu2019lm} parameter's partial information gain for the 22x20 \textbf{1perday} lightcurves of the \textbf{ManySims} set, in \textsc{Bu2019lm} inferences using prior sets \textbf{A1} and \textbf{A0.5}. \textbf{Right:} Distribution of each \textsc{Bu2025MLP} parameter's partial information gain for the 100 \textbf{1perday} lightcurves of the \textbf{25MLPSims} set, in \textsc{Bu2025MLP} inferences using prior sets \textbf{A'1} and \textbf{A'0.5}. A red horizontal line denotes the 0.2 nats threshold below which a parameter's distribution gained no significant constraint.}
    \label{fig:ParGainViolins}
\end{figure*}

We show in Figure~\ref{fig:ParGainViolins} the expected partial gains of each model parameters for \textsc{Bu2019lm} and \textsc{Bu2025MLP} analyses of simulated events with \textbf{1perday} $grizyJHK$ cadences. We notice in the $\sigma_{\rm sys}=1$ mag case that all of the \textsc{Bu2019lm} parameters are typically constrained more strongly than our 0.2-nat significance threshold, whereas for \textsc{Bu2025MLP} this is only reliably the case for the wind ejecta mass and electron fraction and the luminosity distance. All parameters however pass the significance threshold in the $\sigma_{\rm sys}=0.5$ mag case.\\

\begin{figure*}
    \centering
    \includegraphics[width=0.49\columnwidth]{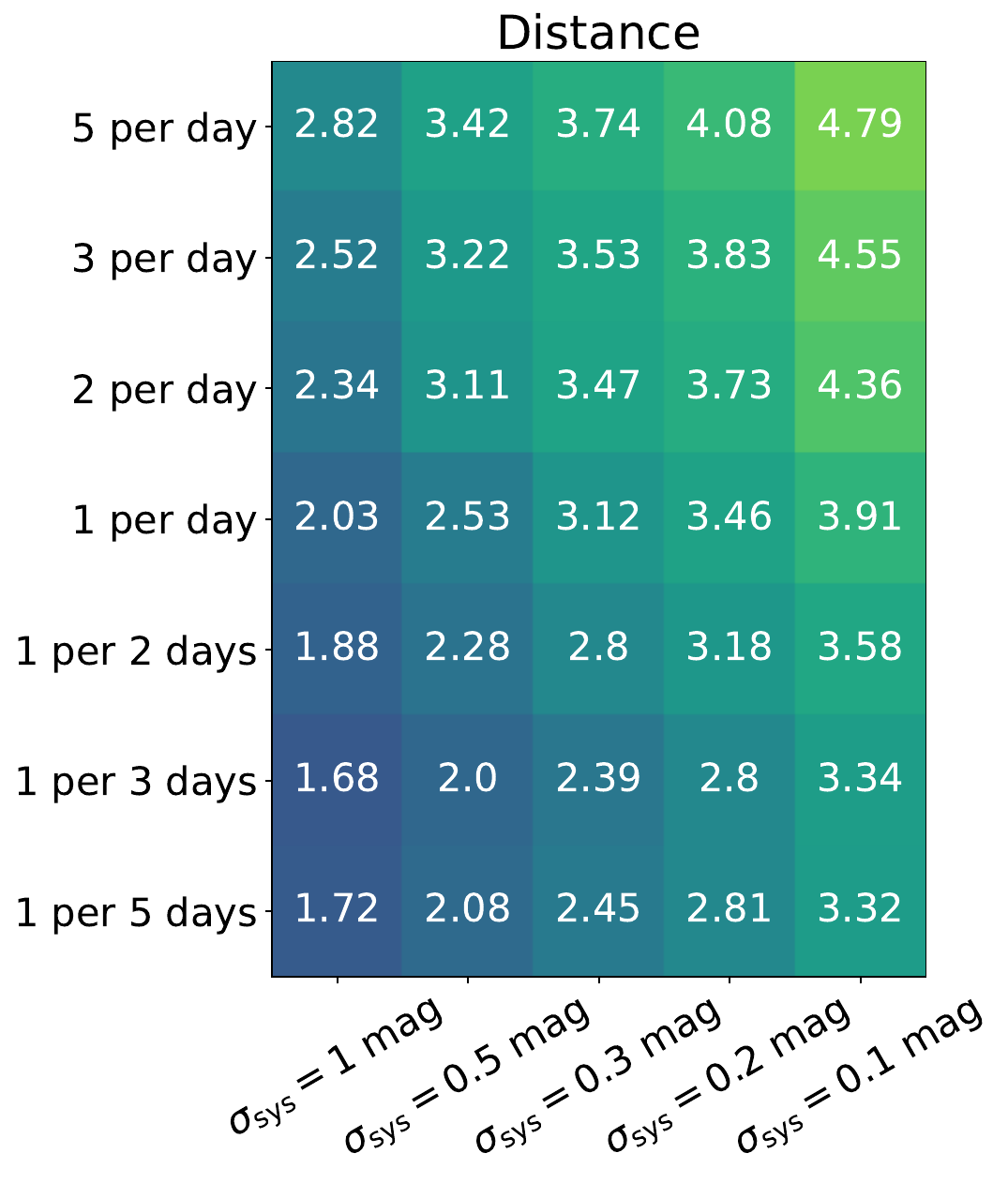}
    \includegraphics[width=0.49\columnwidth]{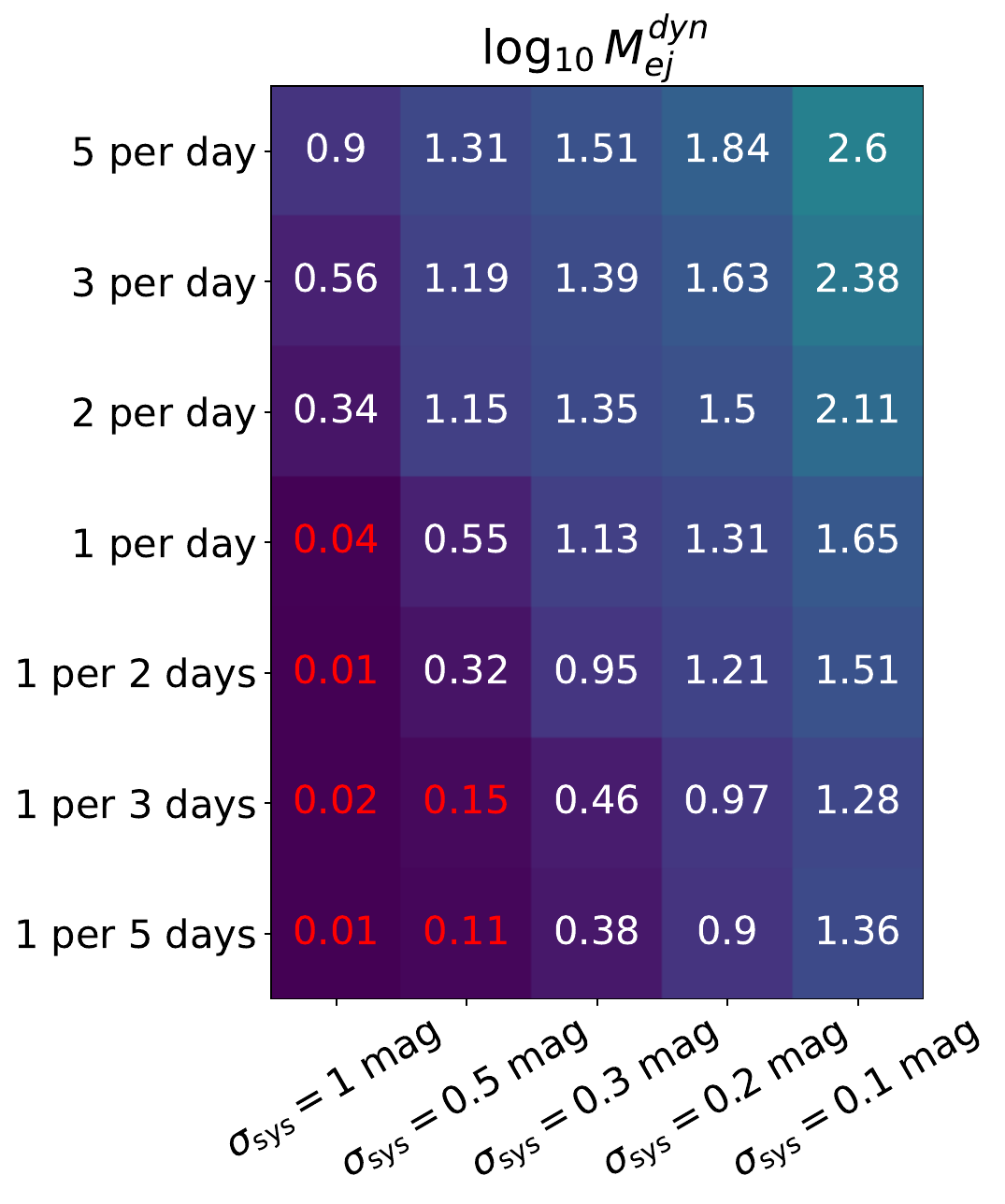}
    \includegraphics[width=0.49\columnwidth]{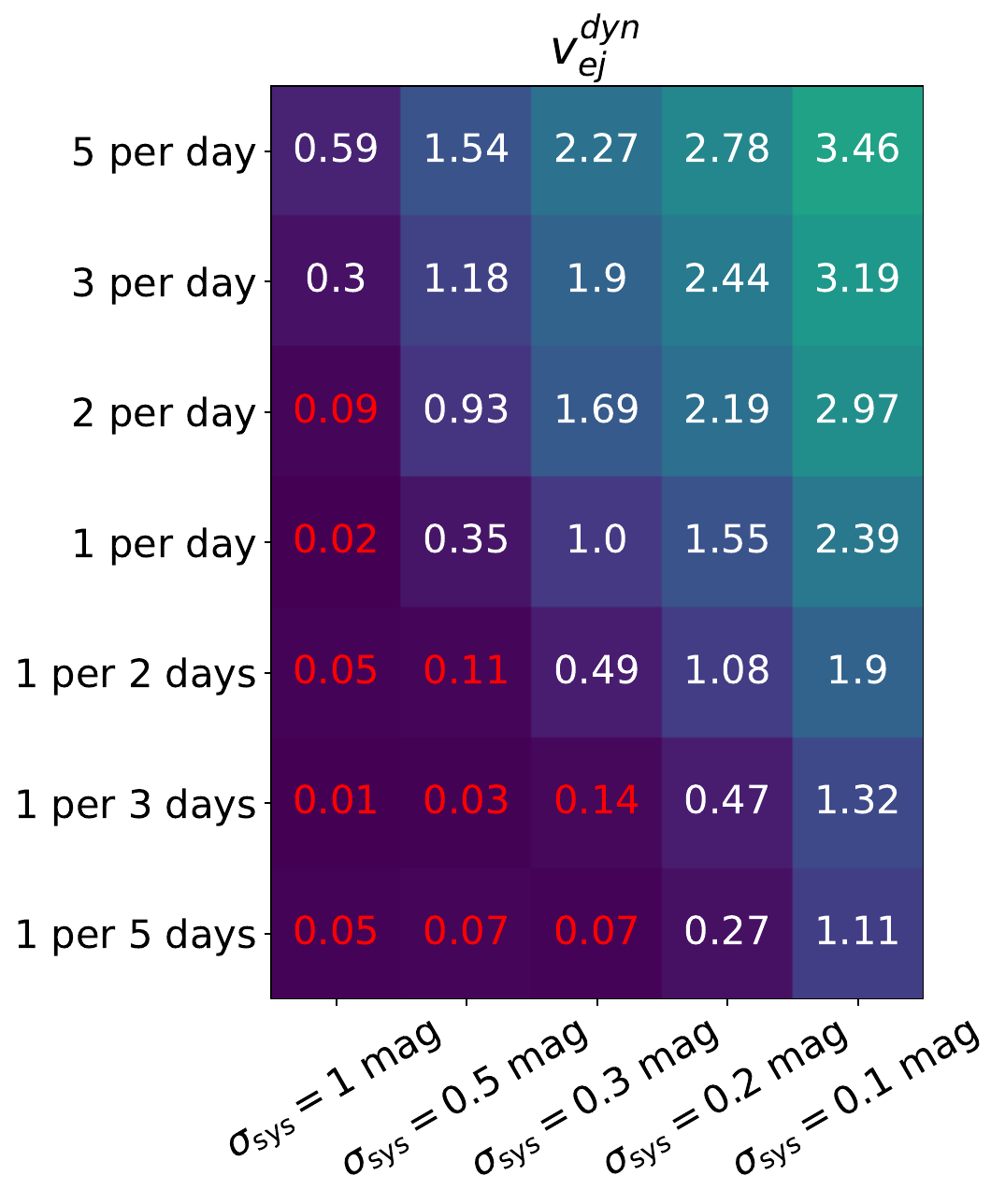}
    \includegraphics[width=0.49\columnwidth]{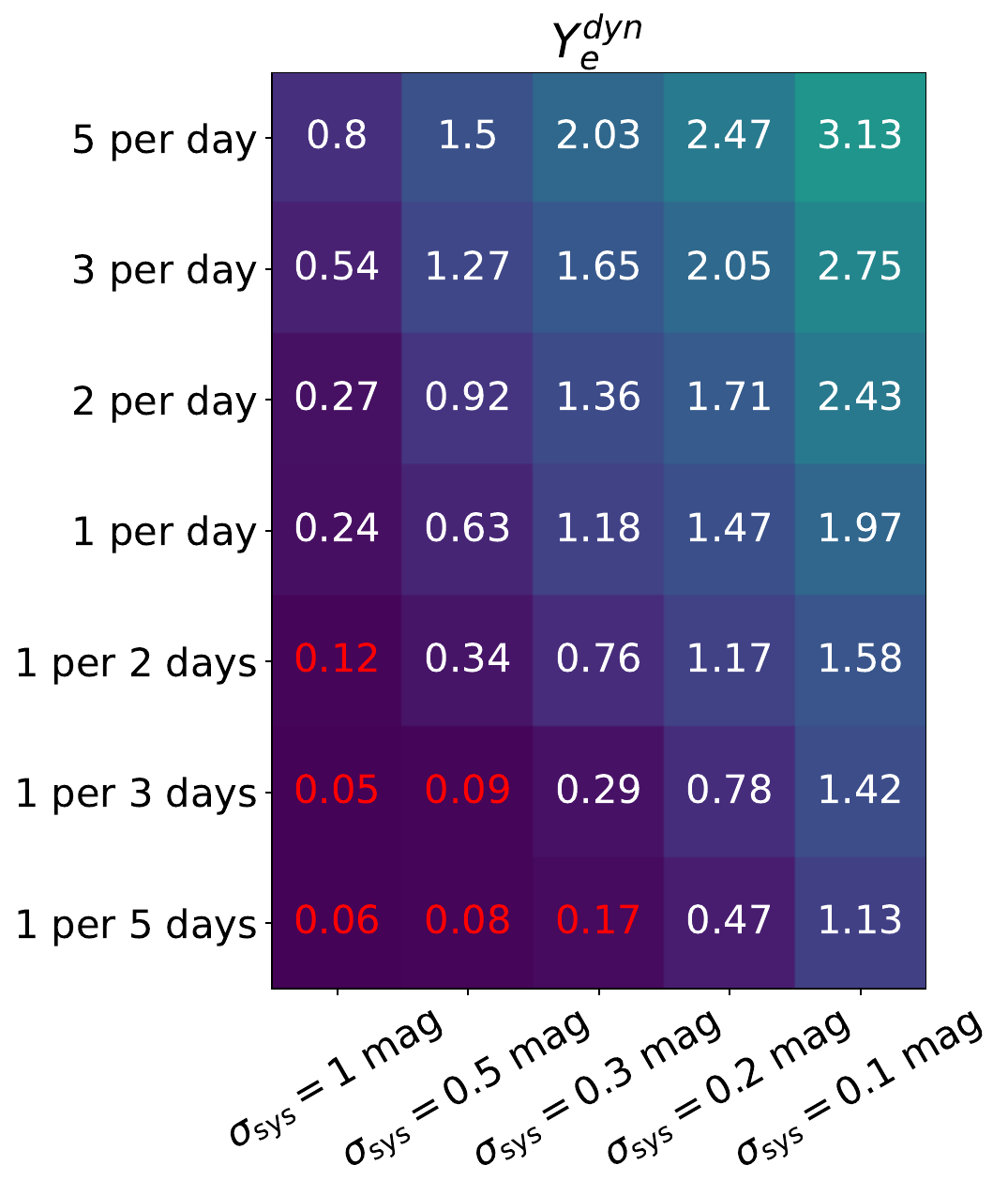}
    \caption{Partial information gains when analyzing \textbf{Injection5} lightcurves with \textsc{Bu2025MLP} prior set \textbf{A'1} (from Table~\ref{tab:PriorsBu25}); varying the observation cadences and the systematic uncertainty. We highlight with in red the configurations in which the partial gains are below 0.2 nats, i.e. when that parameter's distribution gained no significant constraint from the inference.}
    \label{fig:25Partial}
\end{figure*}

\begin{figure*}
    \centering
    \includegraphics[width=0.49\columnwidth]{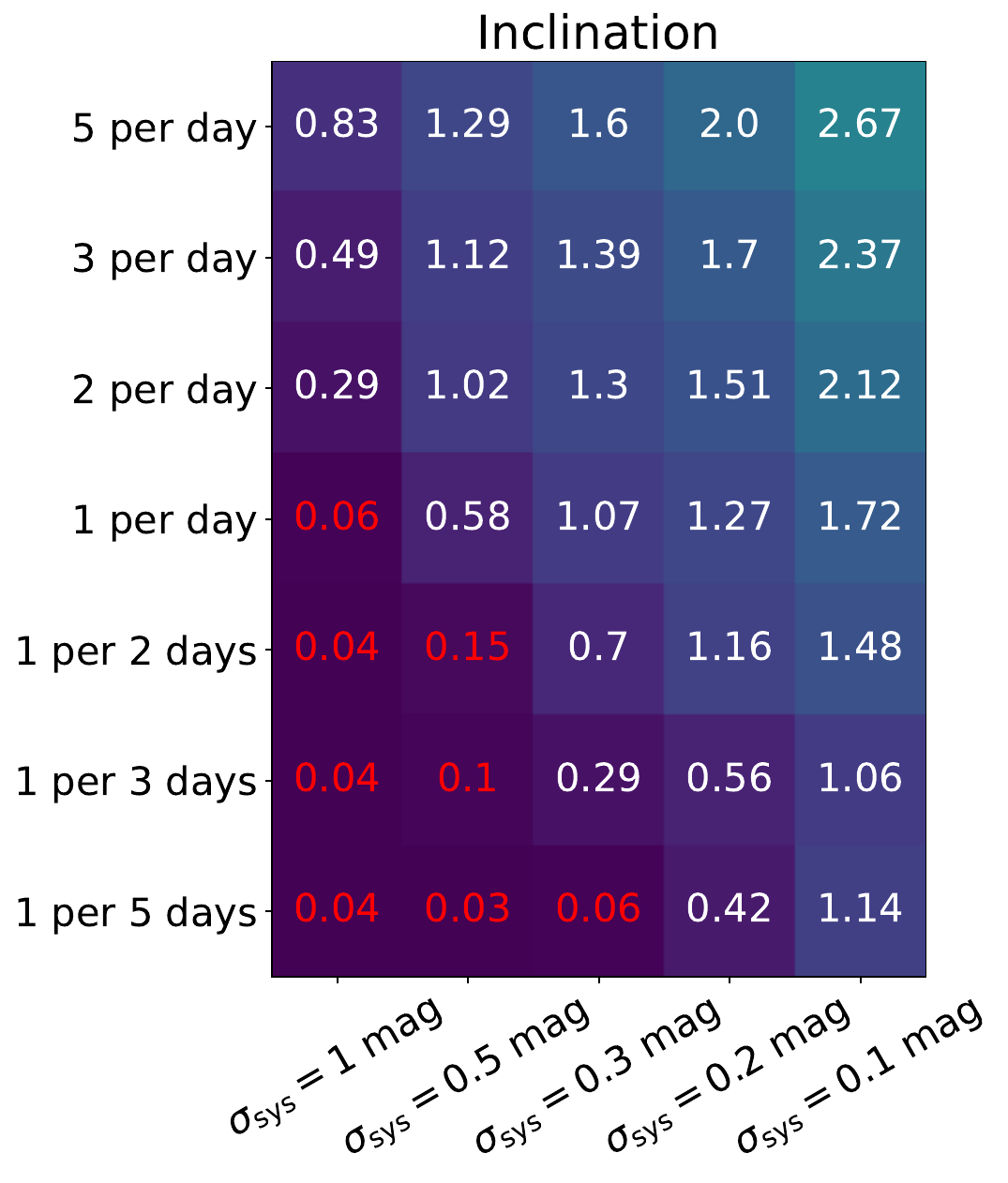}
    \includegraphics[width=0.49\columnwidth]{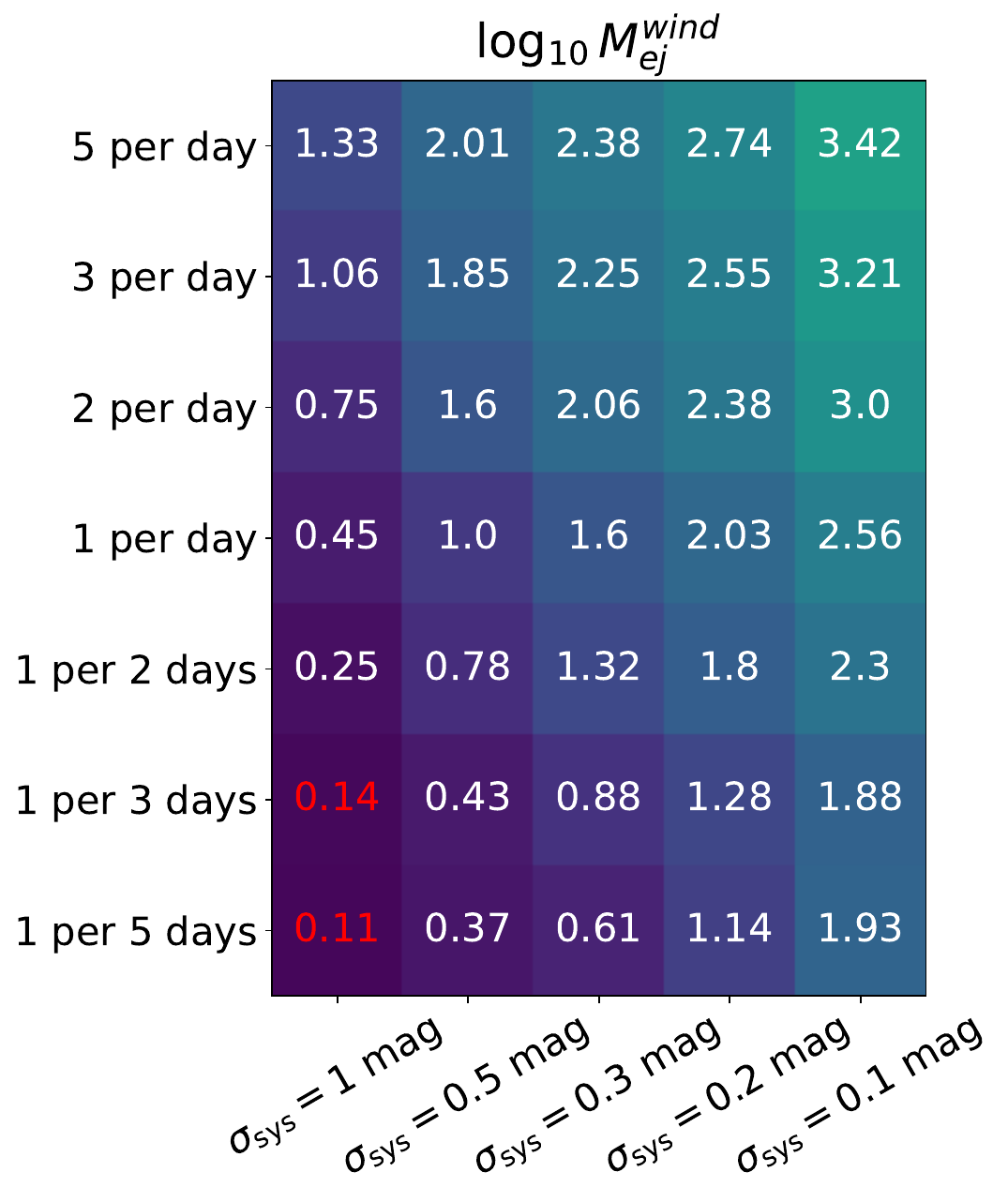}
    \includegraphics[width=0.49\columnwidth]{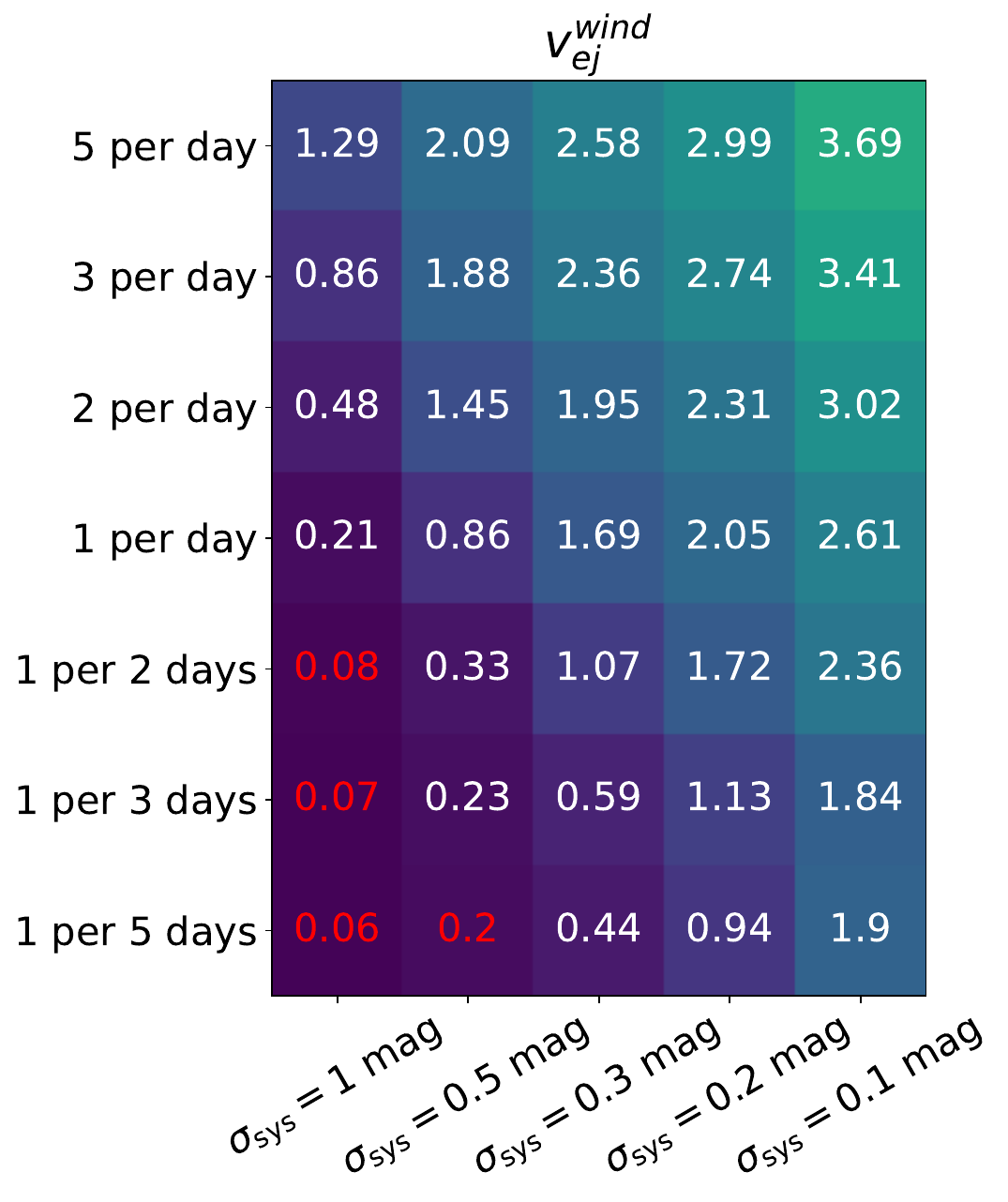}
    \includegraphics[width=0.49\columnwidth]{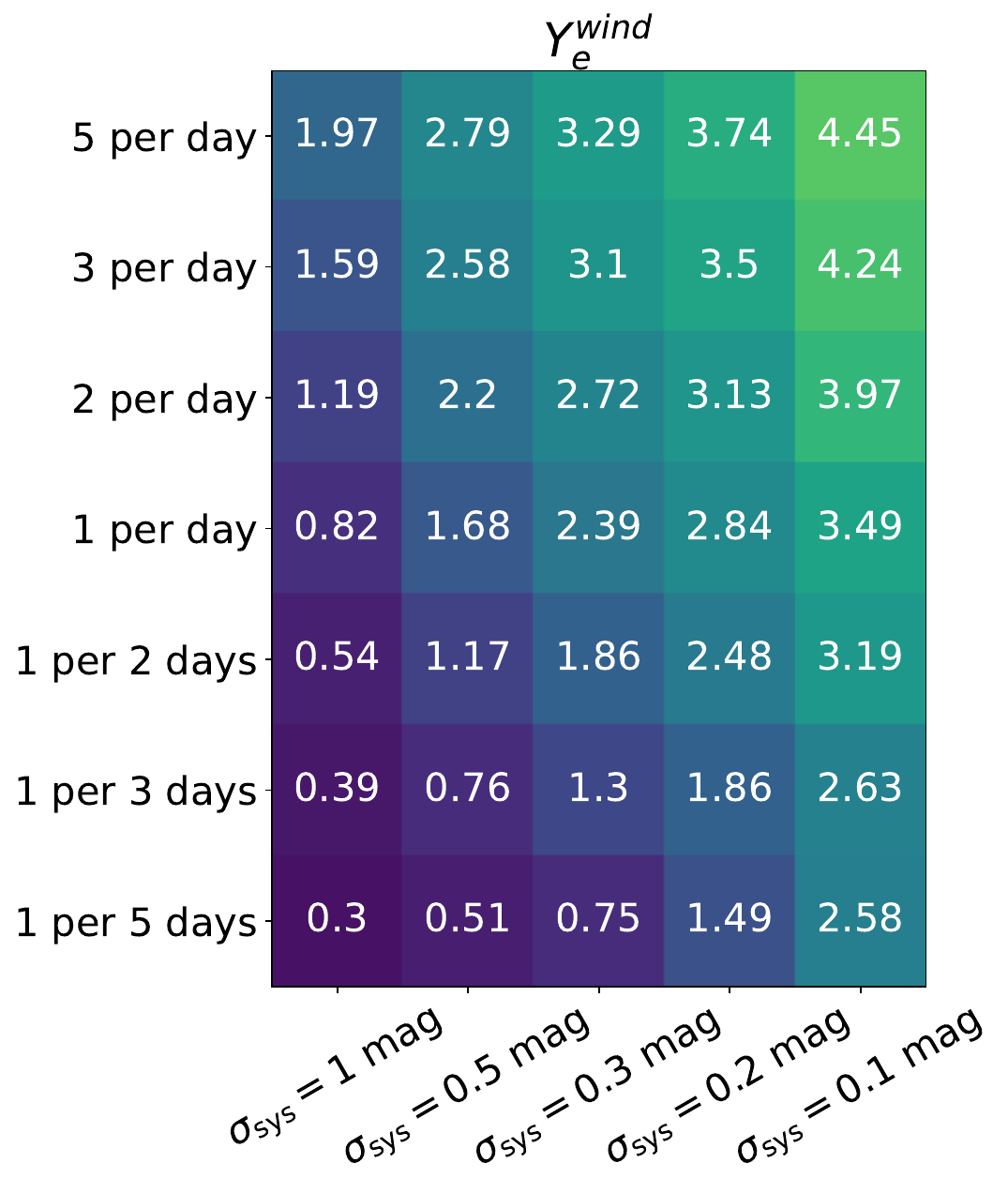}
    \caption{Partial information gains when analyzing \textbf{Injection5} lightcurves with \textsc{Bu2025MLP} prior set \textbf{A'1} (from Table~\ref{tab:PriorsBu25}); varying the observation cadences and the systematic uncertainty. We highlight with in red the configurations in which the partial gains are below 0.2 nats, i.e. when that parameter's distribution gained no significant constraint from the inference.}
    \label{fig:25Partial2}
\end{figure*}

We detail in Figures \ref{fig:25Partial} and \ref{fig:25Partial2} typically expected partial gains of \textsc{Bu2025MLP} model parameters for $grizyJHK$ inference analyses using varying cadences and $\sigma_{\rm sys}$ values.

\bibliography{references}{}
\bibliographystyle{JHEP}

\end{document}